\documentclass[traditabstract]{aa} % for the abstract without structuration 
\usepackage{graphicx}
\usepackage{aalongtable,lscape}
\usepackage{exscale}
\usepackage{amsmath}
\usepackage{txfonts}
\usepackage{natbib}
\newcommand\arcdeg{\mbox{$^\circ$}}%
\newcommand\farcsec{\mbox{$.\!\!^{\prime\prime}$}}
\newcommand\farcmin{\mbox{$.\!^{\prime}$}}%
\begin{document}

\title{Dynamics of the NGC\,4636 globular cluster system II  \thanks{Based on observations made with ESO Telescopes at the  Paranal Observatories under programme ID 075.B-0762}}
\subtitle{Improved constraints from a large sample of globular cluster velocities}
  
\author{Y.~Schuberth
    \inst{1,2}
    \and
    T.~Richtler \inst{2}
    \and 
   M.~Hilker \inst{3}
    \and
    R.~Salinas \inst{2,4}
    \and
    B.~Dirsch \inst{5}   
    \and{S.S~Larsen} \inst{6}      
}

\offprints{tom@astroudec.cl}

\institute{Argelander-Institut f\"ur Astronomie,
Universit\"at Bonn 
Auf dem H\"ugel 71, D-53121 Bonn, Germany 
\and
Universidad de Concepci\'on, Departamento de Astronom\'{\i}a, Casilla
160-C, Concepci\'on, Chile 
\and  
European Southern Observatory,
Karl-Schwarzschild-Str.~2, D-85748 Garching, Germany 
\and  
Finnish Centre for Astronomy with ESO (FINCA), University of Turku,
V\"ais\"al\"antie 20, FI-21500 Piikki\"o, Finland
\and
Friedrich-Ebert-Gymnasium Bonn, Ollenhauerstr.\,5, 53113 Bonn, Germany
\and
Department of Astrophysics, IMAPP, Radboud University Nijmegen, PO Box 9010, 6500 GL Nijmegen, The Netherlands
}  
\date{Received xx; accepted xx}
   
% \abstract{}{}{}{}{} % 5 {} token are mandatory 
\abstract{We present new radial velocities for 289 globular clusters around 
NGC 4636, the southernmost giant elliptical galaxy of the Virgo cluster. The 
data were obtained with FORS2/MXU at the Very Large Telescope. Together with 
data analysed in an earlier study (Schuberth et al.~2006), we now have a 
sample of 460 globular cluster velocities out to a radius of 12 arcmin (60 kpc) 
available -- one of the largest of its kind. This new data set also provides a 
much more complete angular coverage. Moreover, we present new kinematical data 
of the inner stellar population of NGC 4636. We perform an updated Jeans 
analysis, using both stellar and GC data, to better constrain the dark halo 
properties. We find a stellar M/L-ratio of 5.8 in the R-band, higher than 
expected from single stellar population synthesis. We model the dark halo by 
cored and cuspy analytical halo profiles and consider different anisotropies 
for the tracer populations. Properties of NFW halos lie well within the 
expected range of cosmological simulations. Cored halos give central dark 
matter densities, which are typical for elliptical galaxies of NGC 4636's 
luminosity. The surface densities of the dark matter halos are higher than 
those of spiral galaxies. We compare the predictions of Modified Newtonian 
Dynamics with the derived halo properties and find satisfactory agreement. 
NGC 4636 therefore falls onto the baryonic Tully-Fisher relation for
spiral galaxies. The comparison with the X-ray mass profile of Johnson et al. 
(2009) reveals satisfactory agreement only if the abundance gradient of hot 
plasma has been taken into account. This might indicate a general bias towards
higher masses for X-ray based mass profiles in all systems, including
galaxy clusters, with strong abundance gradients.}
\maketitle
 
\section{Introduction}
\label{sect:intro}
NGC 4636 is a remarkable elliptical galaxy. Situated at the Southern
border of the Virgo galaxy cluster, and thus not in a very dense
environment, its globular cluster system (GCS) exhibits a richness, which
one does not find in other galaxies of comparable luminosity in a
similar environments \citep{kissler94,dirsch05}.
Therefore, NGC 4636 offers the opportunity to employ globular clusters (GCs)
as dynamical tracers to investigate its dark halo out to large radii,
which is rarely given. In our first study (\citealt{schuberth06},
hereafter Paper\,I), we measured 174 GC radial velocities to confirm
the existence of a dark halo, previously indicated by X-ray analyses
(e.g.~\citealt{loewenstein03}) and tried to constrain its
mass. Paper\,I also gives a summary of the numerous works related to
NGC 4636 until 2006, which we do not want to repeat
here. \cite{dirsch05} presented a wide-field photometry in the
Washington system of the globular cluster system of NGC 4636, which
was the photometric base for Paper\,I as well as for the present
work. Also there, the interested reader will find a summary of earlier works.
Noteworthy peculiarities of NGC 4636 include the appearance of the supernova 1937A
(a bona--fide Ia event, indicative of presence of an intermediate--age
population), the high FIR--emission \citep{temi03,temi07b}, the
chaotic X--ray features in the inner region \citep{jones02,baldi09},
pointing to feed--back effects from supernovae or enhanced nuclear
activity in the past. Table\,\ref{tab:4636basic} summarises the basic
parameters related to NGC 4636.

Since NGC 4636 is X--ray bright, it has been the target of numerous
X--ray studies (for earlier work see
\citealt{matsushita98,jones02,loewenstein03} and references therein),
offering the possibility to compare X--ray based mass profiles with
stellar dynamical mass profiles, using a data base, which is not found
elsewhere for a non-central elliptical galaxy.  

The most recent X-ray based analysis of the mass profile is from 
\cite{johnson09} whose work is based on Chandra X-ray data.
The new feature in their work
is the inclusion of an abundance gradient in the X--ray gas. We will
show that, if this gradient is accounted for, the resulting X--ray
mass profile is in good agreement with the one derived from our GC
analysis, which may indicate the need to respect abundance gradients,
if present, in any X--ray analysis.

Two recent publications on the NGC 4636 GCs are from \cite{park10} and 
\cite{lee10}, who present about 100 new Subaru spectra of NGC 4636 GCs and 
discuss their kinematics in combination with the data from \cite{schuberth06}.    
For the sake of brevity, we postpone a detailed
comparison of the different samples to a later publication.

Our data from Paper\,I have also been used by \cite{chakra08} who
employed their own code to obtain a mass profile.

\par In the analysis presented in Paper\,I,
the most important source of uncertainty is the sparse data at large
radii: The dispersion value derived for the outermost radial bin
changes drastically depending on whether two GCs with extreme
velocities are discarded or not. If these data are included, the
estimate of the mass enclosed within $30\,\textrm{kpc}$ goes up by a
factor of $\sim\!1.4$ and the inferred dark halo has an extremely
large $(\ga 100\,\textrm{kpc)}$ scale radius. \par Moreover, the
data presented in Paper\,I have a very patchy angular coverage since
the observed fields were predominantly placed along the photometric
major axis of NGC\,4636 (see Fig.~\ref{fig:4636spatial}, right
panel). To achieve a more complete angular coverage and to better
constrain the enclosed mass and the shape of the NGC\,4636 dark halo,
we obtained more VLT FORS\,$2$ MXU data, using the same instrumental
setup as in our previous study. \par

In our present study, an important difference to Paper\,I is the
distance, which is of paramount importance in the dynamical
discussion. In Paper\,I, we adopted a distance of 15\,Mpc, based on
surface brightness fluctuation (SBF) measurements \citep{tonry01}, but
already remarked that we consider this value to be a lower limit. A
more recent re--calibration of the SBF brings the galaxy even closer
to 13.6\,Mpc \citep{jensen03}. 

On the other hand, the method of
globular luminosity functions revealed a distance of 17.5 Mpc
\citep{kissler94,dirsch05}. We do not claim a superiority
of GCLFs over SBFs, but one cannot ignore the odd findings which
result from adopting the short distance. The specific frequency of GCs
would assume a value rivalling that of central cluster galaxies
\citep{dirsch05}. Moreover, the stellar M/L for which in Paper\,I we
already adopted a very high value of 6.8 in the $R$-band, would climb
to 7.5 for a distance of 13.6\,Mpc, which almost doubles the value
expected for an old metal-rich population
(e.g. \citealt{cappellari06}).  This is under the assumption that the
total mass in the central region, for which we present independent
kinematical data, is dominated by the stellar mass (the `maximal disk'
assumption).  The alternative, namely that NGC\,4636 is dark matter
dominated even in its centre, would be intriguing, but would make
NGC\,4636 a unique case. To adopt a larger distance is the cheapest
solution. The explanation for the discrepancy between SBF and GCLF
distance might lie in some effect enhancing the fluctuation signal,
either by an intermediate-age population (which in turn would require
a lower M/L) or another small-scale structure, which went so far
undetected. In the following, we adopt a distance of 17.5\,Mpc.  Thus,
1\arcmin is $\sim\!5.1\,\textrm{kpc}$ and 1\arcsec=85\,pc.\par
\par This paper is organised as follows: 
In Sect.~\ref{sect:4636obs}, we give a brief description of the
spectroscopic observations and the data reduction.  In
Sect.~\ref{sect:4636dataset}, we present the combined data set, before
discussing the spatial distribution and the photometric properties of
the GC sample in Sect.~\ref{sect:4636sample}.
In Section\,\ref{sect:subsamp}, we define the subsamples used for the
dynamical study.  The line--of--sight velocity distributions of the
different subsamples are presented in Sect.~\ref{sect:4636losvd}.  In
Sect.~\ref{sect:4636rot}, the subsamples are tested for rotation, and
in Sect.~\ref{sect:4636veldisp}, the velocity dispersion profiles are
shown. 
In Sect.~\ref{sect:4636stars} we 
present new stellar kinematics
for NGC\,4636 and in Sect.~ \ref{sect:4636models} we summarise the
theoretical framework for the Jeans modelling. The mass profiles are 
shown in Sect.~11.
Sections 12 and 13 give the discussion and the conclusions, respectively.

%===== Figure (1) ==========
\begin{figure*}
\centering
\includegraphics[width=0.49\textwidth]{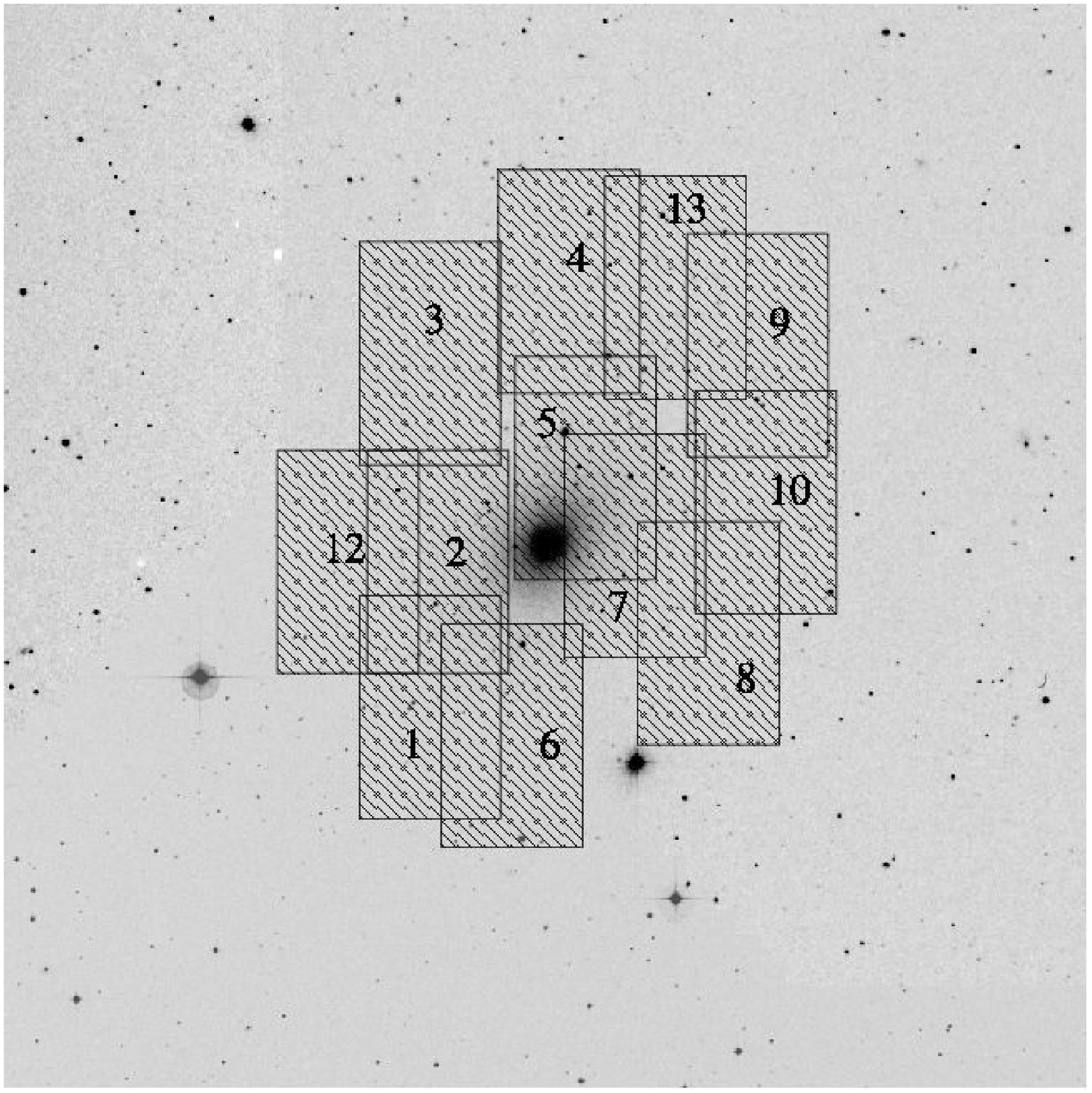}
\includegraphics[width=0.49\textwidth]{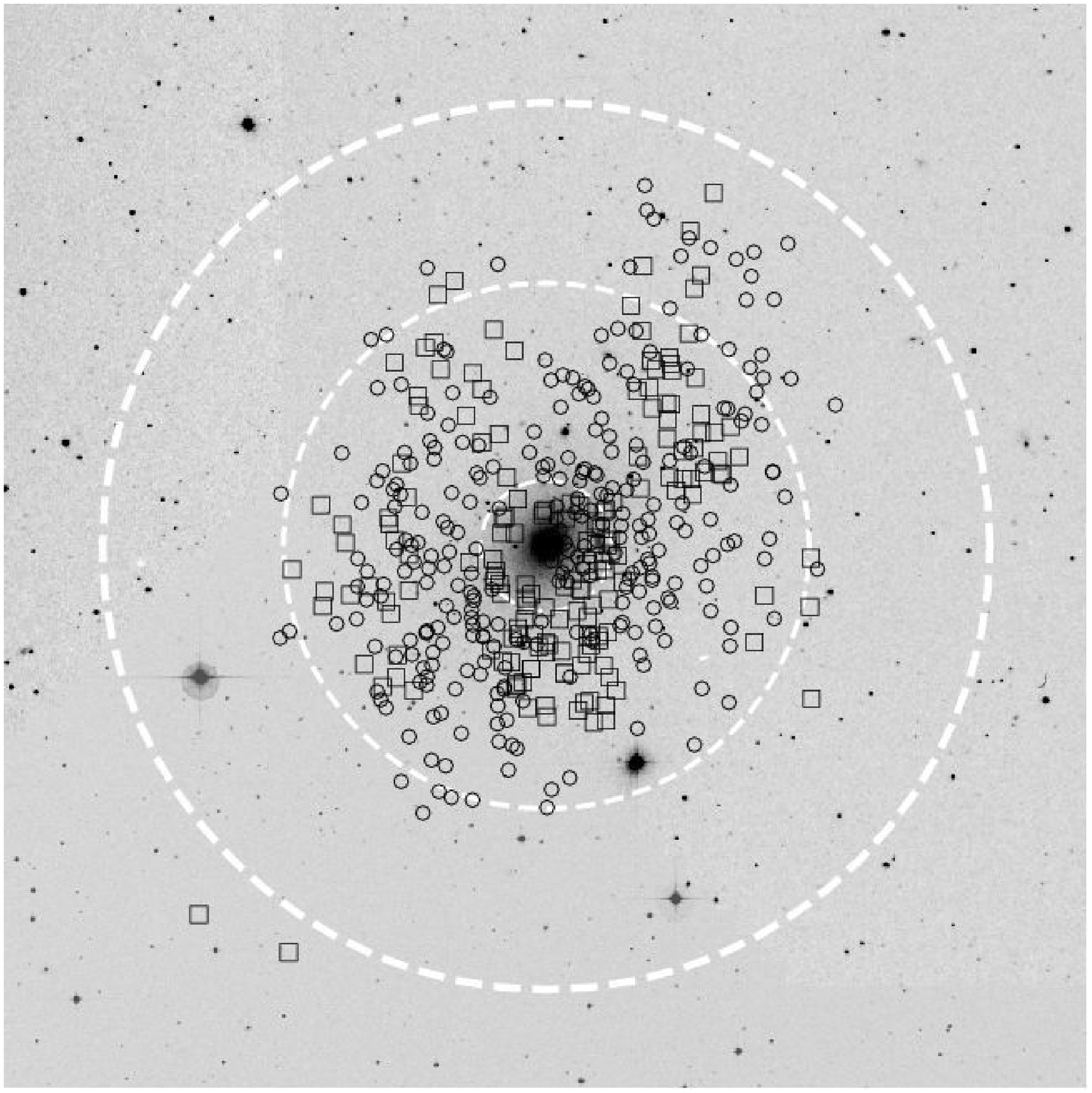}
\caption[]{NGC\,4636 GC spectroscopic data. The
$33\arcmin\times33\arcmin$ DSS image shown in both panels is centred
on NGC\,4636, North is to the top and East is to the
left. \textbf{Left:} Location of the new fields. Note that only the
inner $4\farcmin3\times6\farcmin8$ of the FORS\,$2$ field--of--view
are shown, since the positions of the slits are confined to this
area. \textbf{Right:} All NGC\,4636 velocity--confirmed GCs. Squares
show the data set from Paper\,I, circles the new data. The white
dashed circles have radii of $2\farcmin0$, $8\farcmin0$ and
$13\farcmin5$, corresponding to approximately $10$, $41$, and
$69\,\textrm{kpc}$ (at $D=17.5\,\textrm{Mpc}$), respectively. }
\label{fig:4636dssGCs}
\label{fig:lookup}
\end{figure*}\noindent

%===== TABLE (1) ==========
\begin{table}
\caption[NGC\,4636 basic data]{NGC\,4636 basic data}
{\small
\begin{tabular}{llr}
\hline
\hline
\multicolumn{3}{c}{\normalsize{{NGC\,4636 basic data}\rule[-1ex]{0ex}{3ex}}}\\\hline
Other names &UGC\,07878,  VCC\,1939& NED\rule[-0.5ex]{0ex}{3ex}\\ \hline
Position (2000) & $12\,\fh\, 42\, \fm \,49\, \fs8\quad$  $+02\, \fdg 41 \arcmin 16 \arcsec$& NED\\
                & $190.707792\qquad 2.687778$&\rule[-0.5ex]{0ex}{3ex}\\
Galactic coordinates       &     $ l=297.75\arcdeg\qquad$  $b = 65.47\arcdeg$& \rule[-0.5ex]{0ex}{3ex}\\
\hline
Galactic extinction &$\mathrm{A_{B}} = 0.050$ & \rule[-0.5ex]{0ex}{3ex}\nocite{1982AJ.....87.1165B}(1)\\
                    &$\mathrm{A_{B}} = 0.118$& \rule[-0.5ex]{0ex}{3ex}\nocite{1998ApJ...500..525S}(2)\\
\hline
Distance & $D=17.5\,\mathrm{Mpc}$&\rule[-0.5ex]{0ex}{3ex} (Sect.\,1)\\
Scale & 1\arcsec = 85\,pc$\quad$  1\arcmin = 5.09\,kpc &\rule[-0.5ex]{0ex}{3ex}\\
Distance modulus &$(m-M)= 31.22$\rule[-0.5ex]{0ex}{3ex}\\ 
\hline
Heliocentric velocity&$\varv_{\textrm{helio}}= 906 \pm 7\,\mathrm{km\,s}^{-1}$&\rule[-0.5ex]{0ex}{3ex}\nocite{schuberth06}(3)\\
Hubble type & E0+ & \rule[-0.5ex]{0ex}{3ex}R3C\\
Ellipticity &$\epsilon = 0.15$ &\rule[-0.5ex]{0ex}{3ex}\nocite{BSG94}(4)\\
Position angle  & $PA=150\arcdeg$& \rule[-0.5ex]{0ex}{3ex}R3C \\
Effective radius &${R_{e}}=101\farcsec 7 \,(=8.64\,\mathrm{kpc})$ &\rule[-0.5ex]{0ex}{3ex}\nocite{BSG94}(4)\\
\hline
Age \rule[-1ex]{0ex}{3ex} & $13.5\,\pm3.6\,\textrm{Gyr}$  &\nocite{annibali07}(5)\\
Metallicity & $Z = 0.023\pm0.006$  &\nocite{annibali07}(5)\\
$\alpha/\textrm{Fe}$ & $[\alpha/\textrm{Fe}]=0.29\pm0.06$&\nocite{annibali07}(5) \\
\hline
Total blue mag & $b_T=9.78$&\rule[-1ex]{0ex}{3ex}\nocite{prugniel98}(6)\\ 
Absolute blue mag & $B_T= -21.05$& \\ \hline
Colours& $U\!-\!B=0.50$ &\rule[-0.5ex]{0ex}{3ex}\nocite{prugniel98}(6) \\
& $B\!-\!V=0.87$  \\
& $V\!-\!R=0.67$  \\
& $V\!-\!I=1.30$  \\
\hline
Stellar pop $M/L$ & $M/L_I = 3.74$ &\rule[-0ex]{0ex}{2ex}\nocite{maraston03}(7) \\ 
Dynamical $M/L$ & $M/L_B=12.2$ & \nocite{kronawitter00}(8) \\ 
\hline
X--ray luminosity&$L_{X}=1.78\pm 0.10\times\!10^{\,41}\, \textrm{ergs/s}$& \rule[-0.5ex]{0ex}{3ex}\nocite{forman85}(9)\\
Nuclear X--ray emission & $\leq 2.7\times 10^{\,38}\, \textrm{ergs/s}$ & \rule[-0.5ex]{0ex}{3ex}\nocite{2001ApJ...555L..21L}(10)\\
Central black hole & $M_{\textrm{SMBH}}\sim\!8\times 10^{\,7}\, M_{\odot}$ &\rule[-0.5ex]{0ex}{3ex}{\nocite{2001MNRAS.320L..30M}(11)}\\
\hline
\hline
\end{tabular}
}
\label{tab:4636basic}
\note{ 
UGC = Uppsala
General Catalogue of Galaxies \citep{1973UGC...C...0000N}, VCC = Virgo
Cluster Catalogue \citep{binggeli85}, NED = NASA/IPAC Extragalactic
Database \texttt{(http://nedwww.ipac.caltech.edu)}, R3C = Third Reference
Catalogue of Bright Galaxies \citep{1991trcb.book.....D}.
References:  (1)\,\cite{1982AJ.....87.1165B}; 
(2) \cite{1998ApJ...500..525S}; 
(3) \nocite{schuberth06} Paper\,I; 
(4) \cite{BSG94};
(5) \cite{annibali07};
(6) \cite{prugniel98}; 
(7) \cite[]{maraston03};
(8) \cite{kronawitter00};
(9) \cite{forman85};
(10) \cite{2001ApJ...555L..21L};
(11) \cite{2001MNRAS.320L..30M}.
}

\end{table}

\section{Observations and data reduction}
\label{sect:4636obs}
The data were acquired using the same instrumental setup as described
in Paper\,I. Therefore, we just give a brief description here.
\subsection{VLT--observations}
The observations were carried out in service mode at the European
Southern Observatory (ESO) Very Large Telescope (VLT) facility on
Cerro Paranal, Chile. We used the FORS\,2 (FOcal Reducer/low
dispersion Spectrograph) equipped with the Mask EXchance Unit (MXU).
The programme ID is \mbox{075.B-0762(B)}.\par Pre--imaging of twelve
fields, shown in the left panel of Fig.~\ref{fig:lookup}, was obtained
in April\,2005. The GC candidates were selected from the photometric
catalogue by \cite{dirsch05}.  The targets have colours in the range
$0.9 <C\!-\!R < 2.1$ and $R$--band magnitudes brighter than
$22.5\,\rm{mag}$.  \par The spectroscopic masks were designed using
the ESO FORS Instrumental Mask Simulator (FIMS) software.  
In contrast to
Paper\,I, where object and sky spectra were obtained from different
slits, we chose to observe the GC candidates through longer slits and
to extract the background from the same slit. The slits
for point--sources have a width of $1\arcsec$ and a length of
$4\arcsec$.  
The spectroscopic observations were carried out in service mode in
the period May $2^\textrm{nd}$ to June $7^\textrm{th}$, 2005.  We used
the Grism\,600\,B which gives a resolution of $\sim\!3\,\rm{\AA}$.  A
total of twelve masks were observed with exposure times between 3600
and 5400 seconds. To minimise the contamination by cosmic--ray hits,
the observation of each mask was divided into two or three exposures.
A summary of the MXU observations is given in
Table\,\ref{tab:4636obs}.  

\begin{table*}
\caption[Summary of NGC\,$4636$ VLT FORS$2$/MXU observations]{Summary of
 NGC\,$4636$ VLT FORS$2$/MXU observations} 
\centering
\begin{tabular}{llllllll}\\ \hline \hline
 Mask ID & \multicolumn{2}{c}{Centre Position}  & Obs.~Date & Seeing & \# Exp. & Exp.~Time & \# Slits \\ 
 \, & \multicolumn{2}{c}{(J\,2000)}  & \, & \, & \, & (sec) & \, \\ \hline
 \rule[2.2ex]{0ex}{0ex}F\,01 &  12:43:04.0 & +02:36:20.0 & 2005-05-04 & $0\farcsec73$ &  3 & 5400 & 57 \\
 \rule[2.2ex]{0ex}{0ex}F\,02 &  12:43:03.0 & +02:40:46.0 & 2005-05-05& $0\farcsec90$ & 3  & 5400 & 67 \\
 \rule[2.2ex]{0ex}{0ex}F\,03 &  12:43:04.0 & +02:47:07.0 & 2005-05-02 & $1\farcsec00$ &  3 &  5400  & 65\\
%      F\,03 &  12:43:04.0 & +02:47:07.0 & 2005-05-06 & seeing &  3 &  5400  & \\
%      F\,03 &  12:43:04.0 & +02:47:07.0 & 2005-05-07 & seeing &  3 &  5400  & \\
%       \rule[2.2ex]{0ex}{0ex}F\,04 &  12:42:47.0 & +02:49:18.0 & 2005-05-02 & seeing &  2 & 3600 & \\
 \rule[2.2ex]{0ex}{0ex}F\,04 &  12:42:47.0 & +02:49:18.0 & 2005-05-07 & $0\farcsec90$ &  2 & 3600  &64  \\
  \rule[2.2ex]{0ex}{0ex}F\,05 &  12:42:45.0 & +02:43:37.0 & 2005-05-05 & $0\farcsec71$ &  3 & 5400 & 63 \\
  \rule[2.2ex]{0ex}{0ex}F\,06 &  12:42:54.0 & +02:35:28.0 & 2005-05-31 & $0\farcsec71$ &  2 & 3600 & 71\\
  \rule[2.2ex]{0ex}{0ex}F\,07 &  12:42:39.0 & +02:41:16.0 & 2005-05-31 & $0\farcsec90$ &  2 & 3600 & 75\\
  \rule[2.2ex]{0ex}{0ex}F\,08 &  12:42:30.0 & +02:38:35.0 & 2005-05-05 & $0\farcsec66$ &  2 & 3600 & 67\\
  \rule[2.2ex]{0ex}{0ex}F\,09 &  12:42:24.0 & +02:47:22.0 & 2005-05-05 & $0\farcsec63$ &  3 & 5100 & 63\\
  \rule[2.2ex]{0ex}{0ex}F\,10 &  12:42:23.0 & +02:42:35.2 & 2005-06-01 & $0\farcsec58$ &  3 & 5400 & 68  \\
%      F\,10 &  12:42:23.0 & +02:42:35.2 & 2005-06-07 & seeing &  3 & 5400 &  \\
  \rule[2.2ex]{0ex}{0ex}F\,12 &  12:43:14.0 & +02:40:46.0 & 2005-06-02 & $0\farcsec50$ &  3 & 5400  & 62\\
  \rule[2.2ex]{0ex}{0ex}F\,13 &  12:42:34.0 & +02:49:07.0  & 2005-06-07 & $0\farcsec76$ &  3 & 5400 & 68 \\ \hline \hline
%      NGC\,4636--F01 & &centre coordinates & 2005-05-04 & seeing & \# 3 & $3\times 1800$ & $\quad$ \\
%      NGC\,4636--F13 & &centre coordinates & 2005-06-07 & seeing & \# 3 & $3\times 1800$ & $\quad$ \\ \hline \hline
\normalsize
\end{tabular}
\note{ESO program ID 75.B-0762(B). The seeing values were measured
from the acquisition images obtained prior to the corresponding
spectroscopic observations.}
\label{tab:4636obs}
\end{table*}

\subsection{Data reduction}

%===== Figure (2) ==========
\begin{figure*}
\centering
\includegraphics[width=0.32\textwidth]{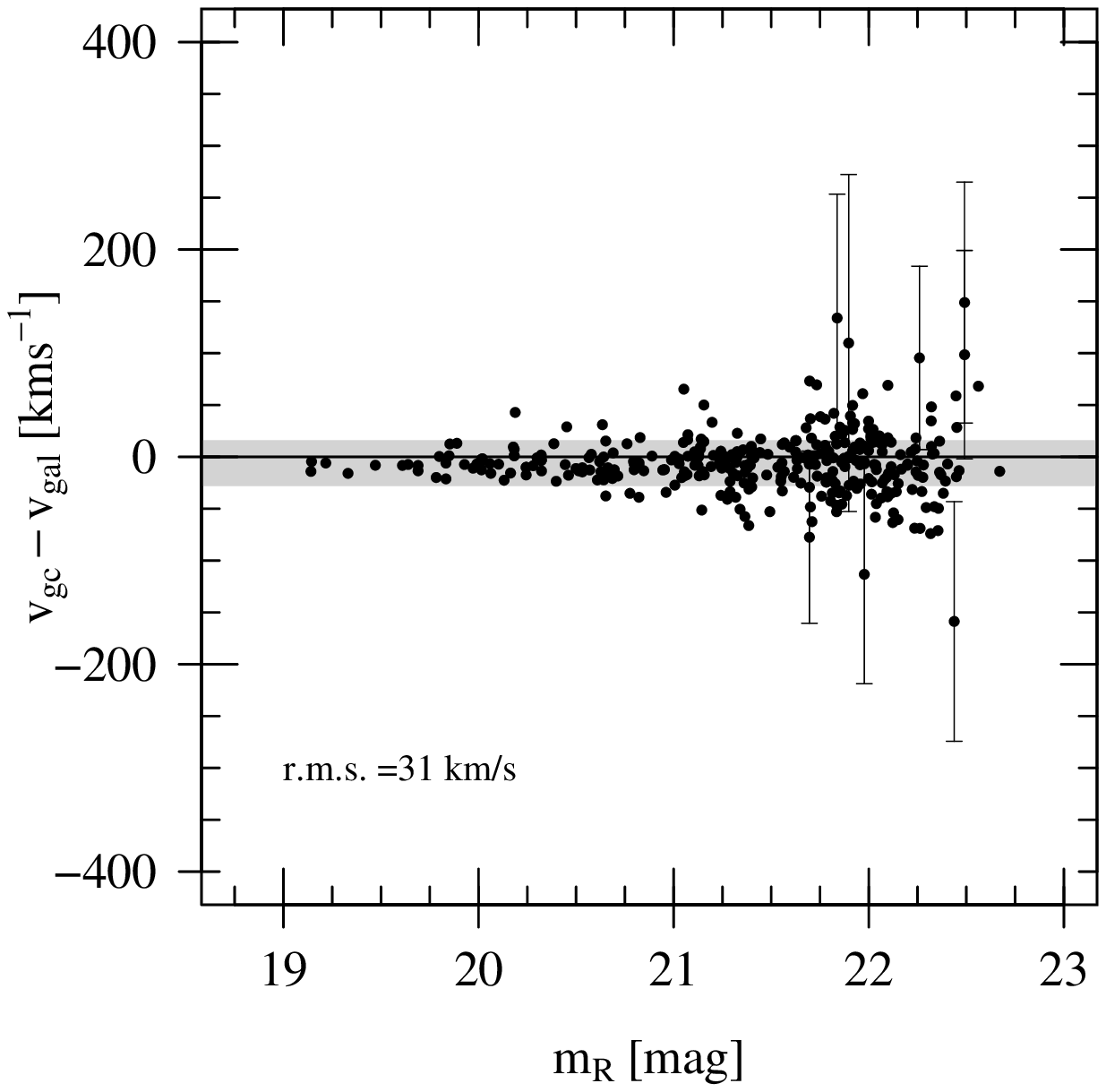}
\includegraphics[width=0.32\textwidth]{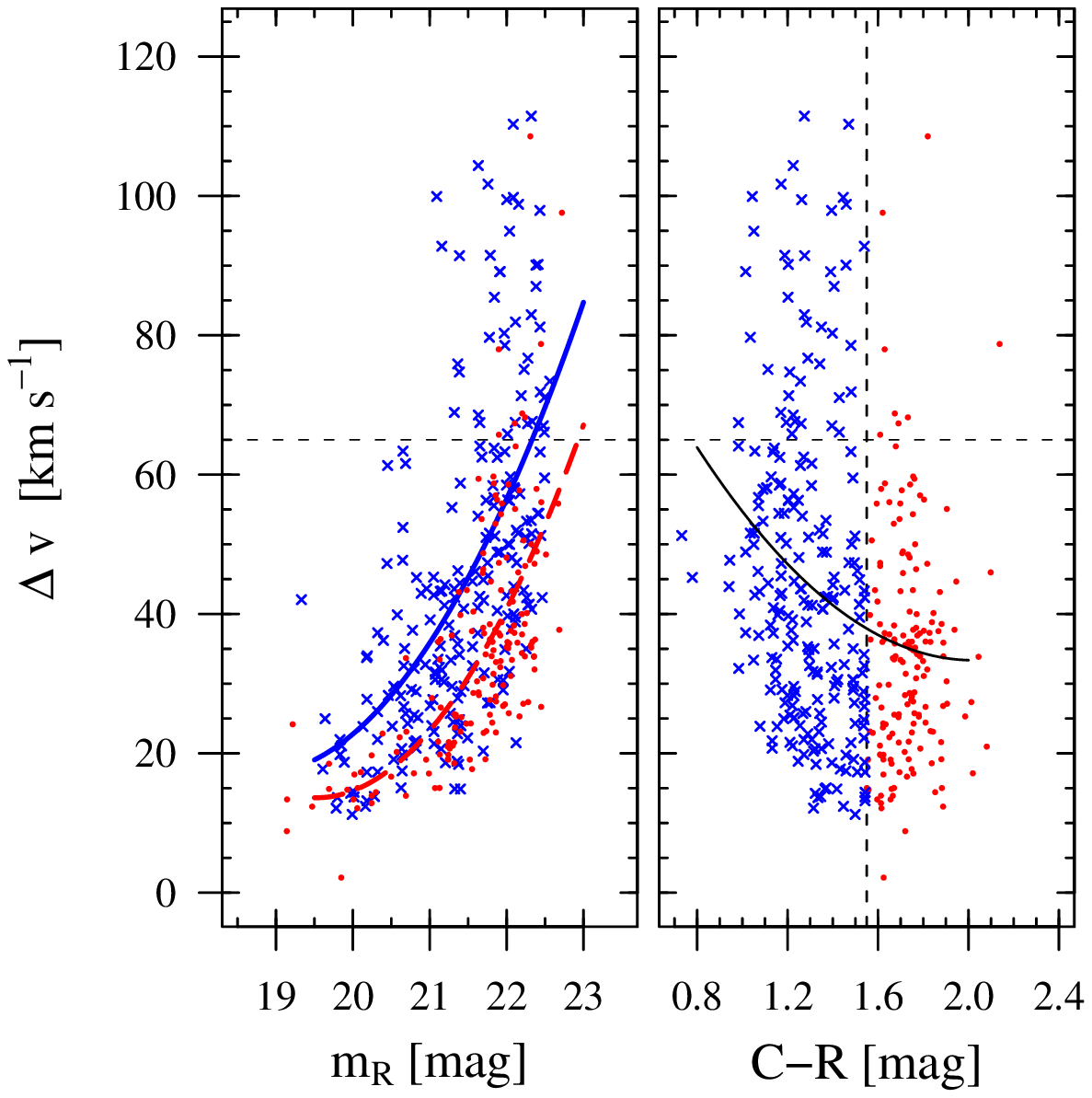}
\includegraphics[width=0.32\textwidth]{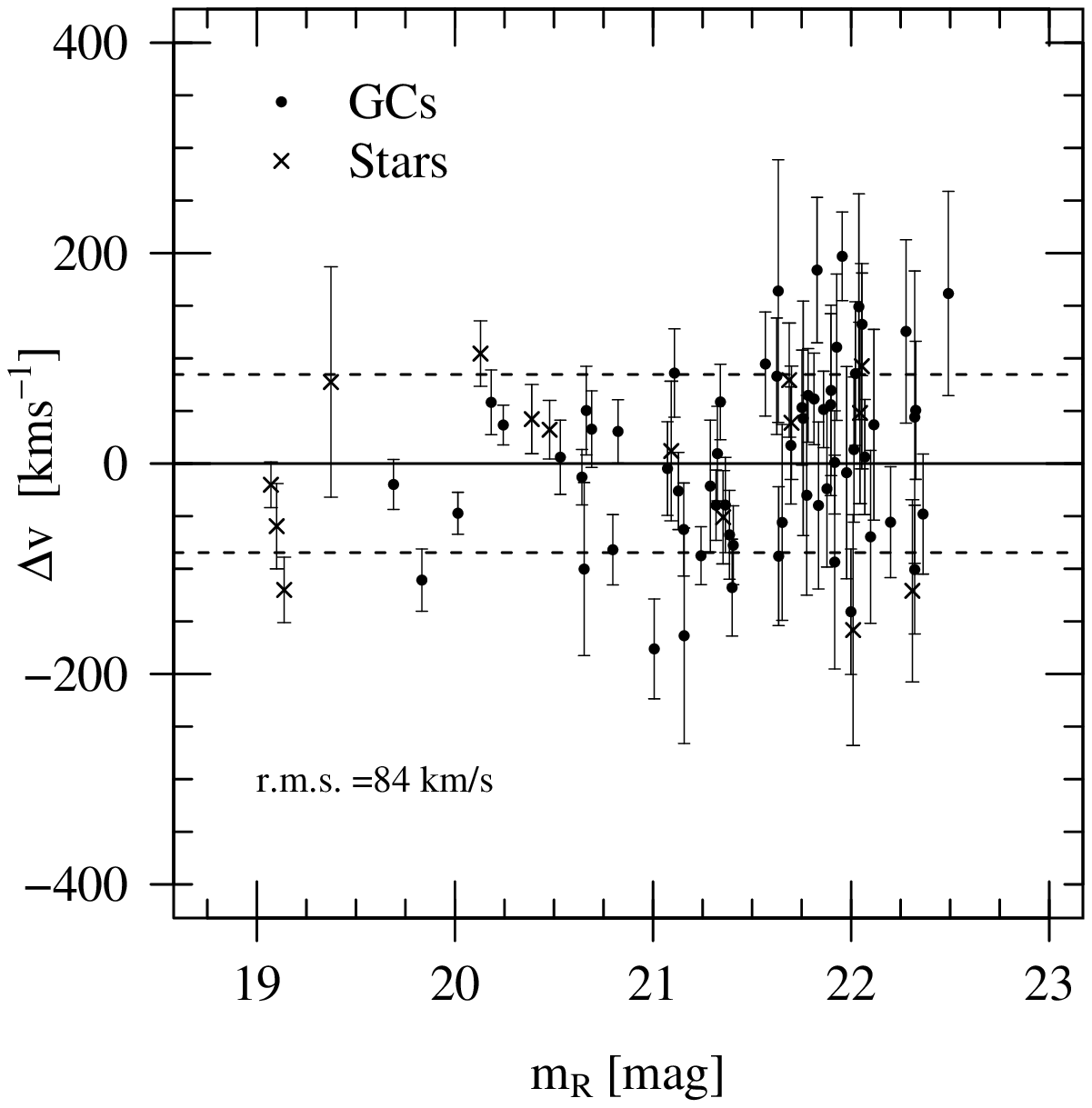}
\caption[]{Velocity uncertainties. \textbf{Left:} Comparison of the GC
  velocity estimates obtained with the two different template
  spectra. For graphic convenience, the error-bars (the two
  \emph{fxcor}--uncertainties added in quadrature) are only shown for
  objects where the estimates differ by more than
  $75\,\textrm{km\,s}^{-1}$. The grey area shows the region
  encompassing 68 per cent of the data points. 
\textbf{Middle:} Velocity uncertainties as determined by
  \texttt{fxcor}. \emph{Left sub--panel}: \texttt{fxcor}--uncertainty
  vs.~$R$--magnitude. Crosses and dots represent blue and red GCs,
  respectively.  The solid and dashed curve show a quadratic fit to
  the blue and red GC data, respectively.  \emph{Right sub--panel}:
  \texttt{fxcor}--uncertainty vs.~colour. The dashed line at
  $C\!-\!R=1.55$ indicates the colour dividing blue from red GCs.  The
  solid line illustrating the increase of the uncertainties towards
  bluer colours is a 2nd order polynomial fit to the data.  The dashed
  lines at $\Delta v=65\,\rm{km\,s}^{-1}$ show the limit adopted
  for the sample definition in Sect.~\ref{sect:4636subsamples}.
 \textbf{Right:}
  Comparison of the duplicate velocity measurements. Dots and crosses
  represent GCs and stars, respectively. The dashed lines show the
  r.m.s.~scatter of {\bf{84}}\,$\rm{km\,s}^{-1}$ found for the
  GCs.  }
\label{fig:4636fxco}
\label{fig:4636velqual}
\end{figure*}

Prior to bias subtraction, 
the \texttt{fsmosaic}--script, which is part of the FIMS--software was
used to merge the two CCD--exposures of all science and calibration
frames.  The spectra were traced
and extracted using the \textsc{Iraf} \texttt{apall} package.  

\label{sect:skysub}
For the extraction of point--sources, we chose an aperture size of 3
(binned) pixels, corresponding to $0.75\arcsec$.  In each slit,
we interactively defined an emission--free background region, the
pixel values of which were averaged and subtracted off the spectrum
during the extraction.  This sky--subtraction was found to work very
well, only the strongest atmospheric emission lines left residuals.
The spectra were traced using the interactive mode of \texttt{apall},
and the curve fit to the trace was a Chebyshev polynomial of order
3--11.  The use of this wide range of polynomials is motivated by the
fact that the characteristics of the bottom CCD (`slave') deviate from
those of top (`master') CCD.  Besides having a worse point spread
function, the traces of spectra on the `slave' CCD are more
contorted. Thus, the tracing of spectra on the `slave' CCD required
the use of the higher--order polynomials. On the `master' frame, an
order of 3--5 proved to be sufficient.  \par The wavelength
calibration is based on Hg-Cd-He arc--lamp exposures obtained as part
of the standard calibration plan. The  one--dimensional arc spectra
were calibrated using the \textsc{Iraf} task
\texttt{identify}. Typically, 17--22 lines were identified per
spectrum and the dispersion solution was approximated by a
$5^{\textrm{th}}$--order Chebyshev polynomial. The rms--residuals of
these fits were about 0.05\,\AA.\par The wavelength calibrations were
obtained during day--time, after re--inserting the masks and with the
telescope pointed at zenith. Consequently, the offsets introduced by
instrument flexure and more importantly, the finite re--positioning
accuracy of MXU have to be compensated for.  To derive the
corresponding velocity corrections we proceeded as follows: The
wavelength of the strong $[ \textrm{OI} ]\, 5577\,\textrm{\AA}$\,
atmospheric emission line was measured in all raw spectra of a given
mask as a function of the slit's location on the CCD (orthogonal to
the dispersion axis). The correction for each aperture was determined
from a $2^{\textrm{nd}}$ order polynomial fit to the
wavelength--position data.  The average magnitude of this correction
was $\sim\!30\,\textrm{km\,s}^{-1}$. On a given mask, the wavelength
drift between the lower edge of the slave chip to the top edge of the
master chip corresponded to velocity differences between 10 and
$90\,\textrm{km\,s}^{-1}$.

\subsection{Velocity determination}
\label{sect:veldet}
To determine the radial velocities of our spectroscopic targets, we
proceeded as described in Paper\,I. We used the same spectrum of the
dwarf elliptical galaxy NGC\,1396
(\mbox{$v=815\,\pm8\,\rm{km\,s}^{-1}$}, \citealt{dirsch04}) as
template, and the velocities were measured using the \textsc{Iraf}
\texttt{fxcor} task, which is based upon the technique described in
\cite{TD79}. 

\par To double check our results, we also
used the spectrum of one of the brightest globular clusters found in
this new dataset (object f12-24,
$v_{\textrm{helio}}=980\pm15\,\textrm{km\,s}^{-1}$,
$m_R=19.9\,\rm{mag}$, $C\!-\!R=1.62$) as a template. Note that these
are the same templates we used in our recent study of the NGC\,1399
GCS \cite{schuberth09}.
\par The wavelength range used for the correlation was
$\lambda\lambda\,4100$--$5180\, \textrm{\AA}$. The blue limit was
chosen to be well within the domain of the wavelength
calibration. Towards the red, we avoid the residuals from the
relatively weak telluric nitrogen emission line at
$5199\,\textrm\AA$. In the few spectra affected by cosmic ray
residuals or bad pixels, the wavelength range was adjusted
interactively. 

The left panel of Fig.~\ref{fig:4636velqual} shows the difference of
the velocities determined using the the `GC' and the `galaxy' template
as function of $R$--magnitude. As expected, the scatter increases for
fainter magnitudes. The r.m.s of the residuals is
$35\,\rm{km\,s}^{-1}$, which is comparable to the mean velocity
uncertainties returned by \texttt{fxcor}
$(\overline{\Delta v}_{\rm{gc}}=37$,
$\overline{\Delta v}_{\rm{gal}}= 48\,\rm{km\,s}^{-1})$.  For each
spectrum, we adopt the velocity with the smaller
\texttt{fxcor}--uncertainty. \par

The middle panel of Fig.~\ref{fig:4636fxco} shows the
\texttt{fxcor}--uncertainties as function of $R$--magnitude (left
sub--panel).  The solid (dashed) curve is a quadratic fit to the blue
(red) GCs.  While the uncertainties increase for fainter objects, one
also notes that, at a given magnitude, the blue GCs, on average, have
larger velocity uncertainties. The same trend can be seen in the right
sub--panel, where we plot colour--dependence of the uncertainties: The
bluer the objects, the larger the uncertainties. This trend is likely
due to the paucity of spectral features in the spectra of bluest,
i.e.~most metal--poor GCs.

\section{The combined data set}
\label{sect:4636dataset}
In this section, we describe the database used for the dynamical
analysis. The final catalogue as given in Table\,\ref{tab:veltab}
combines the new velocities with the data presented in Paper\,I and
the photometry by Dirsch et al.~(\citeyear[D+05
hereafter]{dirsch05}).\par Our data base has {\bf893} entries,
({\textbf{547}} new spectra, {\textbf{346}} velocities from the
catalogues presented in Paper\,I).  The new spectra come from
{\textbf{463}} unique objects (\textbf{327} GCs and {\textbf{136}}
Galactic foreground stars\footnote{Following Paper\,I, we consider
objects with $v_{\mathrm{helio}}$ below $350\,\textrm{km\,s}^{-1}$
as foreground stars (see Sect.~\ref{sect:4636gcsepvel})}). Of these,
{\textbf{289}} GCs and {\textbf{116}} stars were not targeted in the
previous study.  The spectra of background galaxies were discarded
from the catalogue at the stage of the \texttt{fxcor} velocity
determination, and there were no ambiguous cases.  \\Thus, including
the {\textbf{171}} ({\textbf{170}}) unique GCs (foreground stars),
from Paper\,I, the data set used in this work comprises the velocities
of {\textbf{460}} individual bona--fide NGC\,4636 GCs and
{\textbf{286}} foreground stars.

%===== Figure (3) ==========
\begin{figure*}
\centering
\includegraphics[width=0.49\textwidth]{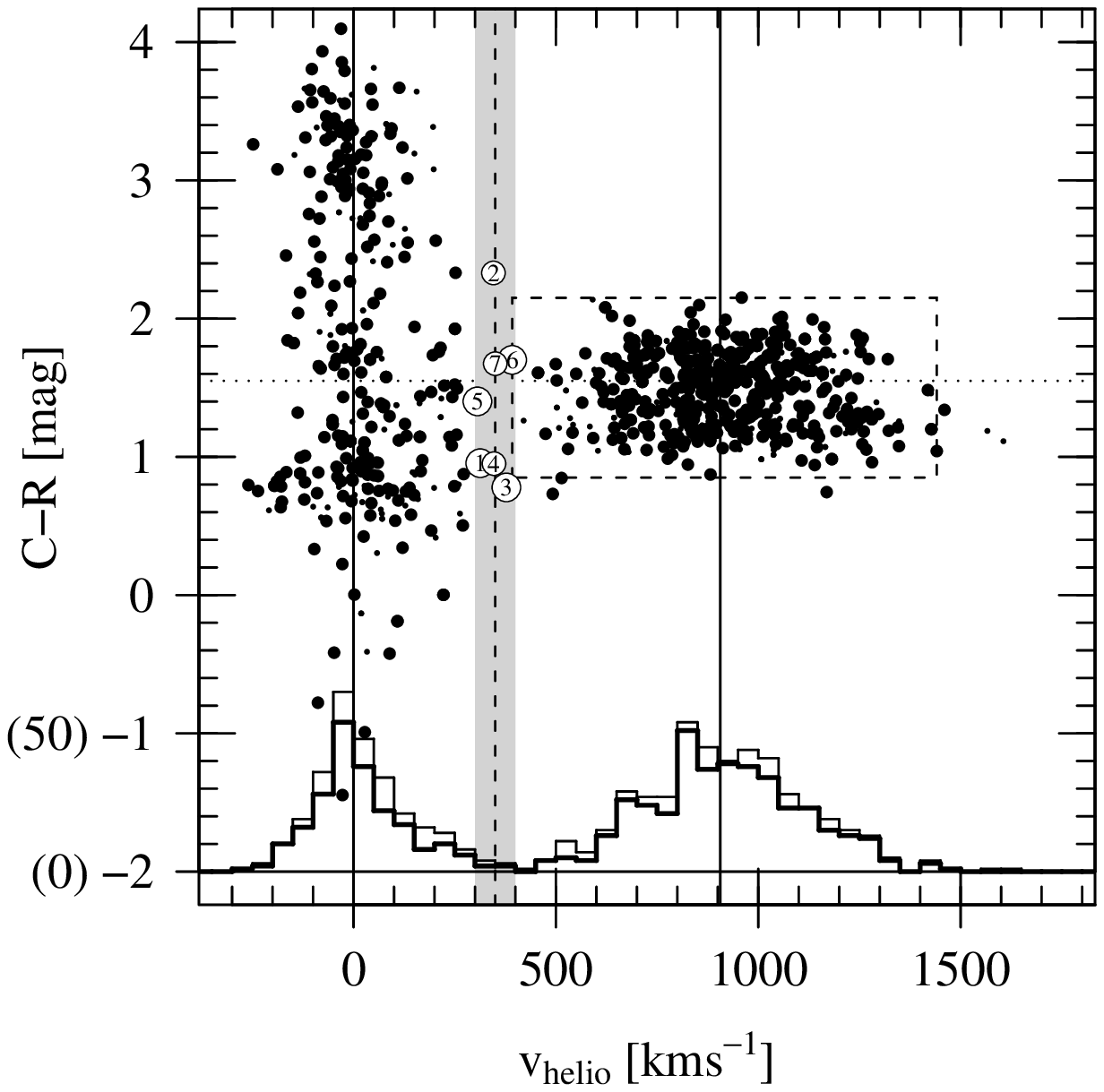}
\includegraphics[width=0.49\textwidth]{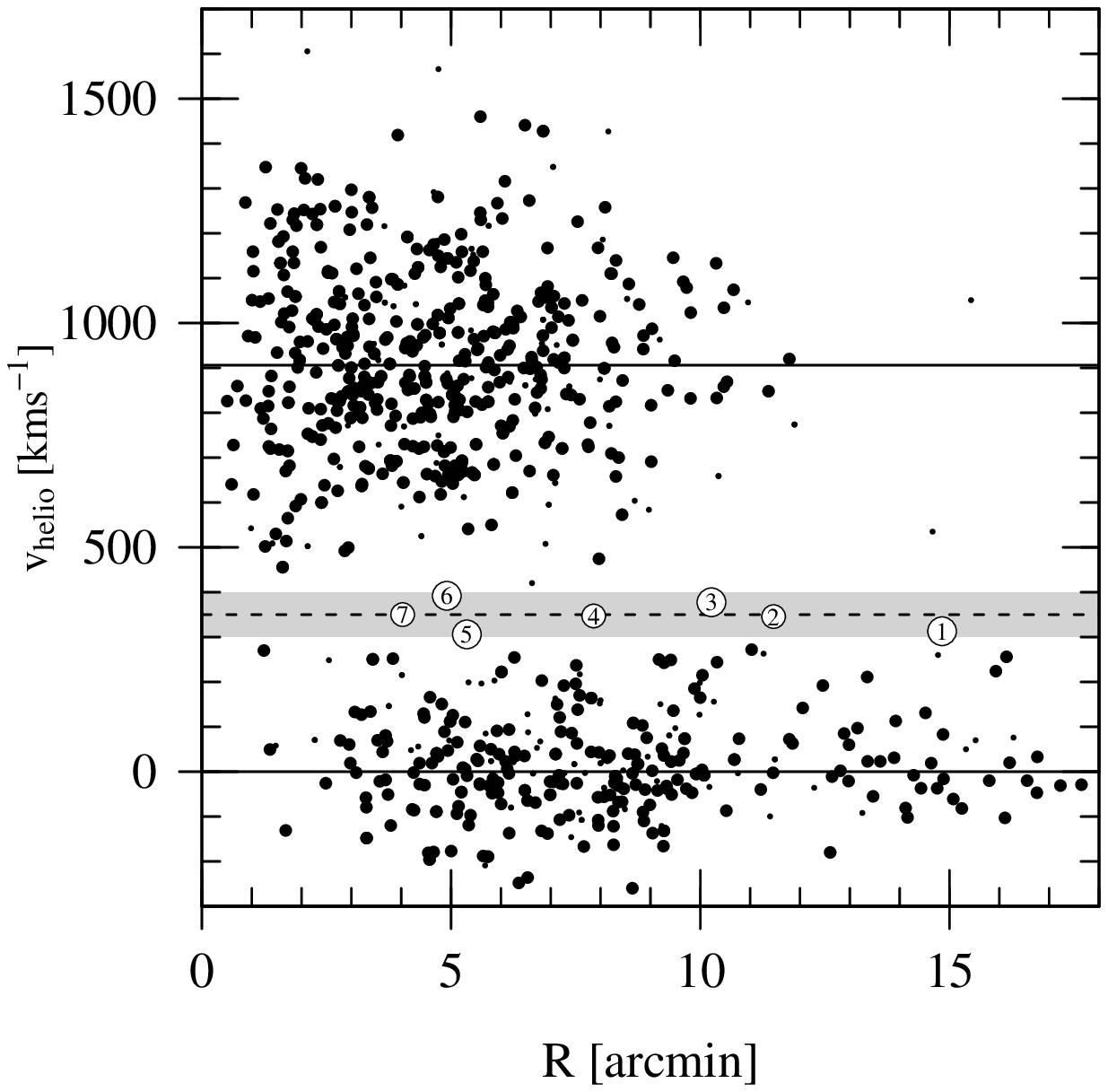}
\caption[Separating GCs from foreground
stars]{Velocity distributions: Separating GCs from foreground
stars.  In both panels, the dashed line at $350\,\textrm{km\,s}^{-1}$
shows the velocity cut used in Paper\,I, and the grey region has a
width of $100\,\textrm{km\,s}^{-1}$. In both panels, objects in this
region are shown as white dots and numbered in decreasing order of the
radial distance.  The solid line indicates the systemic velocity of
NGC\,4636 ($v=906\,\textrm{km\,s}^{-1}$). \textbf{Left:} Colour
vs.~heliocentric velocity.  The dashed box shows the range of
parameters for the GCs analysed in Paper\,I (cf.~Table\,3
therein). Large symbols show objects with velocity uncertainties
$\Delta v\leq65\,\textrm{km\,s}^{-1}$. The histograms ($y$--axis
labels in parenthesis) have a bin width of $50\,\textrm{km\,s}^{-1}$,
and the thick line shows the objects with
$\Delta v\leq65\,\textrm{km\,s}^{-1}$. \textbf{Right:} Heliocentric
velocity vs.~projected galactocentric distance. Again, large symbols
show objects with $\Delta v\leq65\,\textrm{km\,s}^{-1}$.
}
\label{fig:4636gcstars}
\end{figure*}

\subsection{Photometry database}
\label{sect:photodb}
As in Paper\,I, we used the Washington photometry by D+05 to assign
$C\!-\!R$ colours and $R$--band magnitudes to the spectroscopic
targets. \par For some objects with velocities in the range expected
for GCs, however, no photometric counterpart was found in the final
photometric catalogue published by D+05. The reason for this is that
the D+05 catalogue lists the \emph{point sources} in the field, which
were selected using the DAOPhot\,II \citep{stetson87,stetson92}
`$\chi$' and `sharpness' parameters. While this selection, as desired,
rejects extended background galaxies, it also removes those GCs whose
images deviate from point sources. The GCs which were culled from the
photometric data set fall into two categories, the first encompassing
objects whose images have been distorted because of their location
near a gap of the CCD mosaic, a detector defect (e.g.~a bad row) or a
region affected by a saturated star.  The second group are GCs whose
actual sizes are large enough to lead to slightly extended images on
the MOSAIC images which have a  seeing of about $1\farcsec0$.  \par
Using the `raw' version of the photometry database, we were able to
recover colours and magnitudes for {\bf 154} objects, {\bf80} of which are
GCs (according to the criteria defined below in
Sect.~\ref{sect:4636gcsepvel}).  As can be seen from the left and
middle panels of Fig.~\ref{fig:4636velcmd}, many of the brightest GCs
were excluded from the D+05 catalogue.

\subsection{Duplicate measurements}
In our catalogue, we identify {\textbf{131 (94)}} duplicate,
{\textbf{8 (7)}} triple measurements, and {\textbf{607 (359)}} objects
were measured only once (the corresponding numbers for the GCs are
given in parenthesis).  For objects with multiple measurements we
combine the measurements using the velocity uncertainties as weights.
The velocities from the duplicate measurements are compared in the
right panel of Fig.~\ref{fig:4636velqual}. { The r.m.s.~scatter of
$84\,\rm{km\,s}^{-1}$ found for the GCs is about twice the average
velocity uncertainty quoted by the cross--correlation programme.
Random offsets in velocity might be introduced by targets which are
not centred in the slit in the direction of the dispersion direction.
}

\subsection{Separating GCs from Galactic foreground stars}
\label{sect:4636gcsepvel}

In the left panel of Fig.~\ref{fig:4636gcstars}, we plot the
$C\!-\!R$ colour versus heliocentric velocity: The GCS of NGC\,4636
occupies a well--defined area in the velocity--colour plane. 
The foreground stars, clustering around Zero velocity,
have a much broader colour distribution\footnote{The data set includes
foreground stars which do not fulfil the colour criteria used for the
GC candidate selection because \textbf{all} spectra, including those
of the bright stars used for mask alignment, were extracted.}.  \par
As in Paper\,I, we adopt
$v_\textrm{helio}=350\,\textrm{km\,s}^{-1}$ as lower limit for
bona--fide NGC\,4636 GCs.
Although both classes of objects are  well separated in velocity space,
there is a number of uncertain cases: The seven objects with
velocities in the range $300$--$400\,\textrm{km\,s}^{-1}$ (indicated
by the grey areas in both panels of Fig.~\ref{fig:4636gcstars}) merit
closer scrutiny. These objects are discussed in
Sect.~\ref{sect:outrej} where we remove the likely outliers from our sample.

The following section gives a description of the photometric
properties and the spatial distribution of the 460 bona--fide
NGC\,4636 GCs in our velocity catalogue.

\section{Properties of the GC sample}
\label{sect:4636sample}

%===== Figure (4) ==========
\begin{figure*}
\centering
\includegraphics[width=0.32\textwidth]{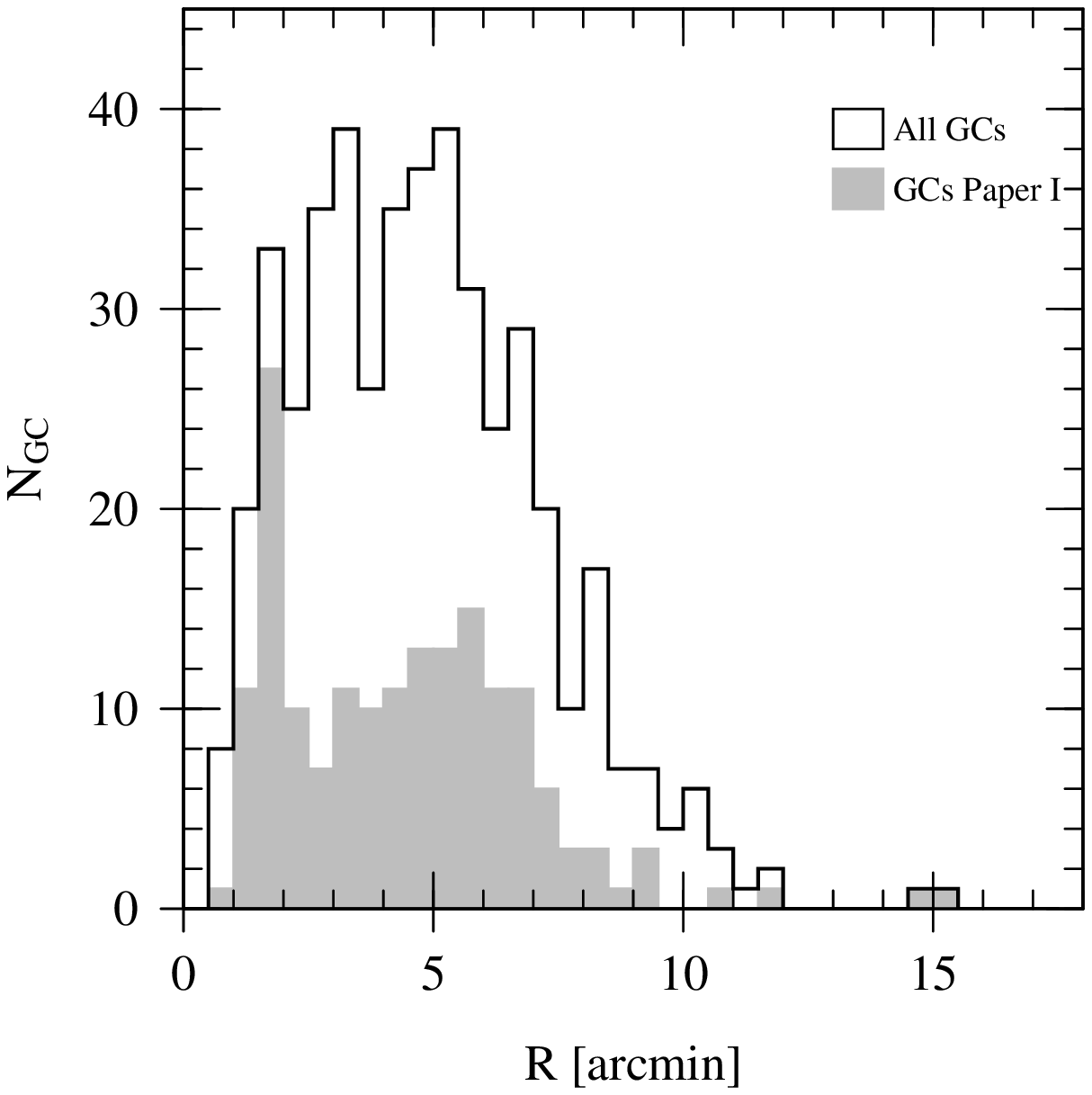}
\includegraphics[width=0.32\textwidth]{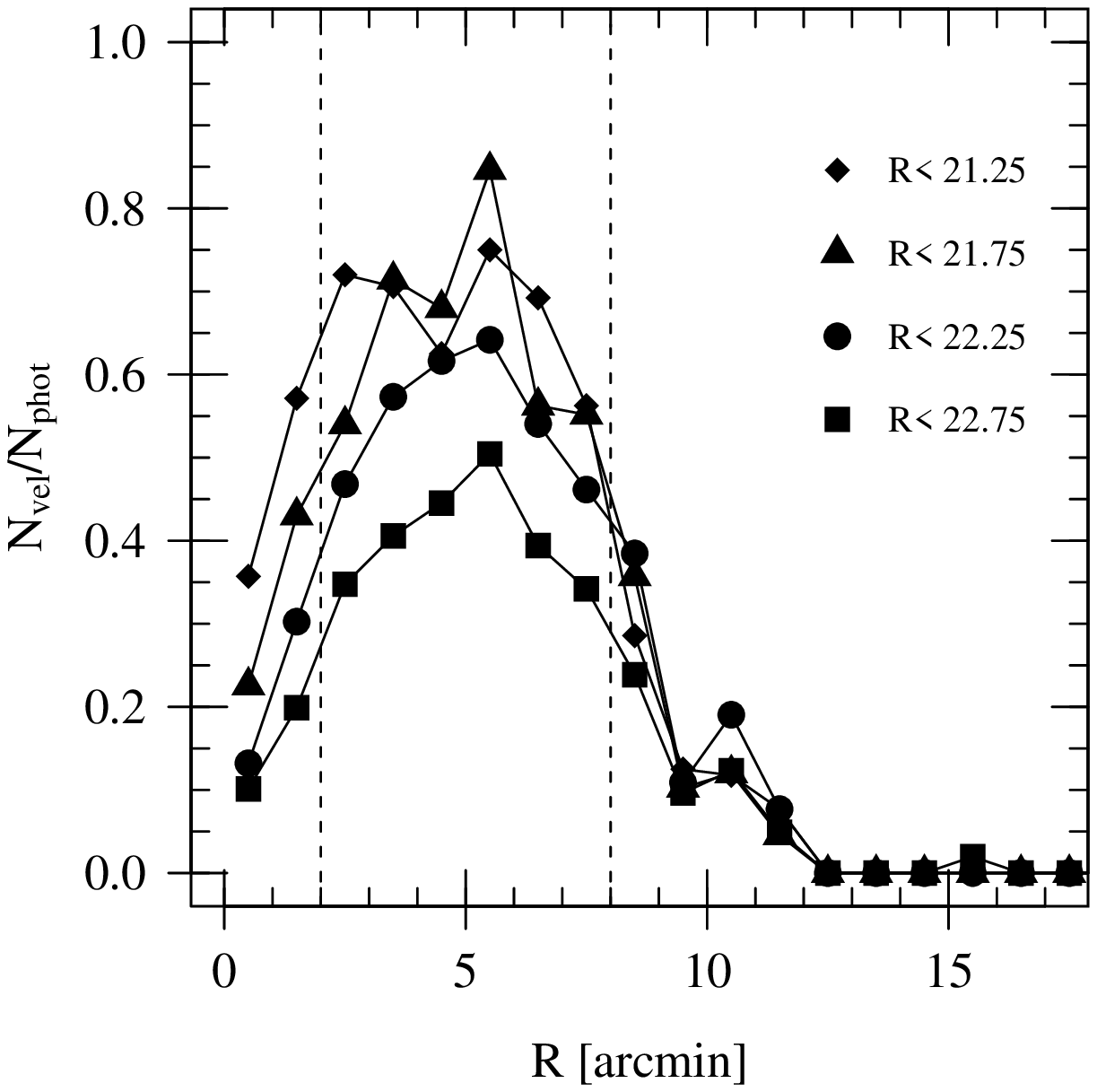}%&
\includegraphics[width=0.32\textwidth]{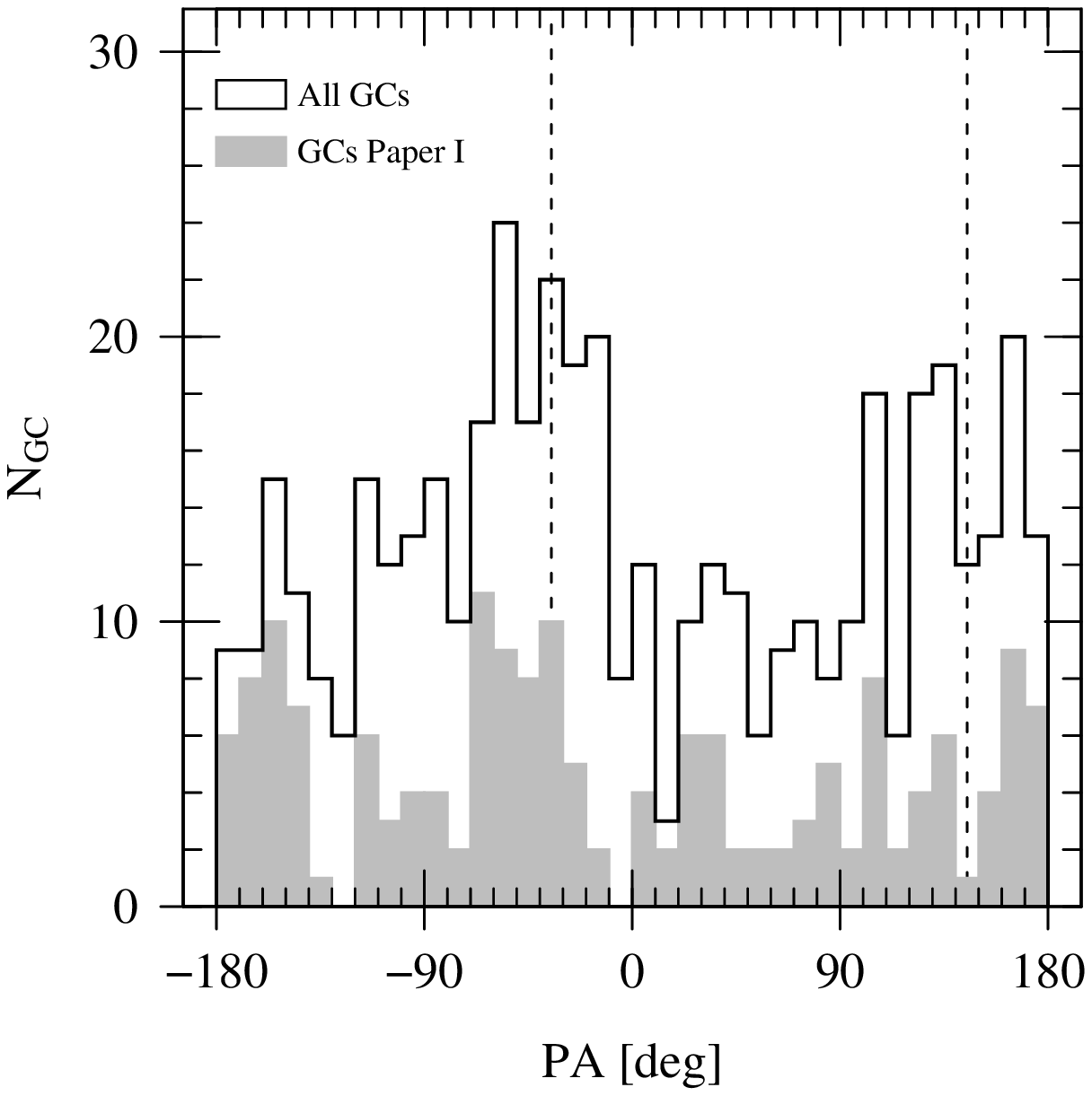}
\caption[NGC\,4636: Radial and azimuthal distribution of the GC
sample]{Spatial distribution of the velocity--confirmed
  GCs. \textbf{Left panel:} Radial distribution of the spectroscopic
  GC sample. The unfilled and grey histograms show the full sample and
  the data from Paper\,I, respectively.  \textbf{Middle panel:} Radial
  completeness for different faint--end magnitude limits: For the GCs
  from the D+05 catalogue, we plot the ratio of GCs with velocity
  measurements with respect to the total number of GC candidates with
  colours in the range ($0.73\leq C\!-\!R \leq2.15 $) found for the
  kinematic sample (cf.~Fig.~\ref{fig:4636velcmd}, left panel).
  \textbf{Right panel:} Azimuthal distribution of the GCs. The
  position angle (PA) is measured North over East and the dashed lines
  indicate the location of the photometric major axis of
  NGC\,4636. The histogram styles are the same as in the left panel.}
\label{fig:4636complete}
\label{fig:4636spatial}
\end{figure*}

Our kinematic sample now comprises a total of $460$ GCs, which is more
than $2.5$ times the number of GCs used in Paper\,I. After NGC\,$1399$
(\citealt{richtler04,richtler08,schuberth09}) with almost 700 GC
velocities, Cen A (\citealt{woodley10}) with about 560 velocities, and M87
(\citealt{romanowsky12}) with 488 velocities, this 
is the fourth--largest GC velocity sample to date.
\par
\subsection{Spatial distribution}
\label{sect:spatial}
The spatial distribution of the GCs with velocity measurements is
shown in Fig.~\ref{fig:4636spatial}.  The left panel plots the
distribution of the galactocentric distances of the GCs as unfilled
histogram. The azimuthal distribution is shown in the right
panel. Compared to Paper\,I (grey histogram), we have now achieved a
more homogeneous coverage, especially filling the gaps in the radial
range $2\arcmin$--$5\arcmin$, in the area near the minor axis of
NGC\,4636.\par Figure\,\ref{fig:4636complete} (middle panel) shows our
estimate of the radial completeness of the kinematic GC sample: For
radial bins of $1\arcmin{}$ width, we compute the ratio of GCs with
velocity measurements to the number of candidate GCs for the published
D+05 catalogue (for consistency, GCs not listed in D+05 are not
considered here).  Since bright GC candidates were preferred over
faint ones for the spectroscopic observations, the completeness level
changes significantly depending on the faint--end limiting magnitude,
and the corresponding curves are shown with different symbols in the
middle panel of Fig.~\ref{fig:4636complete}.  In the innermost bins,
the completeness is very low, since the mask positions were chosen to
avoid these parts where the light of NGC\,4636 would dominate the GC
spectra. As can already be seen from Fig.~\ref{fig:4636dssGCs} (right
panel), the spatial coverage in the area between about $2\arcmin$ and
$7\arcmin$ is very good, and indeed the completeness peaks around a
radial distance of $5\farcmin5$.  Beyond about $8\arcmin$, the number
of GCs with velocity measurements becomes very small, hence the
rapidly declining completeness. Apart from the sparse spatial
coverage, this is also due to the fact that the total number of GCs
expected in the outer regions is small, owing to the steeply declining
number density profiles. This is also illustrated by
Fig.~\ref{fig:4636gcstars} (middle panel), where, for radii beyond
10\arcmin{} we find a number of foreground stars but almost no
GCs.\par

\subsection{Luminosity distribution}

%===== Figure (5) ==========
\begin{figure*}
\centering
\includegraphics[width=0.32\textwidth]{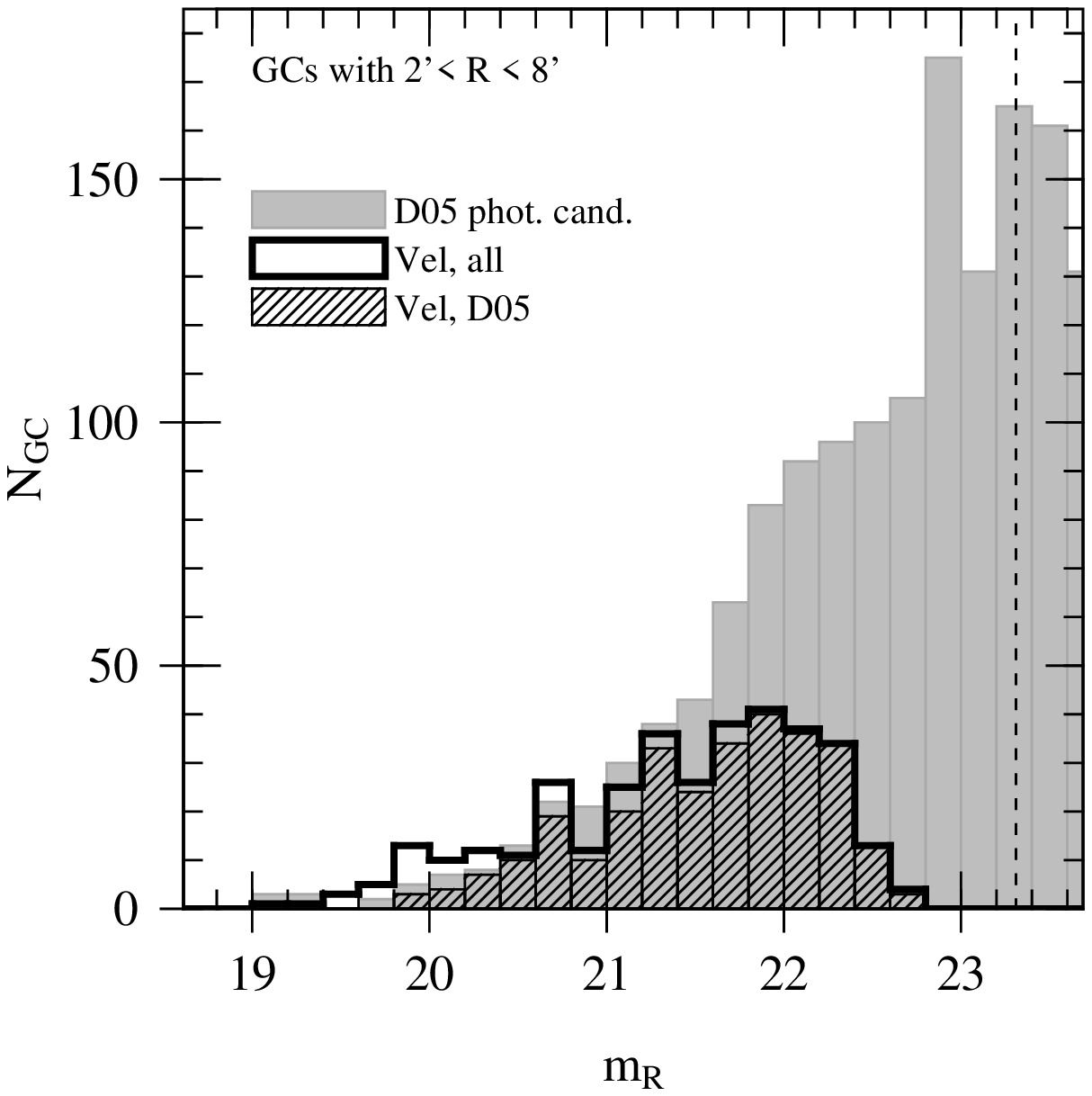}%&
\includegraphics[width=0.32\textwidth]{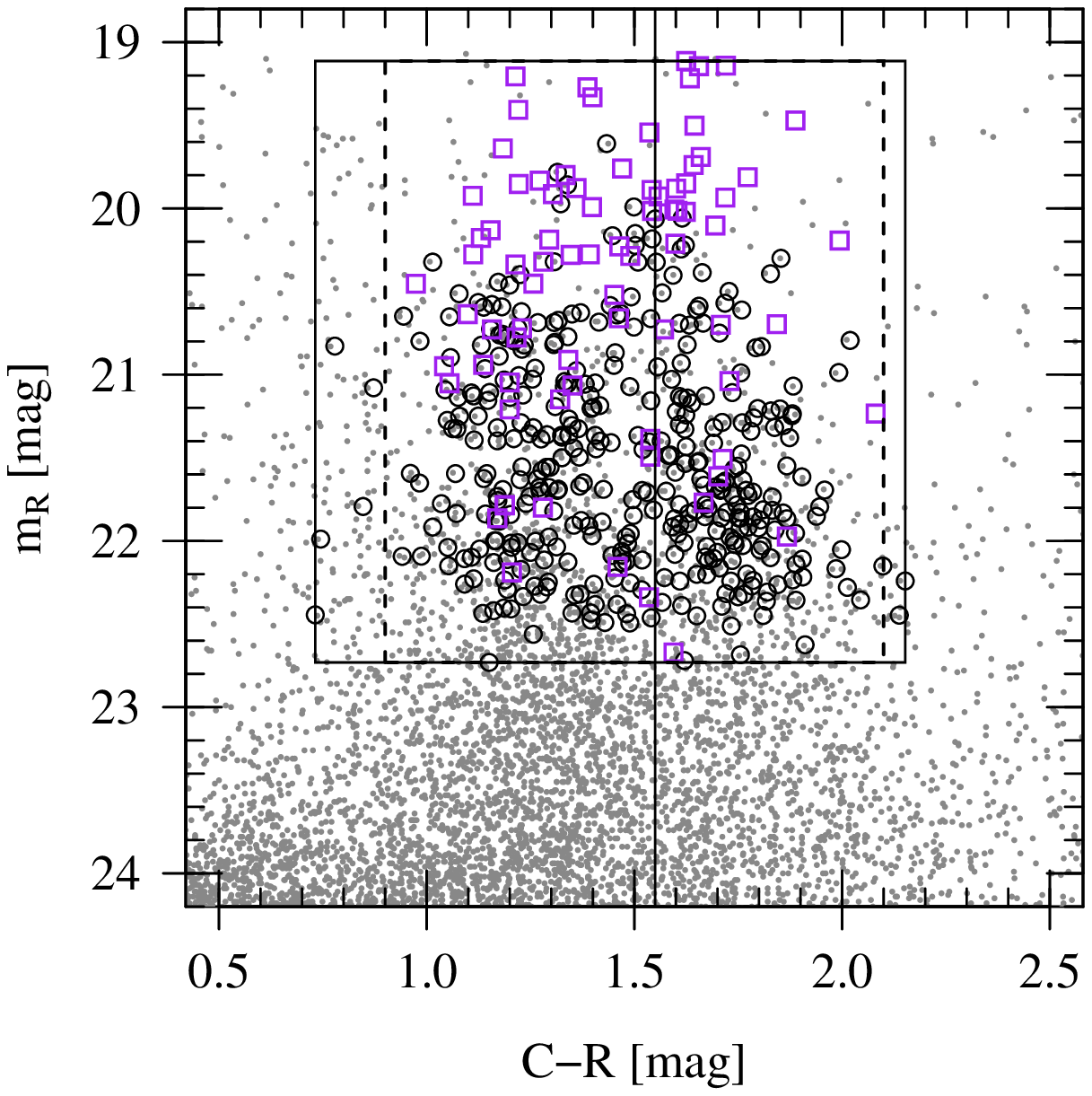}%&
\includegraphics[width=0.32\textwidth]{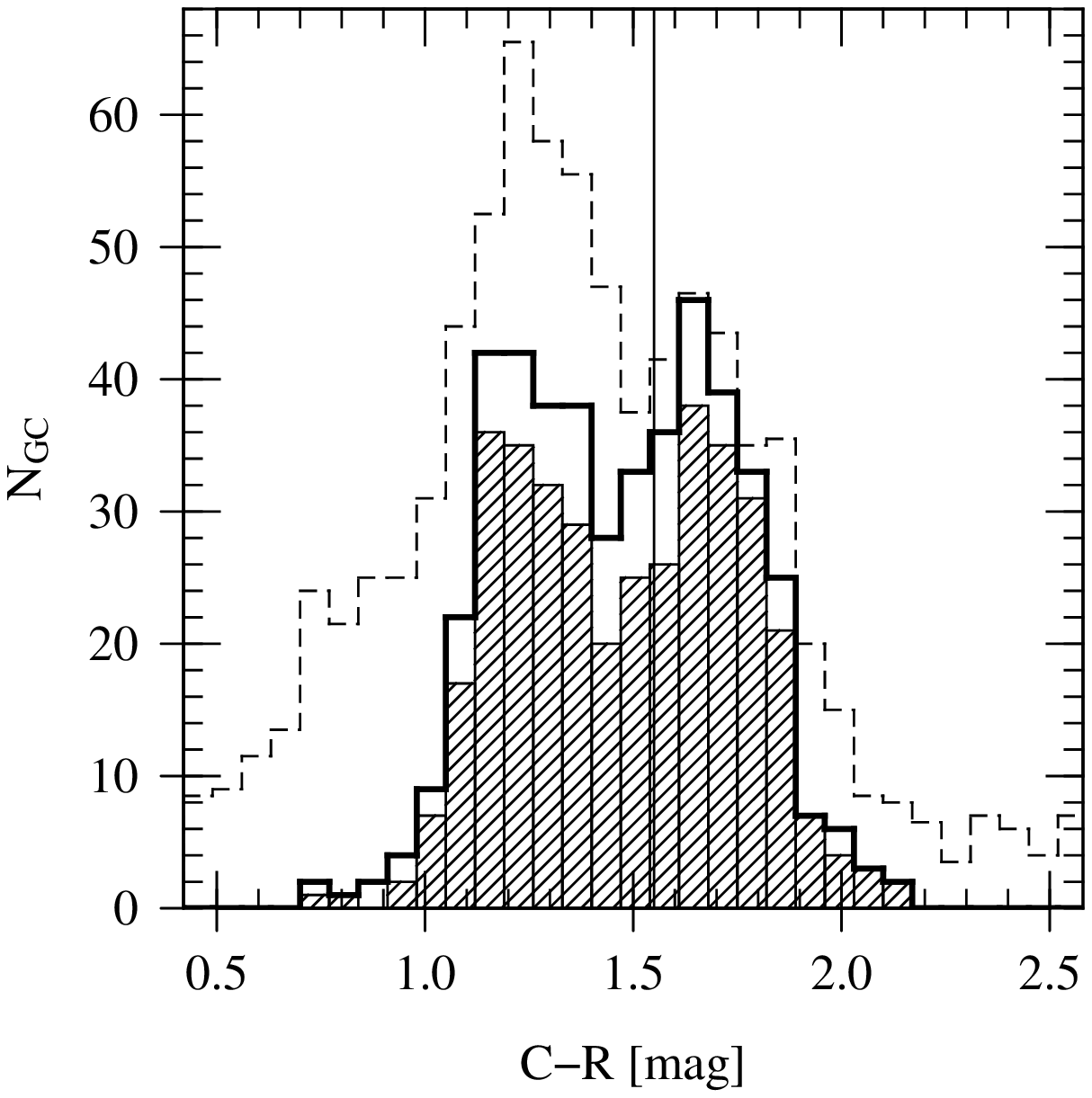}%&
\caption[]{Photometric properties of the GC sample.  \textbf{Left:}
Luminosity distribution of the GCs in the radial range $2\arcmin < R<
8\arcmin$ (indicated by the dashed lines in the middle panel of
Fig.~\ref{fig:4636complete}). The grey histogram shows the
distribution of the GC candidates from the D+05 catalogue, the dashed
bars are those D+05 GCs with velocity measurements. The unfilled
histogram (thick black line) shows all GCs with $2\arcmin < R <
8\arcmin$ and Washington photometry and velocity measurements. The
dashed line at $m_R=23.33$\, mag indicates the turn--over magnitude
(TOM) of the GC luminosity function (D+05).  {\textbf{Middle:}}
Colour--magnitude diagram. Small dots are the point sources from the
D+05 catalogue (only objects within the radial range covered by our
spectroscopic GC sample are shown), and open symbols are GCs with
velocity measurements: Squares indicate those GCs with velocity
measurements which are \emph{not} in the D+05 list, circles are those
identified in the D+05 catalogue.  The solid line at $C\!-\!R=1.55$
shows the colour adopted to divide red from blue GCs. The solid
rectangle shows the area occupied by the velocity confirmed GCs in
this study, and the vertical dashed lines at 0.9 and 2.1 indicate the
colour range adopted by D+05 to identify candidate GCs.
\textbf{Right:} Colour distribution. The unfilled dashed histogram
plots the colour distribution of the D+05 candidate GCs with
magnitudes and radial distances in the range of our
velocity--confirmed GCs. No statistical background has been
subtracted; for graphical convenience, the number counts have been
scaled by a factor of $0.5$. The thick solid line shows the
distribution for all velocity--confirmed GCs with Washington
photometry, and the dashed histogram shows the colours of the
velocity--confirmed GCs with $R\geq2\farcmin5$
(cf.~Sect.~\ref{sect:outer} for details).}
\label{fig:4636lumfunc}
\label{fig:4636velcmd}
\end{figure*}

The left panel of Fig.~\ref{fig:4636lumfunc} compares the
luminosity distribution of the spectroscopic sample to the luminosity
function (LF) of the GCs in the photometric catalogue. For a direct
comparison between the two data sets, we only consider objects in the
radial range $2\arcmin < R < 8\arcmin$, i.e.~where the spatial
completeness of the spectroscopic sample is largest. For GCs brighter
than $m_R\simeq21.4$, the luminosity distributions are almost
indistinguishable indicating a high level of completeness, while, for
fainter GCs, the sampling of the luminosity function becomes
increasingly sparse.  Note, that even the faintest GCs in the
spectroscopic sample are still $\sim \!0.6$\,mag brighter than the turn--over
magnitude ($m_{R,\textrm{TOM}}=23.33$, D+05). 

\subsection{Colour distribution}
\label{sect:outer}

\begin{table*}
\centering
\caption{Colour distribution of the spectroscopic GC sample 
from normal mixture modelling}
\begin{tabular}{llrcl@{--}ll@{--}llllllll} \hline \hline
ID &
&$N$ 
& Model 
& \multicolumn{2}{c}{$m_R$}
& \multicolumn{2}{c}{$R$}
&$N_b$
&$\mu_b$
&$\sigma_b$
&$N_r$
&$\mu_r$
&$\sigma_r$
&$C\!-\!R_{\mathrm{divi}}$
\\ 
(1)&
&(2)  
&(3)
& \multicolumn{2}{c}{(4)}
& \multicolumn{2}{c}{(5)}
&(6)
&(7)
&(8)
&(9)
&(10)
&(11)
&(12)
\\ 
\hline 
1 &all &458 & E & 19.11 & 22.73 &  0.51 & 15.43 & 226 &  1.25 &  0.16 & 232 &  1.69 &  0.16 &  1.46\\
2& $R>2\farcmin5$& 372 & E & 19.14 & 22.73 &  2.53 & 15.43 & 183 &  1.26 &  0.15 & 189 &  1.70 &  0.15 &  1.47 \\
%%$2 a$&187 & E & 19.14 & 21.59 &  2.40 & 11.79 & 104 &  1.25 &  0.14 & 83 &  1.66 &  0.14 &  1.46  \\
%%$2 b$&190 & E & 21.61 & 22.73 &  2.42 & 15.43 & 79 &  1.25 &  0.15 & 111 &  1.73 &  0.15 &  1.47 \\
%$2 a $& $R>2\farcmin5$ and $\rm{m}_R\leq20.5${} & 46 & E & 19.14 & 20.50 &  2.53 &  9.46 & 20 &  1.26 &  0.11 & 26 &  1.62 &  0.11 &  1.40  \\
%$2 b$& $R>2\farcmin5$ and $\rm{m}_R\leq21.6${} &184 & E & 19.14 & 21.59 &  2.53 & 11.79 & 102 &  1.25 &  0.14 & 82 &  1.66 &  0.14 &  1.45 \\%
%$2 c$&  $R>2\farcmin5$ and $\rm{m}_R> 21.6${}&188 & E & 21.61 & 22.73 &  2.54 & 15.43 & 81 &  1.25 &  0.15 & 107 &  1.74 &  0.15 &  1.48  \\
%%2a&125 & E & 19.14 & 21.21 &  2.46 & 11.79 & 71 &  1.24 &  0.14 & 54 &  1.64 &  0.14 &  1.45  \\
%% 2b &125 & E & 21.21 & 21.87 &  2.40 & 14.66 & 55 &  1.24 &  0.12 & 70 &  1.72 &  0.12 &  1.44  \\
%%2c&127 & E & 21.88 & 22.73 &  2.50 & 15.43 & 60 &  1.27 &  0.18 & 67 &  1.73 &  0.18 &  1.49  \\
%%2&316 & E & 20.01 & 22.00 &  0.51 & 14.66 & 160 &  1.23 &  0.15 & 156 &  1.67 &  0.15 &  1.45 \\
%%3&252 & E & 20.01 & 22.00 &  2.40 & 14.66 & 128 &  1.24 &  0.13 & 124 &  1.70 &  0.13 &  1.46  \\ \hline
%\hline
3& $R\leq2\farcmin5$ & 86 & X & 19.11 & 22.45 &  0.51 &  2.50 & 86 &  1.42 &  0.26 & \ldots &    \ldots &    \ldots &    \ldots  \\
 &$R\leq2\farcmin5$& 86 & $\rm{V}^{\dagger}$ & 19.11 & 22.45 &  0.51 &  2.50 & 38 &  1.22 &  0.21 & 48 &  1.57 &  0.19 &  1.36  \\
4& $2.5<R\leq3\farcmin75$& 86 & E & 19.64 & 22.62 &  2.53 &  3.72 & 49 &  1.26 &  0.16 & 37 &  1.72 &  0.16 &  1.50  \\
\hline \hline
\multicolumn{7}{l}{\rule{0ex}{3ex}$^\dagger$
\protect{\footnotesize{model type and number of components fixed.}}}
\end{tabular}
\note{The
parameters were obtained using the \texttt{MCLUST} normal mixture
modelling algorithm.  The first Column describes and labels the
subsets for which the analysis was performed.  The total number of GCs
is given in Col.~2, the type of model fit to the data is given in
Col.~3, where `E' refers to equal--variance (homoscedastic)
2--component models, and `X' to a single--component model. `V' is a
heteroscedastic 2--component model. The range of $R$--band magnitudes
and radial distances are listed in Cols.~4 and 5,
respectively. Columns 6 through 8 show the number of blue GCs, the
position and width of the blue peak. Columns 9 though 11 are the same
for the red GCs. Col.~12 is the colour dividing blue from red GCs.}
\label{tab:bimo}
\end{table*}

%===== Figure (5) ==========

The middle panel of Fig.~\ref{fig:4636velcmd} shows the
colour--magnitude diagram of the NGC\,4636 GCS: Open symbols are the
GCs from our spectroscopic study. Circles show GCs matched to the D+05
final catalogue, and rectangles are the mostly bright GCs which are
\emph{not} in the D+05 catalogue (cf.~Sect.~\ref{sect:photodb}). Small
dots show the objects from the D+05 catalogue (from the radial range
covered by the spectroscopically confirmed GCs). The solid rectangle
indicates the range of parameters of the GCs with velocities, and the
dashed lines at $C\!-\!R=0.9$ and $2.1$ show the colour interval D+05
used to identify the photometric GC candidates. All but seven of our
velocity--confirmed GCs have colours in the range used by D+05,
confirming their choice of parameters.  \par
\par The right panel of Fig.~\ref{fig:4636lumfunc} compares the colour
histogram for the spectroscopic sample (thick solid line) to the
distribution of GC colours from the D+05 photometry (dashed unfilled
histogram). Also for the smaller spectroscopic sample, the bimodality
is readily visible. The location of the peaks agrees well with the
photometric GC candidates. For our sample, the heights of the blue and
red peak are very similar, while, for the photometric sample, the blue
peak is more prominent. This is due to the fact that the dashed
histogram shows all D+05 photometric GC candidates for the full radial
interval covered by our study: There is a stronger contribution from
larger radii -- where blue GCs dominate (see Sect.~\ref{sect:numdens})
-- than for our kinematic sample which becomes increasingly incomplete
as one moves away from NGC\,4636 (cf.~Fig.~\ref{fig:4636complete},
middle panel).

\subsection{GC colour distribution as a function of galactocentric
  radius}

%===== Figure (6) ==========
\begin{figure}
\centering
\includegraphics[width=0.47\textwidth]{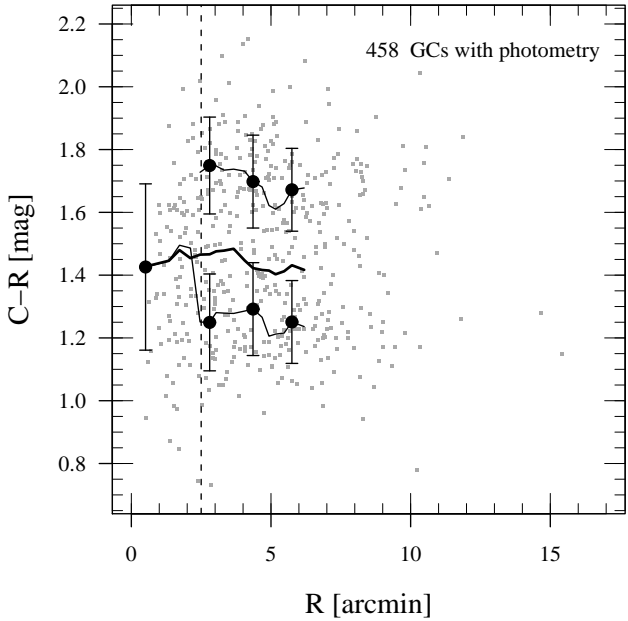}
\caption[]{ Colour
distribution of the GCs as function of radius. The moving window
contains 105 GCs, and the data points show the independent bins. The
number of components to fit was determined by the \texttt{mclust}
software.  The `error-bars' give the widths of the respective
distributions. The thick solid line shows the colour of the combined
light of GCs within the sliding window. The vertical dashed line at
$2\farcmin5$ ($\simeq$13 kpc) indicates the galactocentric distance where
the colour distribution becomes bimodal.}
\label{fig:4636-mclust}
\end{figure}

Due to the different spatial distributions of metal--poor and
metal--rich GCs (and the resulting differences with regard to the
kinematics), a meaningful dynamical analysis requires a robust
partition of the spectroscopic GC sample into blue and red GCs.

D+05 used the minimum ($C\!-\!R=1.55$) of the colour distribution of
the GCs in the radial range $3\farcmin6\la R\la 8\farcmin1$
to separate the two populations.

We use the model--based mixture modelling software provided in the 
\texttt{MCLUST} package 
 \footnote{\texttt{MCLUST} is implemented in the \textsf{R} language
and environment for statistical computing and graphics
(\texttt{http://www.r-project.org}).} \citep{mclust,mclustMAN} to
study the colour distribution of the NGC\,4636 spectroscopic GC
sample: \texttt{MCLUST} fits the sum of Gaussians (via
maximum--likelihood) to the $C\!-\!R$ data. The number of Gaussian
components fit can either be specified by the user, or \texttt{MCLUST}
computes the Bayesian Information Criterion (BIC,
Schwarz\,\citeyear{schwarz78}) to find the optimal number of
components.  
\par In  Fig.~\ref{fig:4636-mclust}, we
plot the colours of the velocity--confirmed GCs versus their
galactocentric distance (small dots). For a sliding window containing
105 GCs, we use \texttt{MCLUST} (in BIC mode) to determine the number
of Gaussian components and plot the positions of the peaks as function
of radius\footnote{The derived quantities are plotted against the
\emph{lower boundary} of the radius interval covered by a bin.}  (thin
solid line). The large data points show the values obtained for the
independent bins, and the vertical bars indicate the width(s) of the
distribution(s). The mean colour of the GCs is shown as thick solid
line. Only for $R \ga 2\farcmin5$ (dashed vertical line), the
colour distribution becomes bimodal.\par In Table\,\ref{tab:bimo}, we
list the parameters derived for various subsamples of our data. Unless
otherwise indicated, the number of components fit to the data was
determined using the BIC. Apart from the GCs with $R<2\farcmin5$
all distributions 
were found to be bimodal.

\subsection{Photometry of the GCs within 2\farcmin5:}

%===== Figure (7) ==========
\begin{figure}
\centering
\includegraphics[width=0.47\textwidth]{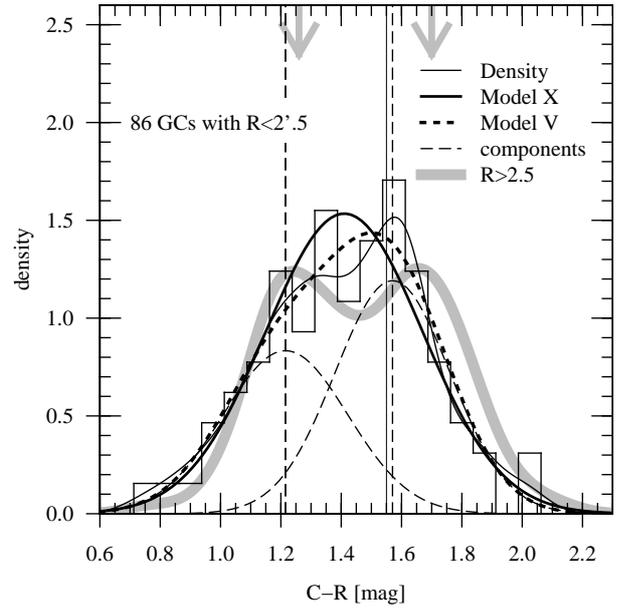}
\caption[]{Colour distribution of the GCs within $2\farcmin5$ (subset 3 in
Table\,\ref{tab:bimo}).  The thin solid curve is the Gaussian kernel
density estimate for a bandwidth of 0.075\,mag (same as the bin width
of the histogram). The thick solid curve is the one--component model
(`X'), and the short--dashed curve is the unequal--variance
2--component model (`V'). The two components are shown with
long--dashed curves, the corresponding peaks are indicated by the
long--dashed vertical lines.  For comparison, the thick grey line
shows the kernel density estimate for the colour distribution of the
GCs with $R>2\farcmin5$ (number\,2 in Table\,\ref{tab:bimo}), and the
arrows mark the positions of the peaks.  For reference, the colour
used to separate blue from red GCs ($C\!-\!R=1.55$, D+05) is shown as
solid vertical line. }
\label{fig:inner2.5}
\end{figure}

As can be seen from Fig. \ref{fig:inner2.5}, the colour distribution
of the GCs with galactocentric distances smaller than $2\farcmin5$
(sample\,3 from Table\,\ref{tab:bimo}) does not look bimodal, and
\texttt{MCLUST} finds the distribution is best described by a single
Gaussian (shown as thick solid line). Compared to the GCs with
$R>2\farcmin5$ (shown as thick grey curve), the distribution appears
to be shifted towards the blue. When fitting a 2--component
heteroscedastic model (`V' in Table\,\ref{tab:bimo}) to the GCs within
$2\farcmin5$, the fitted distributions (shown as thin dashed curves)
are broader than for any of the other samples considered, and the
peaks are offset to the blue.  

To test whether the \texttt{MCLUST} results for the GCs within
2\farcmin5 are due to the small sample size of 86 GCs, we apply the
algorithm to the 86 GCs in the radial range $2\farcmin5 < R \leq
3\farcmin75$ (sample\,4 in Table\,\ref{tab:bimo}). For this sample of
equal size, however, \texttt{MCLUST} determines that the most likely
distribution is indeed bimodal with peak positions consistent with
those found when considering all GCs outside 2\farcmin5 (sample\,2).

We suspect that the photometry of the point
sources within the central 2.5 arcmin is worse than that of sources
outside that region because of issues with the subtraction of the
galaxy light: The shape of the colour distributions is broadened by
larger photometric errors, and a small offset in the continuum
subtraction results in a shift towards bluer colours.  In the central
2.5 arcmin, it is therefore not possible to separate the blue and red
GC populations. Consequently, the dynamical analysis of the
subpopulations has to be restricted to galactocentric distances beyond
$2\farcmin5\, (\simeq 12.7\,\rm{kpc})$.

\par Regarding the spatial distribution and the photometric properties
discussed above, we conclude that our spectroscopic sample is a good
representation of the GCs surrounding NGC\,4636 for galactocentric
distances between 2\farcmin5{} and 8\arcmin{} ($12.7\la
R\la40.7\,\textrm{kpc}$).

\section{Definition of the subsamples}
\label{sect:subsamp}

As was demonstrated in Paper\,I, the presence of interlopers
can severely affect the derived line--of--sight velocity
dispersion profile and, by consequence the inferred mass profile. 
In the following paragraphs, we describe our outlier rejection
technique.  The final subsamples to be used in the dynamical analysis
are defined in Sect.~\ref{sect:4636subsamples}.

\subsection{Interloper rejection}

Objects with velocities that stand out in the velocity
vs.~galactocentric distance plot (Fig.~\ref{fig:4636gcstars}, right
panel) are potential outliers.  Such `deviant' velocities might be due
to measurement errors, a statistical sampling effect, or the presence
of an intra--cluster GC population \citep{bergond07,schuberth08}.  In
the low--velocity domain, possible confusion with Galactic foreground
stars is the main source of uncertainty.

\subsubsection{Contamination by Galactic foreground stars}

\label{sect:fgstars}
In Sect.~\ref{sect:4636gcsepvel}, the division between foreground
stars and bona--fide NGC\,4636 GCs was made at
$v=350\,\rm{km\,s}^{-1}$ (same as in Paper\,I).  There are,
however, seven objects with velocities between 300 and
$400\,\textrm{km\,s}^{-1}$ (shown as labelled white dots in both
panels of Fig.~\ref{fig:4636gcstars}).  In order of decreasing
distance from NGC\,4636 these are:
\begin{enumerate}
\item{7.1:15 ($v=313\pm 20 \,\rm{km\,s}^{-1}$, $m_R=19.83$,
    $C\!-\!R=0.95$) is almost certainly a foreground star: At a
    distance of $14\farcmin8$ we hardly find any NGC\,4636 GCs at all
    (cf.~Fig.~\ref{fig:4636gcstars}, middle panel). Moreover, this
    object would be very blue for a GC of this magnitude
    (cf.~Fig.~\ref{fig:4636velcmd}, middle panel). }

\item{ f09-57 ($v=345\,\pm84\,\rm{km\,s}^{-1}$, $C\!-\!R=2.33$,
$m_R=22.08$) is probably a foreground star: Its very red colour lies
outside the range of colours found for the GCs studied in Paper\,I.  }

\item{Object f09-43 ($v=378\,\pm45\,\rm{km\,s}^{-1}$, $m_R=20.82$)
has an extremely blue colour ($C\!-\!R=0.78$), and its velocity is
offset from the NGC\,4636 GC velocities found at this radial distance
($R=10\farcmin2$), hence it is most likely a foreground star.}
\item{f08-13 ($v=347\,\pm 66\,\rm{km\,s}^{-1}$, $C\!-\!R=0.95$,
$R=7\farcm9$) is very blue and quite faint ($m_R=22.5$).}
\item{2.2:76 ($v=306\,\pm 60\,\rm{km\,s}^{-1}$,
$C\!-\!R=1.40$, $m_R=21.9$, $R=5\farcmin3$) is an ambiguous case.}
\item{1.2:15 ($v=392\,\pm 38\,\rm{km\,s}^{-1}$, $C\!-\!R=1.70$,
$m_R=21.6$, $R=4\farcmin9$) is an ambiguous case.}
\item{f03-09  ($v=350\,\pm 69\,\rm{km\,s}^{-1}$, $C-R=1.67$, 
$m_R=22.2$, $R=4\farcmin0$) is also an ambiguous case.}
%Object f09-43 %
\end{enumerate}
Thus, three of the seven objects in the velocity range
300--400\,$\rm{km\,s}^{-1}$ are almost certainly foreground stars
because of their unlikely combination of extreme velocities and
extreme colour. We remove these objects from the list of GCs.  For the
remaining four objects (No.~4--7) in this velocity interval, the situation
is not as clear, so they remain in the samples to which we apply the
outlier rejection scheme described below.

\subsubsection{Outlier rejection algorithm}
\label{sect:outrej}

\begin{figure*}
\includegraphics[width=0.49\textwidth]{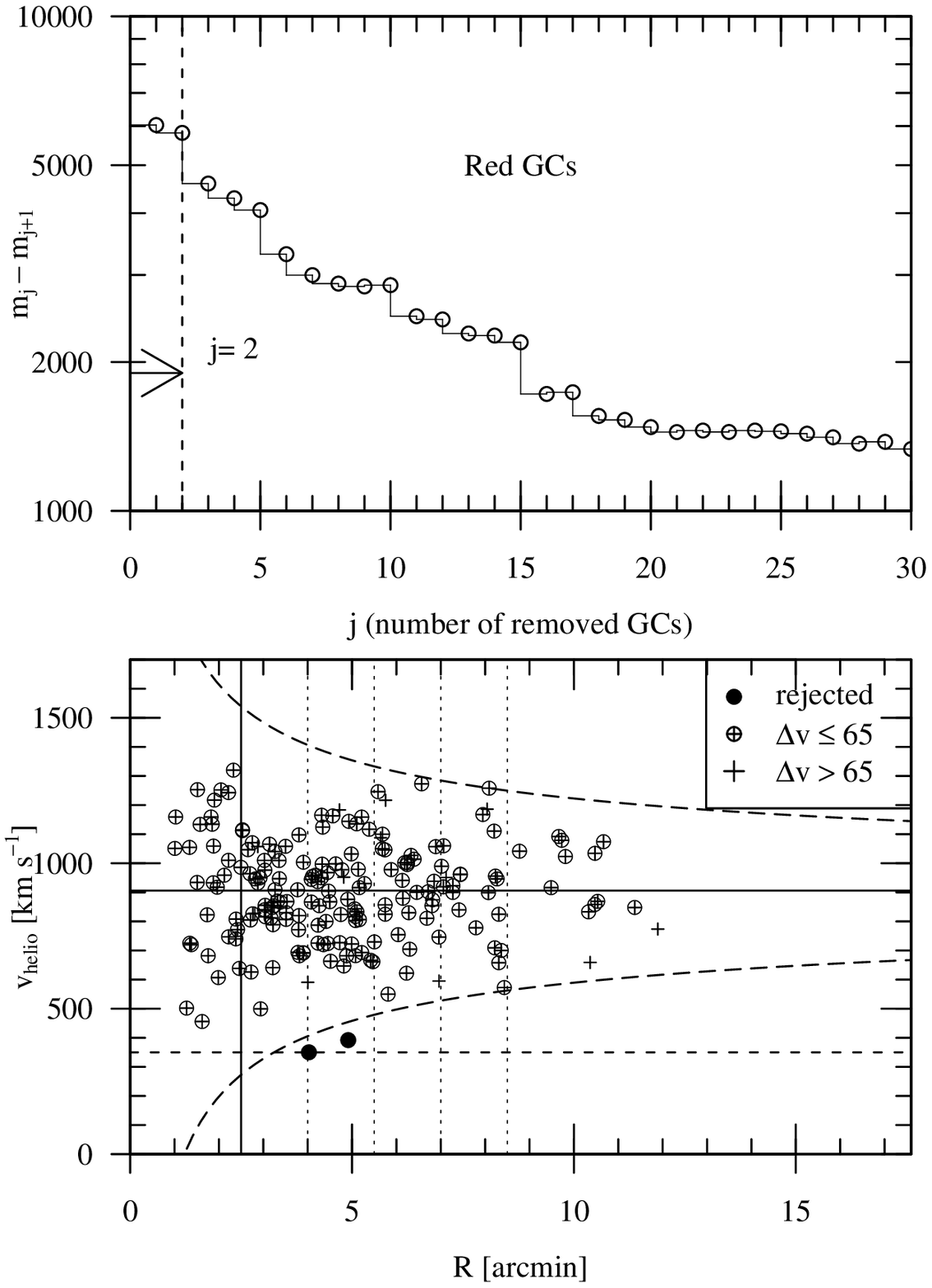}
\includegraphics[width=0.49\textwidth]{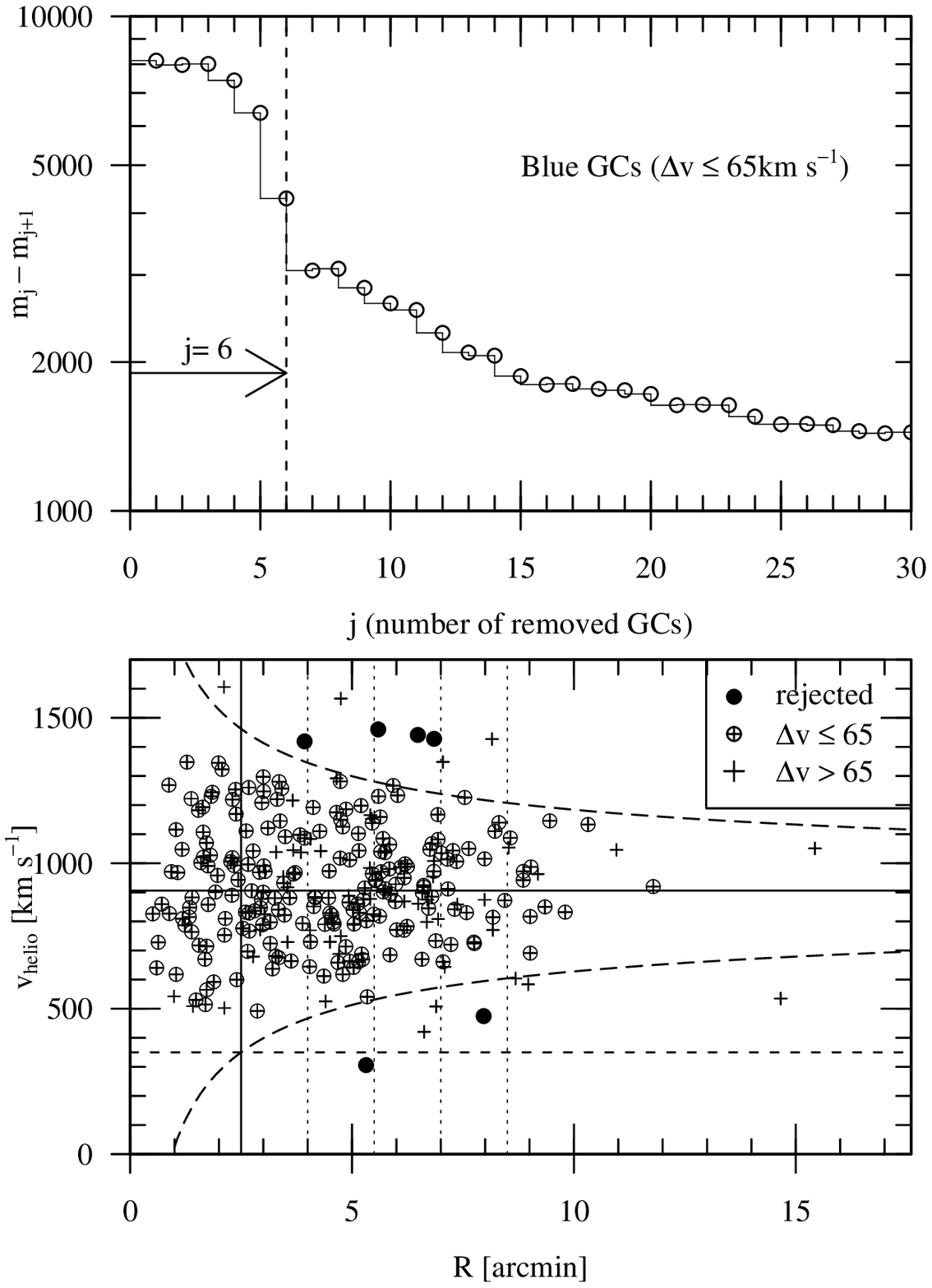}
\caption[]{Outlier removal using the
  $\max(R\cdot v^2)$ algorithm. \textbf{Left:} All red GCs.  In the
  upper panel, we plot the quantity $(m_j-m_{j+1})$ against the index
  $j$. The number of removed GCs ($j=2$) is indicated by the vertical
  dashed line and the arrow.  In the lower sub--panel, we plot the
  heliocentric velocities vs.~projected radius.  Crosses mark GCs with
  velocity uncertainties $\Delta v>65\,\rm{km\,s}^{-1} $, and
  circles indicate GCs with $\Delta v \leq 65\,\rm{km\,s}^{-1}
  $. The rejected GCs are shown as dots.  \textbf{Right:} The same
  algorithm applied to the Blue GCs with
  $\Delta v\leq65\rm{km\,s}^{-1}$. Six GCs are rejected.  In both
  lower sub--panels, the systemic velocity of NGC\,4636 is indicated
  by a solid horizontal line. The vertical solid line at $2\farcmin5$
  indicates the radius inside which the photometry does not permit a
  separation of blue and red GCs. The dotted vertical lines show the
  bins used for the dispersion profiles shown in
  Fig.~\ref{fig:4636disp}.}
\label{fig:outrej}
\end{figure*}

In this section, we apply the same outlier rejection method as
described in our study of the NGC\,1399 GCS \citep{schuberth09} to our
data. The method which is based on the tracer mass estimator by
\cite{evans03} works as follows: For each subsample under
consideration, we start by calculating the quantity
\begin{equation}m_N=\frac{1}{N}\sum_{i=1}^{N} {v_i}^2\cdot R_i\;,
\label{eq:mn}
\end{equation}
where $v_i$ are the relative velocities of the GCs, and $R_i$
their projected galactocentric distances, and $N$ is the number of
GCs.  Now, we remove the GC with the largest contribution to $m_N$,
i.e.~$\max(v^2\cdot R)$ and calculate the quantity in
Eq.~\ref{eq:mn} for the remaining $N-1$  GCs, and so on.  In the upper panels
of Fig.~\ref{fig:outrej} we plot the difference between $m_j$ and
$m_{j+1}$ against $j$, the index which labels the GCs in order of
decreasing $v^2\cdot R$.\par For the blue GCs (shown in the right
panel of Fig.~\ref{fig:outrej}) we consider only objects with velocity
uncertainties $\Delta v\leq 65\,\rm{km\,s}^{-1}$.  The function
$m_j-m_{j+1}$ levels out after the removal of six GCs which are shown
as dots in the lower panel of that figure. The thin solid curves
enveloping the remaining GCs are of the form
\begin{equation} 
v_{\mathrm{max}}(R)=\sqrt{\frac{\mathcal{C}_{\rm{max}}}{{R}}}\;,
\label{eq:velenv}
\end{equation}
where $\mathcal{C}_{\rm{max}}$ is the product $v^2\cdot R$ for the
first GC that is not rejected.  \par For the red GCs, the convergence
of $(m_j-m_{j+1})$ is not as clear.  We reject the two objects with
low velocities which are most likely foreground stars (objects 6 and 7
from the list in Sect.~\ref{sect:fgstars}), and find that the distribution of
the remaining red GCs is symmetric with respect to the systemic
velocity. 

%===== Figure (8) ==========

\subsection{The subsamples}
\label{sect:4636subsamples}

For the analysis of the kinematic properties of the NGC\,4636 GCS, we
use the following four subsamples:
\begin{itemize}
\item{{\it{\textbf{Blue}}} All 209 blue GCs with $R\geq2\farcmin5$}
\item{{\it\textbf{BlueFinal}} 156 Blue GCs with $R\geq2\farcmin5$, $\Delta v\leq65$, six  outliers removed}
\item{{\it\textbf{Red}} All 162 red GCs with $R\geq2\farcmin5$}
\item{{\it\textbf{RedFinal}} 160 Red GCs with $R\geq2\farcmin5$, two outliers
removed\footnote{After removing the two outliers from the red GC sample, 
we find that further restricting the sample objects with velocity 
uncertainties below $60\,\rm{km\,s}^{-1}$ does not significantly change the 
velocity dispersion profile (cf. Table \ref{tab:fixedbin}). Therefore, our 
final red sample still contains 13 GCs with $\mathrm \Delta v > 65 
\rm{km\,s}^{-1}$.}}
\end{itemize}
For reference, we compare these to the full sample
{\it{(\textbf{All})}} of 459 GCs and the sample labelled
{\it{\textbf{AllFinal}}} which combines {\it\textbf{BlueFinal}} and 
{{\it\textbf{RedFinal}} and hence comprises 316 GCs. 
Details on the line--of--sight velocity distributions (LOSVDs) of
these samples are given in the following Section.

\section{The line--of sight velocity distribution}
\label{sect:4636losvd}

%===== TABLE (4) ==========
\begin{table*}
\caption[NGC\,4636: Statistical properties of the spectroscopic GC
sample]{Statistical properties of the NGC\,4636 globular cluster sample}
 \centering
%\resizebox{0.99\textwidth}{!}
{
% Col.~(10) gives the range of
%galactocentric distances in units of arcmins.
\begin{tabular}{lllc c r@{$\pm$}l r@{$\pm$}l r@{$\pm$}ll } \hline \hline
%\multicolumn{1}{c}{(1)} & \multicolumn{1}{c}{(2)} &\multicolumn{1}{c}{(3)} & \multicolumn{1}{c}{(4)} & \multicolumn{2}{c}{(5)} & \multicolumn{2}{c}{(6)} & \multicolumn{2}{c}{(7)} & \multicolumn{1}{c}{(8)} & \multicolumn{1}{c}{(9)}\\
ID & Sample & 
\multicolumn{1}{c}{$N_{\textrm{GC}}$} & \multicolumn{1}{c}{$\bar{v}$} & \multicolumn{1}{c}{$\tilde{v}$} & 
 \multicolumn{2}{c}{$\sigma\pm\Delta\sigma$} &
\multicolumn{2}{c}{Skew} & \multicolumn{2}{c}{$\kappa$} &\multicolumn{1}{c}{$p$ (AD)}
%& \multicolumn{1}{c}{range} 
 \\  
% & 
%\multicolumn{1}{c}{} & \multicolumn{1}{c}{$[\textrm{km\,s}^{-1}]$} & \multicolumn{1}{c}{$[\textrm{km\,s}^{-1}]$} & 
% \multicolumn{2}{c}{$[\textrm{km\,s}^{-1}]$} &
%\multicolumn{2}{c}{} & \multicolumn{2}{c}{} &\multicolumn{1}{c}{$[']$}& \multicolumn{1}{c}{arcmin}  \\  \hline
(1) & \multicolumn{1}{l}{(2)} & \multicolumn{1}{c}{(3)} &\multicolumn{1}{c}{(4)} & \multicolumn{1}{c}{(5)} & \multicolumn{2}{c}{(6)} & \multicolumn{2}{c}{(7)} & \multicolumn{2}{c}{(8)} & \multicolumn{1}{c}{(9)} %& \multicolumn{1}{c}{(10)}
\\
\hline
%[1] "NUMBER 459"
%[1] "VMEAN  916"
%[1] "MEDIAN  909"
%[1] "DISPERS 197 \\pm 6.9"
%[1] "SKEWNESS 0.1737 \\pm 0.115"
%[1] "KURTOSIS 0.0537 \\pm 0.229"
%        Anderson-Darling normality test
%data:  tmp$v.mean 
%A = 0.3318, p-value = 0.5109
{\it{All}} &
& $459$
& $916$
& $909$
& $197 $& $ 7$
& $ 0.17 $&$ 0.12$
& $ 0.05 $&$ 0.23$
& $ 0.15$ \\
%[1] "NUMBER 387"
%[1] "VMEAN  920"
%[1] "MEDIAN  912"
%[1] "DISPERS 181 \\pm 6.8"
%[1] "SKEWNESS 0.1066 \\pm 0.0875"
%[1] "KURTOSIS -0.449 \\pm 0.13"
%        Anderson-Darling normality test
%data:  tmp$v.mean 
%A = 0.5503, p-value = 0.1555
{\it{AllFinal}} & $R\geq2\farcmin5$ \& $\Delta v\leq65$ \& OutRej
& $387$
& $920$
& $912$
& $181 $& $ 7$
& $ 0.11 $&$ 0.09$
& $ -0.45 $&$ 0.13$
& $ 0.15$ \\ \hline
%%%%%%%%%%%%%%%%%%%%%%%%%%%%%%%%%%%%%%%%%%%%%%%
%[1] "NUMBER 209"
%[1] "VMEAN  928"
%[1] "MEDIAN  912"
%[1] "DISPERS 198 \\pm  11"
%[1] "SKEWNESS 0.2976 \\pm 0.152"
%[1] "KURTOSIS 0.031 \\pm 0.276"
%        Anderson-Darling normality test
%data:  tmp$v.mean 
%A = 0.4287, p-value = 0.3077
{\it{Blue}} &$C\!-\!R\leq 1.55$ \& $R\geq2\farcmin5$ 
& $209$
& $928$
& $912$
& $ 198 $& $ 11$
& $ 0.30 $&$ 0.15$
& $ 0.03 $&$ 0.28$
& $ 0.31$ \\
%[1] "NUMBER 156"
%[1] "VMEAN  925"
%[1] "MEDIAN  912"
%[1] "DISPERS 173 \\pm  10"
%[1] "SKEWNESS 0.09811 \\pm 0.124"
%[1] "KURTOSIS -0.691 \\pm 0.175"
%        Anderson-Darling normality test
%data:  tmp$v.mean 
%A = 0.5303, p-value = 0.1730
{\it{BlueFinal}} &$C\!-\!R\leq 1.55$ \& $R\geq2\farcmin5$   \& $\Delta v\leq 65$  \& OutRej 
& $156$
& $925$
& $912$
& $ 173 $& $ 10$
& $ -0.10 $&$ 0.12$
& $ -0.69 $&$ 0.18$
& $ 0.17$ \\ \hline
%%%%%%%%%%%%%%%%%%%%%%%%%%%%%%%%%%%%%%%%%%%%%%%
%[1] "NUMBER 162"
%[1] "VMEAN  893"
%[1] "MEDIAN  900"
%[1] "DISPERS 162 \\pm 9.7"
%[1] "SKEWNESS -0.2874 \\pm 0.199"
%[1] "KURTOSIS 0.117 \\pm 0.393"
%        Anderson-Darling normality test
%data:  tmp$v.mean 
%A = 0.2797, p-value = 0.6411
{\it{Red}}  &$C\!-\!R> 1.55$ \& $R\geq2\farcmin5$ 
& $162$
& $893$
& $900$
& $ 162 $& $ 10$
& $ -0.03 $&$ 0.20$
& $ 0.12 $&$ 0.40$
& $ 0.64$\\
%[1] "NUMBER 160"
%[1] "VMEAN  900"
%[1] "MEDIAN  901"
%[1] "DISPERS 153 \\pm 9.2"
%[1] "SKEWNESS -0.02681 \\pm 0.131"
%[1] "KURTOSIS -0.481 \\pm 0.198"
%        Anderson-Darling normality test
%data:  tmp$v.mean 
%A = 0.2575, p-value = 0.7156
{\it{RedFinal}} &$C\!-\!R> 1.55$ \& $R\geq2\farcmin5$  \& OutRej
& $160$
& $900$
& $901$
& $ 153 $& $ 10$
& $ -0.03 $&$ 0.14$
& $ -0.48 $&$ 0.20$
& $ 0.72$
 \\ \hline \hline
\end{tabular}
}
\normalsize
\label{tab:4636losvd}
\note{Cols.~(1)
and (2) label the samples, and the number of GCs is given in
(3). Cols.~(4) and (5) list the mean and median radial velocity (in
$\textrm{km\,s}^{-1}$). The velocity dispersion in
($\textrm{km\,s}^{-1}$) as computed with the \cite{pm93} formula is
given in Col.~(6). Cols.~(7) and (8) give the skewness and the reduced
kurtosis, the uncertainties were estimated using a bootstrap with 999
resamplings. The $p$--value returned by the Anderson--Darling test for
normality is given in Col.~(9).}
\end{table*}

%===== Figure (9) ==========
\begin{figure}
\includegraphics[width=0.49\textwidth]{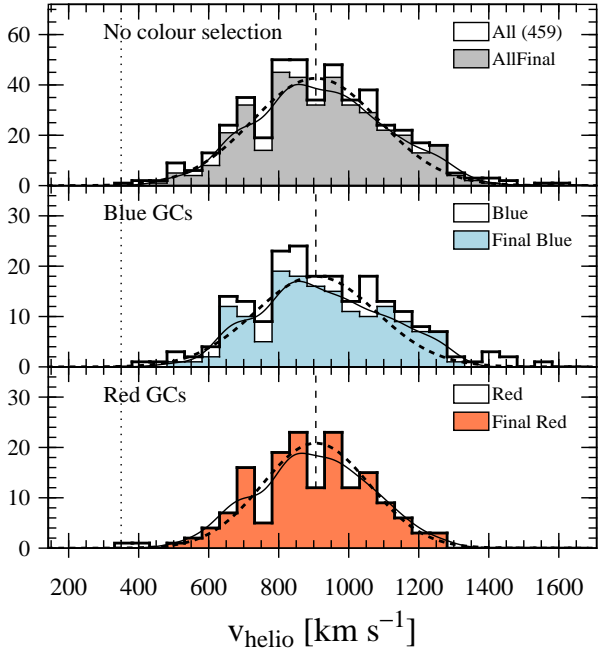}
\caption[]{ {NGC\,4636 line--of--sight velocity distributions. From
top to bottom, the panels show the velocity distribution for the
entire sample, the blue, and the red subsample, as defined in
Sect.~\ref{sect:4636subsamples}. In all panels, the vertical dashed
line shows the systemic velocity of NGC\,4636. The unfilled histograms
show the respective velocity distributions prior to any quality
selection and outlier rejection. The filled histograms show the final
samples. In all panels, the dashed curve shows the Gaussian
corresponding to the dispersion value for the respective final sample
as given in Table\,\ref{tab:4636losvd}. The thin solid line is the
corresponding Gaussian kernel density estimate (for a bandwidth of
$50\,\rm{km\,s}^{-1}$, same as the bin width). }}
\label{fig:4636velhist}
\end{figure}

Figure\;\ref{fig:4636velhist} shows the line--of--sight velocity
distributions (LOSVDs) for the samples defined above in
Sect.~\ref{sect:4636subsamples}.  In each sub--panel, the unfilled
histogram shows the respective initial sample, the filled histogram
bars show the corresponding final samples. Below we comment on the
statistical properties of these distributions which are compiled in
Table\,\ref{tab:4636losvd}.

\subsection{The Anderson--Darling test for normality}

When adopting $p\leq0.05$ as criterion for rejecting the Null
hypothesis of normality, we find that all subsamples are consistent
with being drawn from a normal distribution: The $p$--values returned
by the Anderson--Darling test \citep{stephens74} lie in the range
$0.15\leq p \leq 0.72 $ (cf.~Table\,\ref{tab:4636losvd} Col.~9).
\subsection{The moments of the LOSVD}
 For all samples, the median value of the radial velocities agrees
 well with the systemic velocity of NGC\,4636
($906\pm 7\,\textrm{km\,s}^{-1}$, Paper\,I)\par
\subsubsection{Velocity dispersion}
The velocity dispersion values quoted in Table\,\ref{tab:4636losvd}
(Col.~6) were calculated using the expressions given by \cite{pm93},
in which the uncertainties of the individual velocity measurements are
used as weights.\par The velocity dispersion of the final blue sample
is $20\,\textrm{km\,s}^{-1}$ larger than that of the final red sample.
Although this difference cannot be considered significant (since the
values derived for blue and red GCs marginally agree within their
uncertainties), it will be shown below (Sect.~\ref{sect:4636veldisp}),
that the radial velocity \emph{dispersion profiles} of the two
subpopulations, however, are very different
(cf.~Figs.~\ref{fig:4636disp} and\,\ref{fig:4636modelcompare}).

\subsubsection{Skewness}

The skewness values (Table\,\ref{tab:4636losvd}, Col.~7) for the final
blue and red samples are consistent with being zero,~i.e.~the velocity
distributions are symmetric with respect to the systemic velocity of
NGC\,4636.  The only (significantly) non--zero skewness is found for
the blue GCs prior to weeding out outliers and GCs with large
measurement uncertainties.

\subsubsection{Kurtosis}

Column\,8 of Table\,\ref{tab:4636losvd} lists the \emph{reduced}
kurtosis (i.e.~a Gaussian distribution has $\kappa=0$) of the
respective subsamples. The fourth moment of the LOSVD reacts quite
severely to the treatment of extreme velocities: The removal of only
two clusters from the {\it{Red}} sample changes the kurtosis from
$\kappa=0.12\pm0.40$ to $\kappa=-0.48\pm0.20$ ({\it{RedFinal}}).  The
{\it{FinalBlue}} sample also has a negative kurtosis, meaning that
both distributions are more `flat--topped' than a Gaussian. Indeed,
the kernel density estimates (thin solid lines in
Fig.~\ref{fig:4636velhist}) have somewhat broader wings and a flatter
peak than the corresponding Gaussians (thick dotted curves). These
differences, however, are quite subtle and we suspect that the samples
considered in this work are probably still too small to robustly
determine the $4^{\rm{th}}$ moment of the velocity distributions.
The slightly negative kurtosis values are, however, consistent
with isotropy (a projected kurtosis of zero is expected only for
the isothermal sphere).

We conclude that the final GC samples that will be used for the
dynamical analysis are symmetric with respect to the systemic velocity
of NGC\,4636 and do not show any significant deviations from
normality.

\section{Rotation}
\label{sect:4636rot}

Due to the very inhomogeneous angular coverage of the data, the
results from the search for rotation of the NGC\,4636 GCS presented in
Paper\,I were quite uncertain. Below, we use our enlarged data set for
a re--analysis and compare the findings to the ones in Paper\,I. To
detect signs of rotation, we fit the following relation to the data:
\begin{equation}
v_r{(\Theta)}=v_{\textrm{sys}}+ A\sin\left(\Theta-\Theta_0\right)\;,
\label{eq:rot}
\end{equation}
where $v_r$ is the measured radial velocity at the azimuth angle
$\Theta$, $A$ is the amplitude (in units of $\textrm{km\,s}^{-1}$),
and $\Theta_0$ is the position angle of the axis of rotation (see
\citealt{cote01} for a detailed discussion of the method). 
\par In Table\,\ref{tab:4636rot}, we present the results of the
rotation analysis for the different subsamples of our data: Columns 2
and 3 give the values of $\Theta_0$ and the amplitude $A$ found for
the samples presented in Table\,\ref{tab:4636losvd}. Columns 4 to 6
give the corresponding results obtained for the GCs in the radial
range $2\farcmin5 < R < 8\farcmin0$, where the spatial completeness of
our sample is highest (cf.~Sect.\ref{sect:spatial}).  It appears that
the rotation signal (within the uncertainties) is robust with respect
to the radial range considered and the application of the outlier
rejection algorithm.\par To search for variations of the rotation
signal with galactocentric radius, we plot, in the left and middle
panel of Fig.~\ref{fig:4636rot}, the rotation parameters $\Theta_0$
and $A$ determined for moving bins of 50 GCs.  The following
paragraphs summarise the results for blue and red GCs.

%===== Figure (10) ==========
\begin{figure*}
\centering
\includegraphics[width=0.32\textwidth]{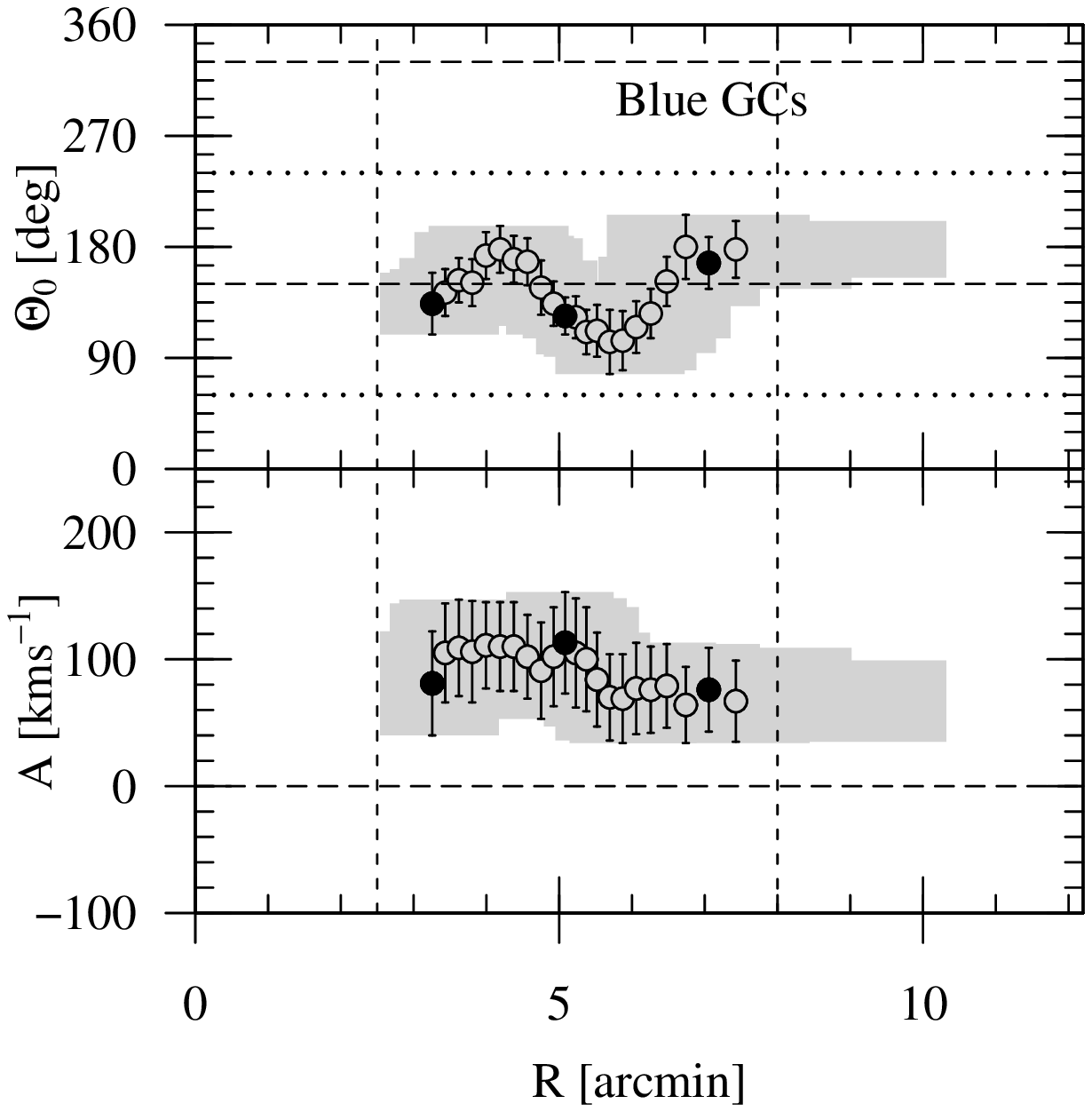}
\includegraphics[width=0.32\textwidth]{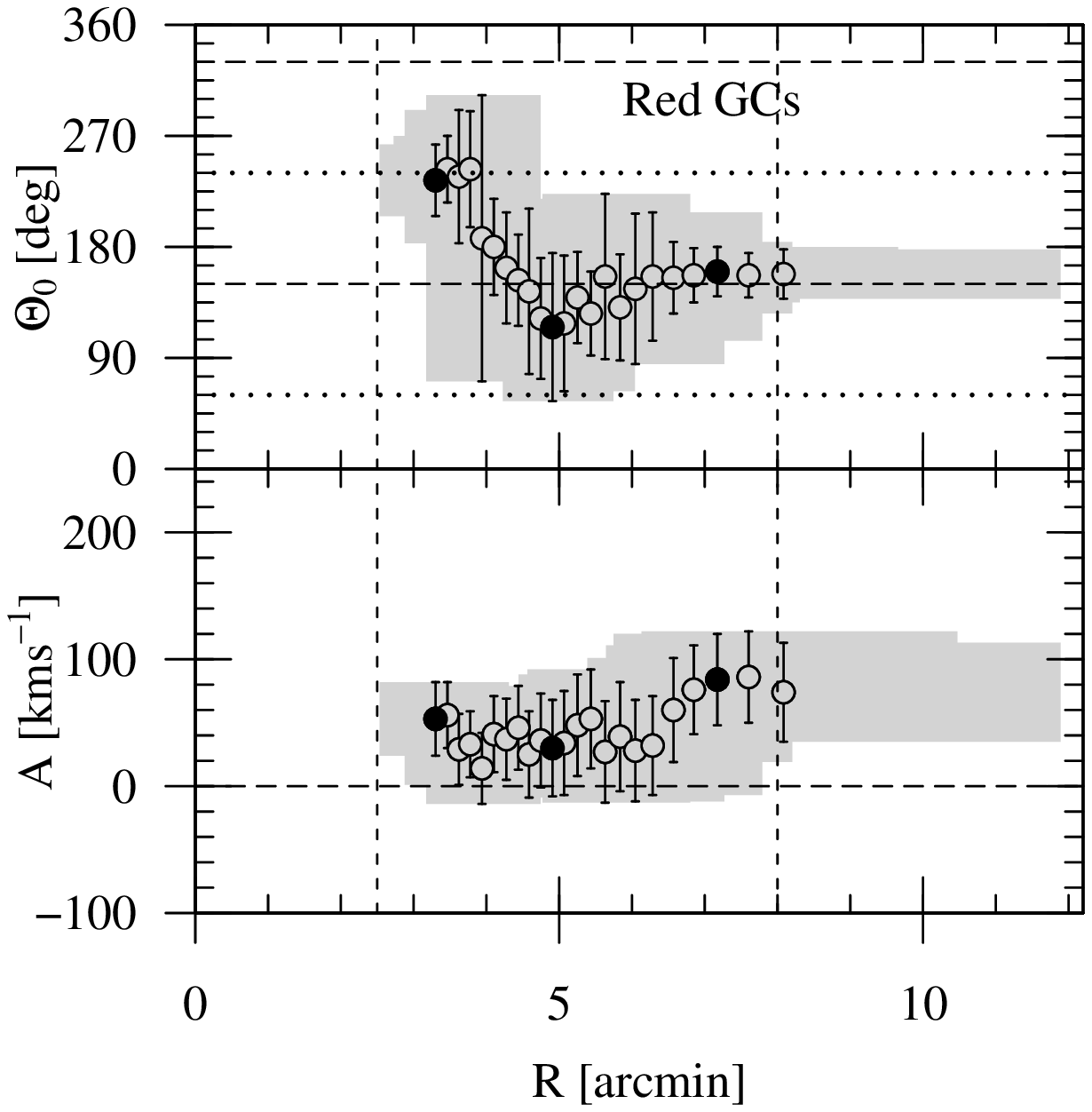}
\includegraphics[width=0.32\textwidth]{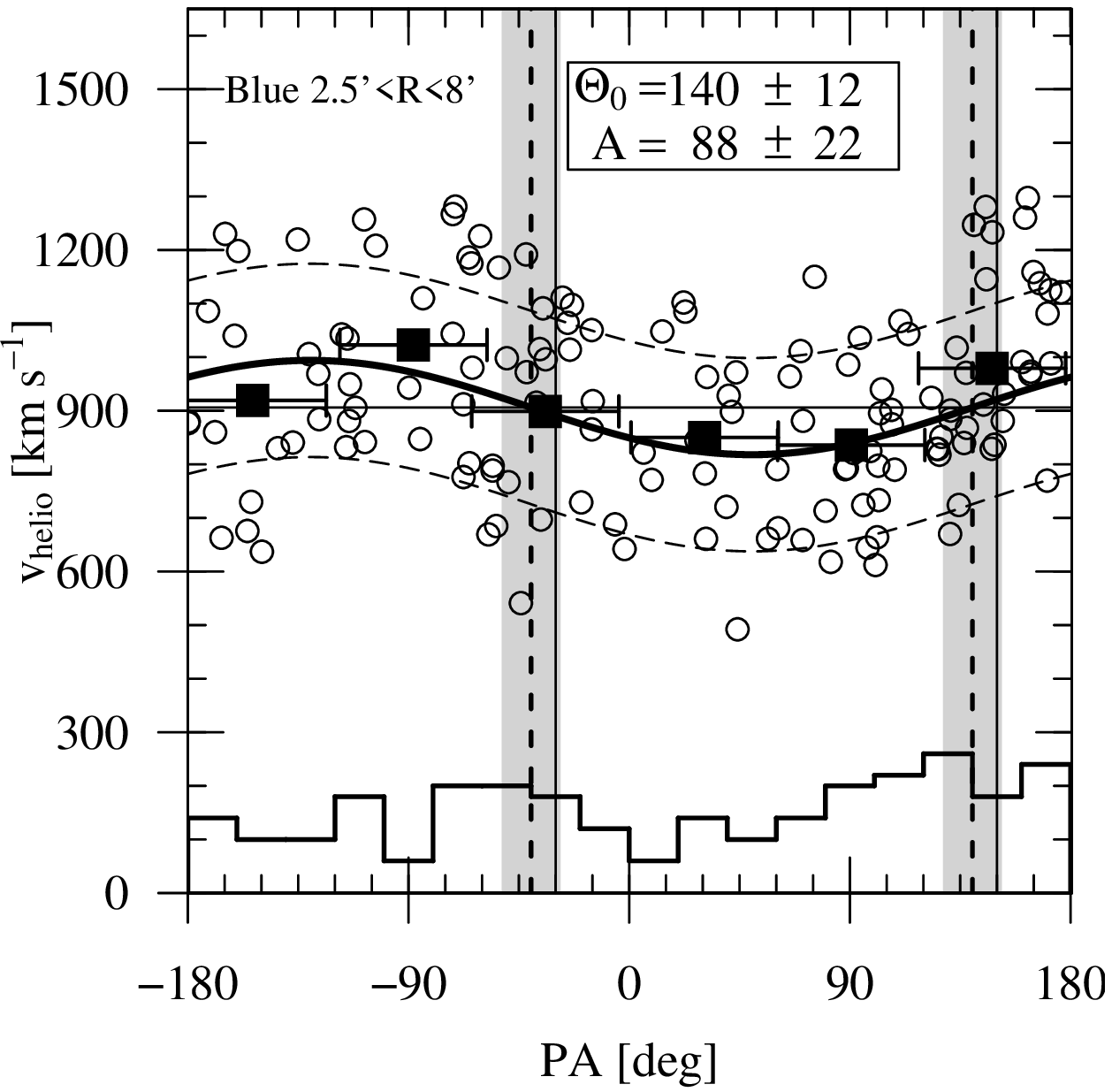}
\caption{Rotation of the NGC\,4636 globular cluster
system. \textbf{Left:} Circles show the rotation of the final blue
sample ({\it{BlueFinal}}) computed for moving bins of 50 GCs with a
step site of 5 GCs. The grey area indicates the radial coverage and
the uncertainties.  Dots indicate independent bins. The upper
sub--panel shows the rotation angle $\Theta_0$ vs.~projected
galactocentric radius. The NGC\,4636 photometric major and minor axis
are shown as long--dashed and dotted horizontal lines,
respectively. The lower sub--panel shows the amplitude $A$ as function
of radius. \textbf{Middle:} The same as the left panel but for the
final red sample ({\it{RedFinal}}). In both plots, the dashed vertical
lines indicate the radial range $2\farcmin5 < R < 8\arcmin$ for which we
have the best spatial coverage of the GCS (see also
Sect:\,\ref{sect:spatial}) \textbf{Right:} Velocity versus position
angle for the 141 blue GCs (circles) from the sample {\it{BlueFinal}}
with distances between $2\farcmin5$ and $8\arcmin$.  The thick solid
curve shows the best fit of Eq.\,\ref{eq:rot} to the data, the thin
dashed curves are offset by the sample dispersion
($\sigma=180\,\textrm{km\,s}^{-1}$). The vertical dashed and solid
lines show the best--fit $\Theta_0$ and the photometric major axis of
NGC\,4636, respectively. The grey areas indicate the uncertainty
$\Delta\Theta_0$.  The squares show the mean velocity for bins of
$60^\circ$.  The unfilled histogram at the bottom shows the angular
distribution of the GCs (the bins have a width of $20^\circ$ and the
counts are multiplied by a factor of 20 for graphic convenience).  }
\label{fig:4636rot}
\end{figure*}

%===== TABLE (6) ==========
\begin{table}
\caption[]{Rotation of the NGC\,4636 globular cluster system}
\centering
\begin{tabular}{lr@{$\pm$}lr@{$\pm$}llr@{$\pm$}lr@{$\pm$}lll} \hline \hline
 & 
\multicolumn{4}{c}{ } & 
\multicolumn{4}{c}{{$ 2\farcmin5 \leq R \leq 8\arcmin$}\rule{0ex}{2ex}} \\   \cline{6-10}
Sample ID & %$N_\textrm{GC}$ &
\multicolumn{2}{c}{$\Theta_0$} &
\multicolumn{2}{c}{ $A$} &
$N_\textrm{GC}$ &
\multicolumn{2}{c}{$\Theta_0$} &
\multicolumn{2}{c}{ $A$} \\
 & %$N_\textrm{GC}$ &
\multicolumn{2}{c}{$[\rm{deg}]$} &
\multicolumn{2}{c}{ $[\rm{km\,s}^{-1}]$} &
 &
\multicolumn{2}{c}{$[\rm{deg}]$} &
\multicolumn{2}{c}{ $[\rm{km\,s}^{-1}]$} 
\\
(1) & 
\multicolumn{2}{c}{(2)} & 
\multicolumn{2}{c}{(3)} & 
(4)& 
\multicolumn{2}{c}{(5)} & 
\multicolumn{2}{c}{(6)} 
\\ \hline
{\it{All}} & $160$  &$21$  & $34$& $14$ &325  &152  &16  &49  &16 \\
{\it{All.Final}} & $150$  &$15$  & $46$& $15$ & 269 & 141 &12  & 59 & 15\\
{\it{Blue}\rule{0ex}{2.5ex}} & $156$  &$18$  & $58$& $21$ & 185 & 144 & 17 & 66 & 23\\
{\it{Blue.Final}} & $144$  &$13$  & $77$& $21$ & 141 & 140 & 12 & 88 & 22\\
{\it{Red}}\rule{0ex}{2.5ex} & $172$  &$27$  & $37$& $19$ & 138 & 175 & 35 & 32 & 20\\
{\it{Red.Final}} & $171$  &$26$  & $36$& $18$ & 136 & 174 & 34 & 30 & 19\\
\hline \hline
\end{tabular}
\note{The position angle of NGC\,4636 is $150^{\circ}$. The first
  column gives the sample identifier (Sect.~\ref{sect:4636subsamples}). Columns 2 and 3 give the axis of
  rotation and the amplitude, respectively. Columns 4 through 6 give the
  number of GCs, rotation angle and amplitude for the samples restricted
  to radii $R\leq 8\arcmin$ (i.e.~the range in which the completeness of
  the spectroscopic GC sample is large).}
\label{tab:4636rot}
\end{table}

\subsection{Rotation of the blue GCs}

Table\,\ref{tab:4636rot} shows that the rotation signature is
strongest for the blue GCs (final sample). 

\par In the right panel of Fig.~\ref{fig:4636rot}, we plot the radial
velocities against the position angle for the final blue sample
restricted to galactocentric distances below 8\arcmin{} (which, in
Table\,\ref{tab:4636rot} is the sample with the strongest rotation
signature). The data for the 141 GCs are shown as circles, and the
 squares mark the mean velocity calculated for $60^\circ$ wide
bins.  The thick solid line shows Eq.~\ref{eq:rot}
($A=88\,\rm{km\,s}^{-1}$ and $\Theta_0=140^\circ$).

\subsection{Rotation of the red GCs}

For the red GC samples, the overall rotation signature as quoted in
Table\,\ref{tab:4636rot} is weaker than that of the blue GCs. Within
$R\la 6\farcmin0$, the amplitude for the final sample plotted in
the middle panel of Fig.~\ref{fig:4636rot} is consistent with being
zero (104 GCs with $\mathrm 2\farcmin05\la R \la 6\farcmin0 :
A=17\pm22 \rm{km\,s}^{-1}, \Theta_0=186\pm73^\circ$).
Only the last few bins suggest an increase with the amplitude
reaching values of $\sim80\,\rm{km\,s}^{-1}$ (32 GCs with $\mathrm 
6\farcmin0\la R \la 8\farcmin0:
A=82\pm37 \rm{km\,s}^{-1}, \Theta_0=160\pm21^\circ$).

Between 3\arcmin and 5\arcmin, the axis of rotation changes from being
aligned with the minor axis to the photometric major axis of
NGC\,4636.  Given the low values of the amplitude at these radii,
however, this change of the axis of rotation remains uncertain. For
radii beyond 5\arcmin, the axis of rotation remains constant and
coincides with the major axis.

\subsection{Comparison to Paper\,I}

The findings in this section deviate significantly from the values
presented in Paper\,I, where no rotation was detected for the blue
GCs, while the strongest signal was found for the red clusters within
4\arcmin{} ($\Theta_0=72\pm25^\circ$, $A=-144\pm
44\,\textrm{km\,s}^{-1}$).  \par The discrepant findings are likely
due to the azimuthal incompleteness of the data in Paper\,I. 

\section{Globular cluster velocity dispersion profiles}
\label{sect:4636veldisp}

\begin{table*}
\caption[]{Velocity dispersion profiles for fixed radial bins}
\centering
\begin{tabular}{ll@{-- }rlr@{$\pm$}llr@{$\pm$}llr@{$\pm$}rlr@{$\pm$}l}  \hline  \hline
&\multicolumn{2}{c}{}& \multicolumn{3}{c}{\it{Blue}}  & 
\multicolumn{3}{c}{\it{BlueFinal}} & 
\multicolumn{3}{c}{\it{Red}} & 
\multicolumn{3}{c}{\it{RedFinal}} 
\\
$\rm{N}^{\rm{o}}$ & \multicolumn{2}{l}{range} & $n$ &\multicolumn{2}{c}{$\sigma$} 
& $n$ &\multicolumn{2}{c}{$\sigma$} 
& $n$ &\multicolumn{2}{c}{$\sigma$} 
& $n$ &\multicolumn{2}{c}{$\sigma$} 
 \\
(1)&\multicolumn{2}{c}{(2)}&
(3)&\multicolumn{2}{c}{(4)}&
(5)&\multicolumn{2}{c}{(6)}&
(7)&\multicolumn{2}{c}{(8)}&
(9)&\multicolumn{2}{c}{(10)}
\\ \hline 
1& $0\farcmin0$ & $2\farcmin5$ & 56 & 249 &25&52&226&23&30&235&32 &30&235&32\\ 
2& $2\farcmin5$ & $4\farcmin0$ & 55 & 199 &21&44&195&22&45&140&16 &45&140&45\\
3& $4\farcmin0$ & $5\farcmin5$ & 60 & 200 &20&45&186&21&50&183&21 &48&154&48\\
4& $5\farcmin5$ & $7\farcmin0$ & 51 & 200 &23&38&146&21&32&169&23 &32&169&32\\
5& $7\farcmin0$ & $8\farcmin5$ & 25 & 206 &32&18&154&28&22&165&27 &22&165&22\\
6& $8\farcmin5$ & $15\farcmin5$& 18 & 166 &33&11&131&34&13&122&31 &13&122&13\\ 
 \hline \hline
\end{tabular}
\note{Column\,1 gives the bin number and Col.\,2 is the radial range covered 
by a given bin. Col.\,3 is the number of GCs in the {\it Blue} subsample,
Col.\,4 the line¿of¿sight velocity dispersion in $\rm{km\,s}^{-1}$. Cols.\,5 
and 6 are the same for the {\it BlueFinal} sample. The corresponding values 
for the red GCs are given in Cols.\,7 through 10. Columns 11 and 12 show the 
values obtained when further restricting the {\it RedFinal} sample to 
velocities with uncertainties below $3\leq65 \rm{km\,s}^{-1}$.}
\label{tab:fixedbin}
\end{table*}

\begin{table}
\centering
\caption[]{GC Velocity dispersion profiles for constant number bins}
\begin{tabular}{cr@{$\farcmin$}lr@{$\pm$}ll|l@{$\farcmin$}rr@{$\pm$}ll} \hline \hline
%BLUE RED
%rbin sig d.sig ngal rbin sig d.sig ngal
&\multicolumn{5}{l|}{Blue GCs ({\it{BlueFinal}})}&  
\multicolumn{5}{l}{Red GCs ({\it{RedFinal}})}\\ 
\multicolumn{1}{c}{No.}&\multicolumn{2}{c}{$\bar{R}$} & $\sigma$ & $\Delta\sigma$ & $n$ & 
\multicolumn{2}{c}{$\bar{R}$}  & $\sigma$ & $\Delta\sigma$ & $n$\\  
(1) & \multicolumn{2}{c}{(2)}  &\multicolumn{2}{c}{(3)} & (4) 
& \multicolumn{2}{c}{(5)}  &\multicolumn{2}{c}{(6)} & (7) \\
\hline
&1&20 &  208&  30& 26 & 
\multicolumn{2}{l}{\ldots} & \multicolumn{2}{c}{\ldots} & \ldots \\
&2&03 &  242&  35&  26 & 1&85& 235& 31 &30  \\ \hline
1&3&05& 202& 24 &39 &3&13& 135& 16 &40\\
2&4&55& 179& 21& 39& 4&44& 152& 19& 40\\
3&5&75& 162& 20& 39& 5&81& 172& 21& 40\\
4&7&92& 143& 19& 39& 8&48& 151& 19& 40\\ \hline \hline
\end{tabular}
\note{The first column numbers the bins (for
GCs with $R>2\farcmin5$). Column\,(2) gives the mean galactocentric
distance of the blue GCs, Col.~3 gives the velocity dispersion in
units of $\rm{km\,s}^{-1}$, and Col.~4 is the number of blue GCs in
that bin. Columns 5 through 8 give the corresponding values for the
red GCs. For completeness, the first two rows give the values for GCs
inside $2\farcmin5$ where the colour distribution is not bimodal. }
\label{tab:4636DispMod}
\end{table}

Figure~\ref{fig:4636disp} shows the velocity dispersion profiles
obtained for the blue (upper panel) and red (lower panel) subsamples
defined in Sect.~\ref{sect:4636subsamples}.  The GC velocities were
grouped into six radial bins (shown as dotted lines in the lower
panels of Fig.~\ref{fig:outrej}). The first bin comprises the GCs within
2\farcmin5, and the following four bins (starting at 2\farcmin5,
4\farcmin0, 5\farcmin5 and 7\farcmin0) have a width of 1\farcmin5. The
outermost bin ($8\farcmin5< R \leq 15\farcmin5$) collects the GCs in
the more sparsely populated (and sampled) outer GCS. 
The data are given in Table\,\ref{tab:fixedbin}.
Circles are the
values obtained for the {\it{Blue}} and {\it{Red}} subsamples and dots
represent the dispersion profiles determined for the final samples
({\it{BlueFinal}} and {\it{RedFinal}}). The values from Paper\,I
(Table\,4 therein) are shown as diamonds.

\subsection{Blue GCs}

Compared to the initial sample, the removal of GCs with velocity
uncertainties $\Delta v\geq65\,\rm{km\,s}^{-1}$ and the six
probable interlopers identified in Sect.~\ref{sect:outrej}
significantly reduces the velocity dispersion, in particular in the
4th and 5th bin. The values for the final blue sample (dots) agree
very well with the data from Paper\,I (diamonds). The dispersion
profile of the blue sample {declines} with galactocentric radius.

\subsection{Red GCs}
The velocity dispersion profile of the red GCs 
(cf. Fig.~\ref{fig:4636disp}, lower panel) shows a more complex
behaviour. The sudden drop by almost 100 $\rm{km\,s}^{-1}$
from the first to the second bin was already present in the data for Paper I.
 While the poorer data in Paper I still left the
possibility of a sampling effect, our new enlarged data shows the
same feature which we attribute to the fact that blue and red GCs 
cannot be separated within the central 2\farcmin5: Inside this radius,
the presence of a substancial number of 'contaminating'metal-poor GCs
(which would have a broader velocity distribution) within our 'red' sample
dominates the measurement and leads to the very high dispersion value.

\par To search for colour trends amongst the GCs inside 2\farcmin5, we 
plot in Fig.\ref{fig:4636dispcolour} (upper panel), the  the line-of-sight 
velocity dispersion as a function of colour. Dots represent the 82 GCs
within 2\farcmin5 with velocity uncertainties below 65 $\rm{km\,s}^{-1}$.
The 316 GCs from the {\it Final} sample are shown as squares. Inside
2\farcmin5, there is no discernible trend with colour, except for a
dip near the colour used to divide blue from red GCs. For
GCs outside 2\farcmin5, we observe a constant dispersion for the red
GCs, which then increases towards bluer colours. However,
there is no sign of a 'jump' near the dividing colour as seen
in the NGC1399 GCS \citep{schuberth10}.
For GCs with projected galactocentric distances between 2\farcmin5 and 
4\farcmin0, we find a low dispersion of only $140\,\pm16\,\rm{km\,s}^{-1}$. 
This value then again rises to almost $170\,\rm{km\,s}^{-1}$ between 
5\farcmin5 and 8\farcmin0.

%===== Figure (11) ==========
\begin{figure}
\includegraphics[width=0.49\textwidth]{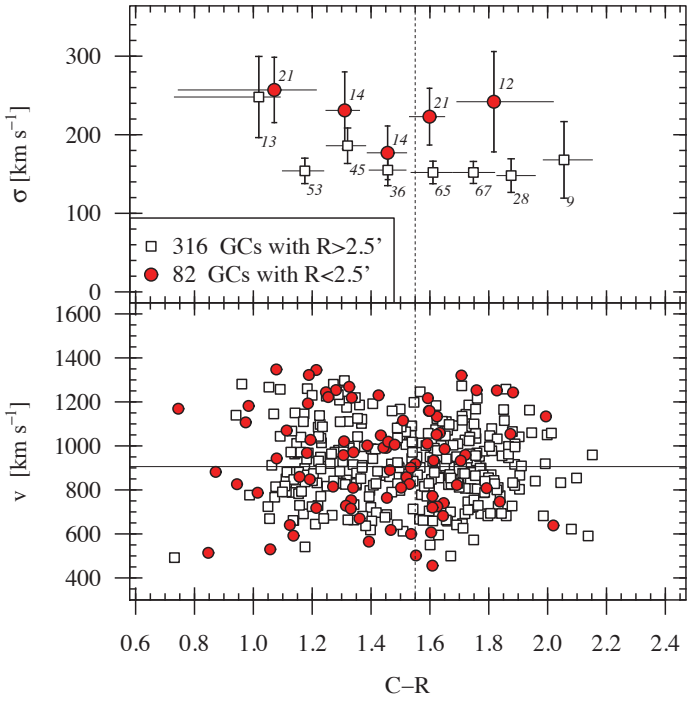}
\caption[]{{\bf Upper panel}: Line-of-sight velocity dispersion as a 
function of colour. Dots represent the values for the 82 GCs within 2\farcmin5
and velocity uncertainties below 65 $\rm{km\,s}^{-1}$, squares show the 
results for the 316 GCs of the {\it AllFinal} sample. The labels give the 
number of GCs per bin. The bin size is 0.15 mag, but the bluest and reddest 
bins have been re¿sized so that no objects are excluded. The horizontal bars 
mark the colour range covered by a given bin. {\bf Lower panel}: Heliocentric 
velocity as a function of colour. The symbols are the same as in the upper
panel.}
\label{fig:4636dispcolour}
\end{figure}

%===== Figure (12) ==========
\begin{figure}
\includegraphics[width=0.49\textwidth]{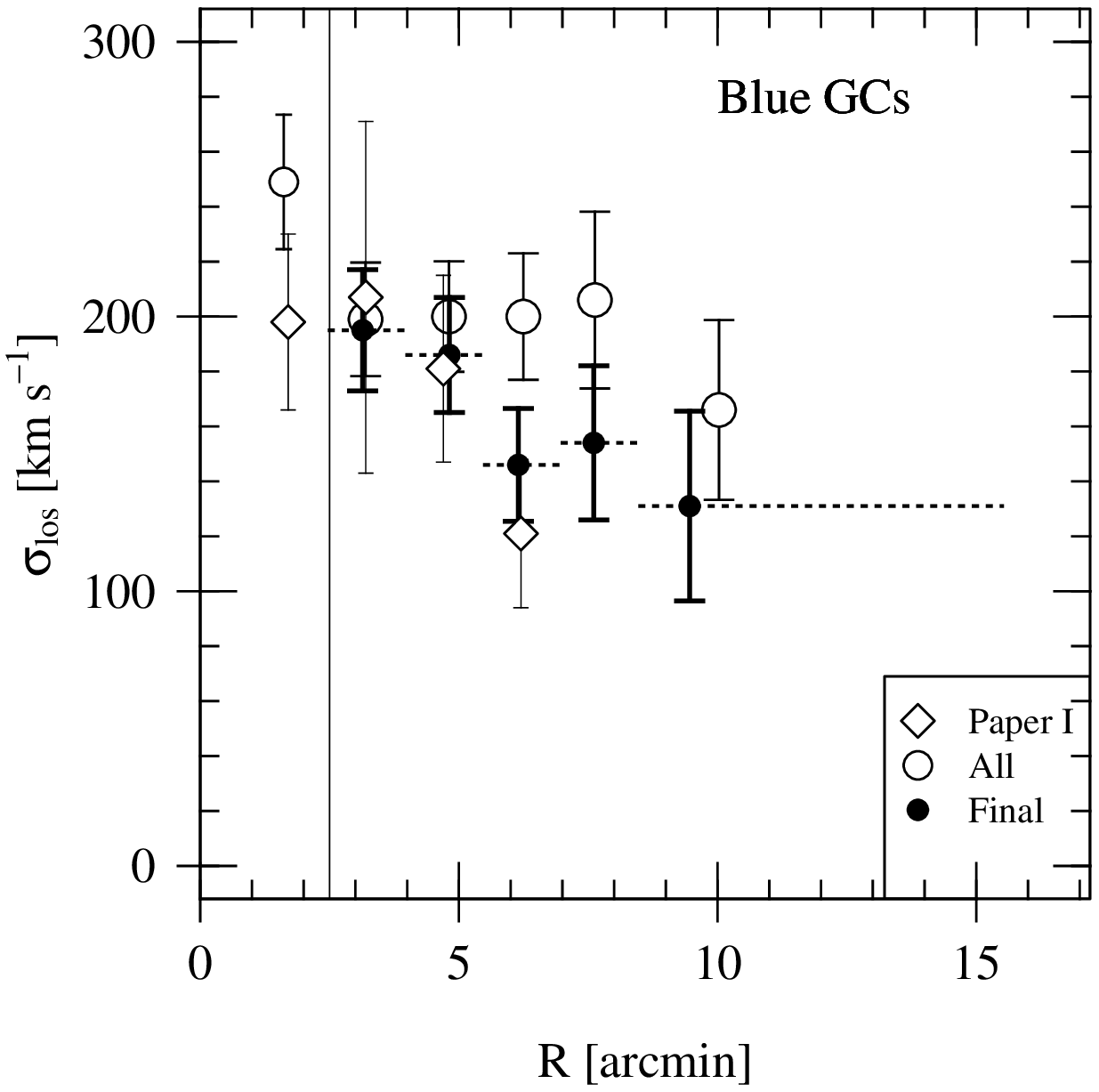} \\
\includegraphics[width=0.49\textwidth]{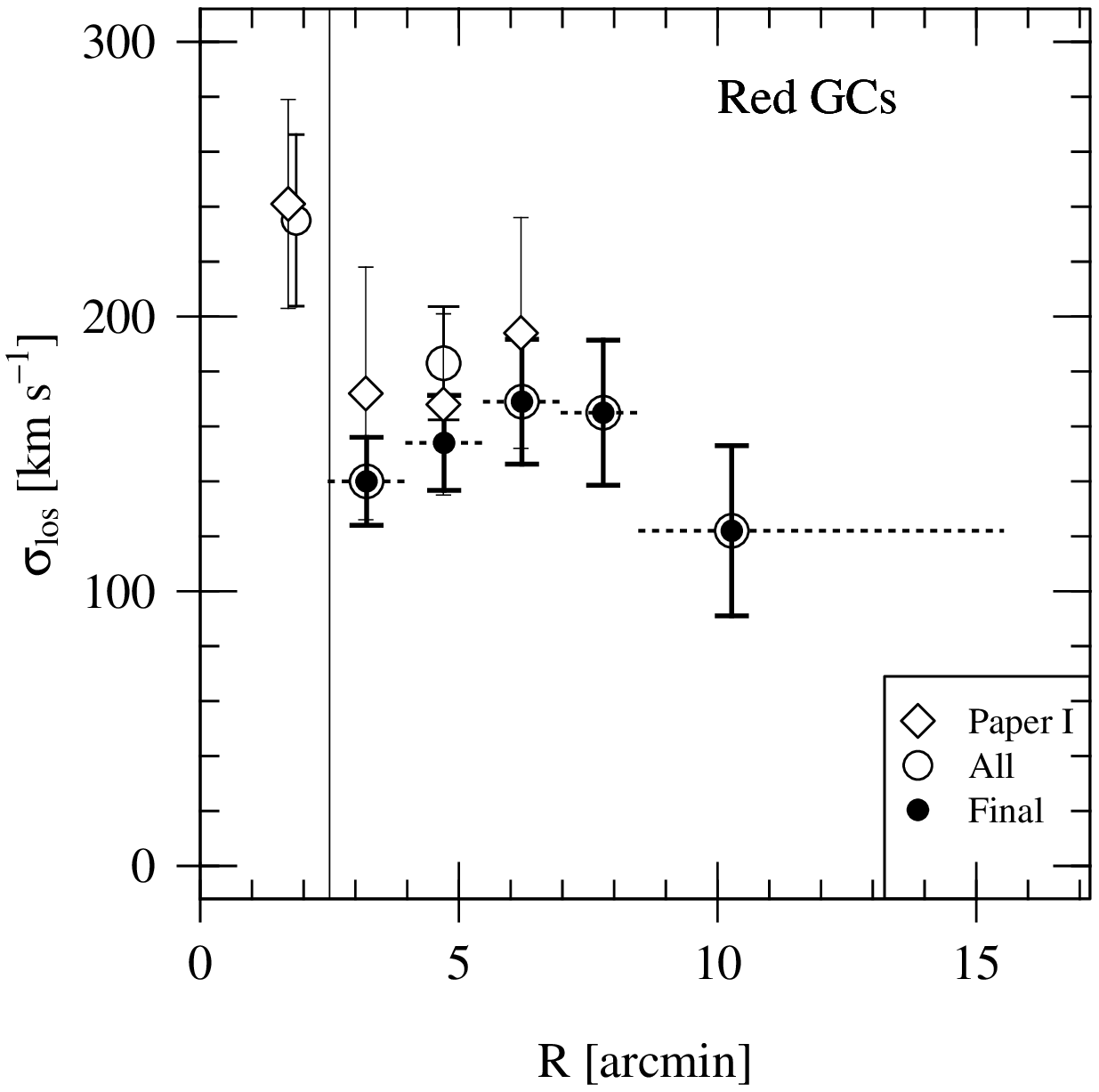}%&
\caption[]{Velocity dispersion profiles. {\textbf{Upper panel:}} Circles
show the dispersion values for the blue GCs (for fixed radial bins
cf.~Table\,\ref{tab:fixedbin}) prior to quality selection and
outlier removal. Dots are the values for the final blue sample. The
dashed horizontal `error bars' indicate the radial range of a given
bin; the bins used here are the ones indicated by the dotted lines
in the lower panels of Fig.\ref{fig:outrej}. The dispersion values
from Paper\,I are shown as diamonds. In both panels, the vertical line
at 2\farcmin5 indicates the first bin which covers the radius inside which
blue and red GCs cannot be distinguished. \textbf{Lower panel:} The same
for the red GCs.}
\label{fig:4636disp}
\end{figure}

%===== Figure (13) ==========
\begin{figure}
\includegraphics[width=0.49\textwidth]{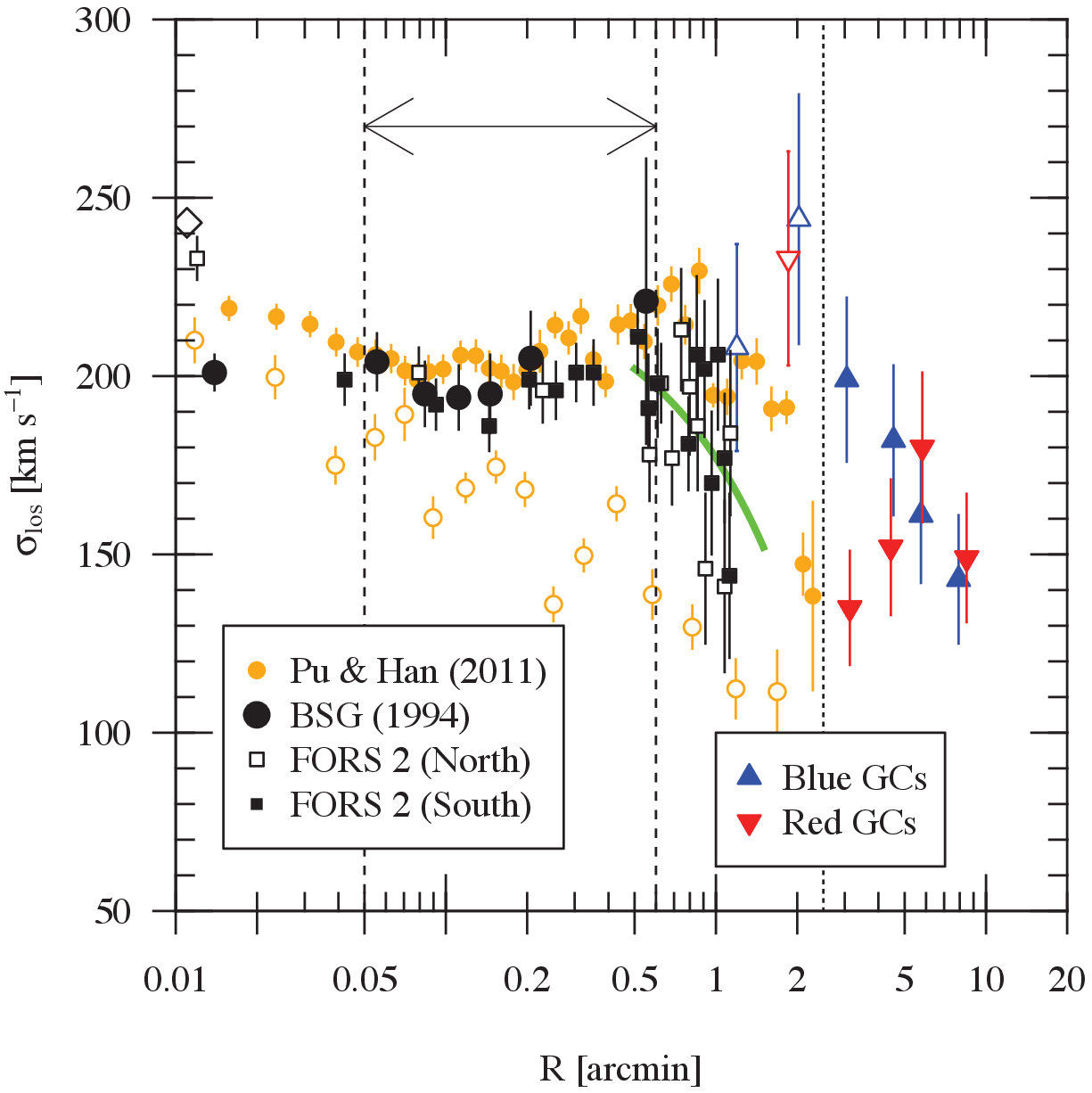}%&
\caption[]{The stellar velocity dispersion profile of NGC\,4636. The diamond
is the central dispersion measured by \cite{proctor02}. Large dots show the 
values from \cite{BSG94}, squares are the values derived from our FORS\,2
spectra (see text for details). Unfilled (filled) squares indicate slits 
North (South) of the galaxy centre. The vertical dashed lines indicate the 
region we use to model the galaxy's velocity dispersion profile
(cf.~Sect.~\ref{sect:4636stellarmod}). The solid green curve is a linear
fit to the data for $R>0\farcmin5$. The major axis measurements by
\citet{pu11} are shown as small (orange) dots. For radii $\leq 0.219'$,
where the long slit extends to either side of the galaxy, we plot the
averaged values; the uncertainties were added in quadrature.
For comparison, the data points
for blue and red GCs for constant number bins (final samples, values from
Table\,\ref{tab:4636DispMod}) are shown as triangles. The vertical
dotted line at $R=2\farcmin5$ marks the radius inside which blue and
red GCs cannot be separated.}
\label{fig:4636stellar}
\end{figure}

\section{NGC\,4636 stellar kinematics}
\label{sect:4636stars}

Long slit spectra of NGC\,4636 were obtained by \cite{BSG94}, and
\cite{kronawitter00} presented detailed modelling based on these
data. Recently, \citet{pu11} presented radial velocities and velocity
dispersions derived from deep long-slit spectra along the major and minor axis.
Their data reach out to $\sim 2\farcmin3$, corresponding to about 11.5 kpc.
\par We will use the stellar kinematics to constrain the halo
models derived for the GCs (see Sect.~\ref{sect:4636stellarmod}). 
Below, in Fig.\,\ref{fig:4636stellar}, we show the \cite{BSG94} data
and compare them to our own measurements obtained in parallel with the
GC observations.

\subsection{FORS\,2 spectra of NGC\,4636}

During the first MXU observations of the NGC\,4636 GCs, we placed 29
slits (along the North--South direction) on NGC\,4636 itself. This
mask (Mask 1\_1 from Paper\,I) has a total exposure time of 2\,hours
(7\,200\,s), and the final image is the co--addition of three
consecutive exposures, thus ensuring a good cosmic--ray rejection. The
data were reduced in the same manner as the slits targeting GCs.  The
sky was estimated from the combination of several sky--slits located
$\sim\!3\arcmin$ from the centre of NGC\,4636.  \par Using the
\texttt{pPXF} (penalised PiXel Fitting) routine by \cite{ppxf}, we
determined the line--of--sight velocity dispersion for the
one--dimensional spectra. The uncertainties were estimated via Monte
Carlo simulations in which we added noise to our spectra and performed
the analysis in the same way as with the original data. The templates
used in the analysis were taken from the \cite{vazdekis99} library of
synthetic spectra. Our data are shown in Fig.~\ref{fig:4636stellar}
where unfilled and filled squares refer to slits placed
to the North and South of the centre of NGC\,4636, respectively. The
data are listed in Table\,\ref{tab:4636table}.
\subsection{The stellar velocity dispersion profile of NGC\,4636}
In  Fig.~\ref{fig:4636stellar}, we compare our
results (squares) with the values from \cite{BSG94} (large
dots). Within the radial range $0\farcmin04\la R
\la0\farcmin6$, the agreement between both data sets is
excellent.  \par For the central velocity dispersion (measured from a
$2\arcsec$ long slit within about $1\arcsec$ of the centre of
NGC\,4636), we find $\sigma_0=233\pm6\,\rm{km\,s}^{-1}$. This is
substantially higher than the value published by BSG94
($\sigma_0=211\pm7\,\rm{km\,s}^{-1}$). However, the high central
velocity dispersion we find is supported by the measurement published
by \cite{proctor02} who found $\sigma_0=243\pm3\,\rm{km\,s}^{-1}$
(shown as diamond in the right panel of Fig.~\ref{fig:4636stellar}).
The low central dispersion quoted by BSG94 is likely due to the
instrumental setup: These authors used a slit of 2\farcsec1 width, and
at this spatial resolution the luminosity weighted dispersion measured
for the centre may be substantially lower than values obtained using a
smaller slit width: \citeauthor{proctor02} used a slit of 1\farcsec25
which provides a similar spatial resolution as our 1\farcsec0 wide MXU
slits, yielding similar dispersion values.\par For the dynamical
modelling, dispersion values at large radial distances are of
particular interest. Unfortunately, due to the low S/N in the small
slits we used, the quality of our data degrades for radial distances
beyond $R\ga0\farcmin7$: The uncertainty of the individual data
points increases and so does the scatter. However, the velocity
dispersion seems to decline as indicated by the solid curve in
Fig.~\ref{fig:4636stellar} (right panel) which shows a linear fit to
$\sigma_{\rm{los}}(R)$ for $R>0\farcmin5$.
This trend is confirmed by the recently
published measurements by \citet{pu11}  (small dots in
Fig. ~\ref{fig:4636stellar}) which clearly show a declining stellar velocity 
dispersion for radii beyond $\sim$ 0\farcmin6 ($\approx$ 3 kpc). These data 
have a higher S/N than our FORS measurements and extend out to
2\farcmin3, i.e. almost into the regime where GC dynamics is available
for both the metal-poor and the metal-rich subpopulation
(shown as filled triangles in Fig. ~\ref{fig:4636stellar}). In this context, it
is interesting to note that the very low velocity dispersion of 135 $\pm$ 16 
$\rm{km\,s}^{-1}$, observed for the red GCs near $\sim$3\farcmin1 is 
consistent with the outermost stellar velocity dispersion value
of 138 $\pm$ 26 $\rm{km\,s}^{-1}$. This might suggest a connection between
stars and metal-rich GCs similar to the one reported for NGC 1399 
\citet{schuberth10}.
\par For the dynamical
modelling of the stellar component of NGC\,4636 we use the \cite{BSG94} data
in the radial range $0\farcmin05$ to $0\farcmin6$ (0.25--3.1\,kpc),
indicated by the dashed lines in the right panel of
Fig.~\ref{fig:4636stellar}. The central data points are not included
in our modelling since the deprojection of the luminosity profile and,
by consequence, the stellar mass profile are only reliable for
$R\ga100\,\rm{pc}\simeq0\farcmin2$ (see Sect.~\ref{sect:dep4636}).

\section{Jeans models for NGC\,4636}
\label{sect:4636models}

In the next paragraphs, we give the relevant analytical expressions
and outline how we construct the spherical, non--rotating Jeans models
for NGC\,4636.\par In Paper\,I, we chose an NFW--halo \citep{nfw97} to
represent the dark matter in NGC\,4636. In this work, we will consider
both NFW halos and two mass distribution with a finite central
density: The cored profile proposed by \cite{burkert95} which has the
same asymptotic behaviour as the NFW halo and the logarithmic
potential which leads to (asymptotically) flat rotation curves.

\subsection{The Jeans equation and the line--of--sight velocity dispersion}
\label{sect:jeans}
\label{sect:jeansana}

The spherical, non--rotating Jeans equation (see
e.g.~\citealt{binneytremaine}) reads:
\begin{equation}
\frac{\mathrm{d}\left(n(r)\, \sigma_{r}^{2}(r)\right)}{\mathrm{d}r} 
+2\, \frac{\beta(r)}{r}\,n(r)\,\sigma_{r}^{2}(r)= - n(r) \,\frac{G\cdot M(r)}{r^{2}} ,
\label{eq:jeans}
\end{equation}
\begin{displaymath}
\textrm{with}\qquad \beta \equiv 1 - \frac{{\sigma_{\theta}}^2}{{\sigma_{r}^2}}\;.
\end{displaymath} 
Here, $r$ is the radial distance from the centre and $n$ is the
spatial (i.e.,~three--dimensional) density of the GCs; ${\sigma_r}$
and ${\sigma_\theta}$ are the radial and azimuthal velocity
dispersions, respectively. $\beta$ is the anisotropy parameter, $M(r)$
the enclosed mass (i.e.~the sum of stellar and dark matter) and $G$ is the constant of gravitation.\par For our
analysis, we use the expressions given by, e.g.~\cite{mamonlokas05}, see also \cite{marelfranx93}.  Given a
mass distribution $M(r)$, a three--dimensional number density of a
tracer population $n(r)$, and a \emph{constant} anisotropy
parameter $\beta$, the solution to the Jeans equation
(Eq.~\ref{eq:jeans}) reads:
\begin{equation}
n(r)\, {\sigma_{r}^{2}(r)} =  \mathrm{G} \int_r^\infty n(s) \, M(s) \frac{1}{s^{2}} \left( \frac{s}{r} \right)^{2 \beta} \mathrm{d} s .
\label{eq:sigl}
\end{equation}
This expression is then projected using the following integral:
\begin{equation}
\sigma_{\mathrm{los}}^{2}(R)= \frac{2}{N(R)} \left[ \int_{R}^{\infty}  \frac{n\sigma_{r}^{2}\, r\, \mathrm{d}r}{\sqrt{r^{2}-R^{2}}}
- R^{2}\int_{R}^{\infty}  \frac{\beta n \sigma_{r}^{2}\,  \mathrm{d}r}{r \sqrt{r^{2}-R^{2}}} \right]\,,
\label{eq:siglos}
\end{equation}
where $N(R)$ is the projected number density of the tracer population,
and $\sigma_{\rm{los}}$ is the line--of--sight velocity dispersion, to
be compared to our observed values. In the following, we discuss the
quantities required to determine $\sigma_{\rm{los}}(R)$.

\subsection{Luminous matter}
\label{sect:dep4636}

%===== Figure (14) ==========
\begin{figure}
\centering
\includegraphics[width=0.49\textwidth]{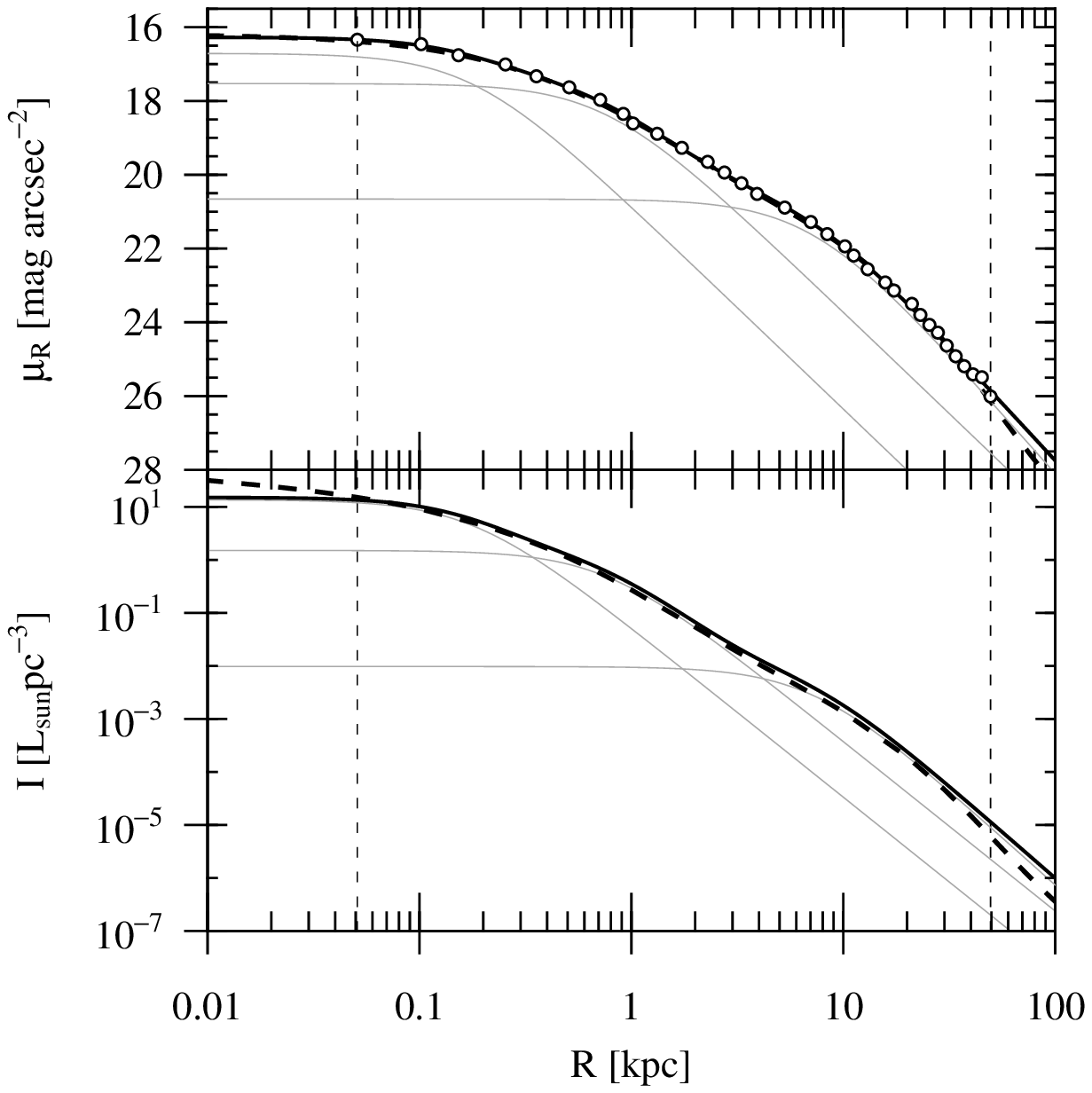}
\caption{Deprojecting the surface brightness profile of
  NGC\,4636. \textbf{Upper sub--panel:} $R$--band surface brightness
  profile. The data points are from D05+ (their Table\,A.4), and the
  dashed line is the fit given by D+05 (Eq.~\ref{eq:fitboris4636}). The
  solid line is the three--component fit given in
  Eq.~\ref{eq:fitylva4636}, and the thin solid lines show the
  individual components. \textbf{Bottom sub--panel:} Luminosity density
  profiles in units of $L_{{\odot}}\,\rm{pc}^{-3}$. The dashed line
  was obtained by numerically integrating
  Eq.~\ref{eq:fitboris4636}. The solid line is the analytical
  deprojection of Eq.~\ref{eq:fitylva4636}; again, the components are
  shown as thin solid lines.  The vertical dotted lines indicate the
  radial range of the data points shown in the upper panel.}
\label{fig:4636Lumpro}
\end{figure}

To assess the stellar mass of NGC\,4636 we need to deproject the galaxy's 
surface brightness profile. Moreover, to consistently model the 
line-of-sight velocity dispersion profile of the stars (cf. 
Fig.\ref{fig:4636stellar}), we require analytical expressions for both, the 
projected and the three-dimensional stellar density.

As in Paper\,I, we use the data
published by D+05 (shown as dots in the upper panel of
Fig.~\ref{fig:4636Lumpro}), for which the authors gave the following
fit:
\begin{eqnarray}
\lefteqn{\mu(R) =} \nonumber\\
  &   -2.5\log\left( 3.3\cdot 10^{-7}
  \left(1+\frac{R}{0\farcm11
}\right)^{-2.2}+
5.5\cdot 10^{-9}
\left(1+\frac{R}{8\farcm5
}\right)^{-7.5}\right)\; .  \nonumber
\\
\label{eq:fitboris4636}
  \end{eqnarray}
Their fit is shown as dashed line in Fig.~\ref{fig:4636Lumpro}.  There
is, however, no analytical solution to the deprojection integral for
this function. We therefore fit the data using the sum of three
Hubble--Reynolds profiles instead:
\begin{equation}
\mu(R)=-2.5\log\left(
\sum_{i=1}^{3} N_{0,i} \left[  1+ \left( \frac{R}{R_{0,i}} \right)^2 \right]^{-\alpha_i}
\right)\;,
\label{eq:fitylva4636}
\end{equation}
where the parameters are given in Table\,\ref{tab:numdens}.  Our fit
is shown as solid black line in Fig.~\ref{fig:4636Lumpro}, and the
thin gray lines indicate the three components.
%: $a_1=2.66\times10^{-7}$,
%\mbox{$a_2=5.3\times10^{-9}$}, $r_1=0\farcm065$, $r_2=1\farcm48$,
%$\alpha_1=0.9$, and $\alpha_2=1.5$.  
The deprojection of
Eq.~\ref{eq:fitylva4636} reads:
\begin{equation}
j(r)\left[ \frac{L_{{\odot}}}{\rm{pc}^{3}} \right]=
\sum_{i=1}^{3} \frac{{N^\prime}_{0,i}}{{R^\prime}_{0,i}\,  \mathcal{B}(\frac{1}{2},\alpha_i)}
  \left[1+\left(\frac{r}{{R^\prime}_{0,i}}\right)^2\right]^{-(\alpha_i+1/2)} \; .
\label{eq:deproylva4636}
\end{equation}
Where $\mathcal{B}$ is the
Beta function and  ${N^\prime}_{0,i} = C_{M_{R}} \cdot N_{0,i}$, where $C_{M_{R}}=2.192\times10^{10}$ is the factor
converting the surface brightness into units of
$L_{{\odot}}\,\rm{pc}^{-2}$ for $M_{{\odot},R} = 4.28$. The radii are
in pc, i.e.~for a distance of $17.5\,\rm{Mpc}$ ${R^\prime}_{0,i}= 5.09
\times10^{3}\cdot R_{0,i}$. \\
The lower panel of Fig.~\ref{fig:4636Lumpro} compares the
deprojection as given in Eq.~\ref{eq:deproylva4636} (solid black
line) to the curve obtained by numerically deprojecting
Eq.~\ref{eq:fitboris4636}. Within the radius interval covered by the
data points, both deprojections agree extremely well, 
and we proceed to use the analytical expression given in
Eq.~\ref{eq:deproylva4636} to represent the density distribution of
the stars in NGC\,4636.\par
The stellar mass profile is then obtained through integration:
\begin{equation}
M(r) = \Upsilon_{\star,R} \cdot 4 \pi
\int_{0}^{r} j (s)\ s^{ 2} {\rm{d}} s \; ,
\label{eq:massInt}
\end{equation}
where $ \Upsilon_{*,R}$ is the $R$--band mass--to--light ratio of the
stellar population of this galaxy (see below in
Sect.~\ref{sect:stellarMtoL}).  \par Since the integral in
Eq.~\ref{eq:massInt} cannot be expressed in terms of simple standard
functions, we use an approximation in our calculations: The inner part
($r\la 45\,\rm{kpc}$) is represented by a sequence of
polynomials, while the behaviour at larger radii is well represented
by an $\arctan$ function. The stellar mass profile is plotted in
Fig.~\ref{fig:masspoly}, and the expressions and coefficients are
given in Appendix\,\ref{sect:masspoly}.

%===== Figure (15) ==========
\begin{figure}
\centering
\includegraphics[width=0.49\textwidth]{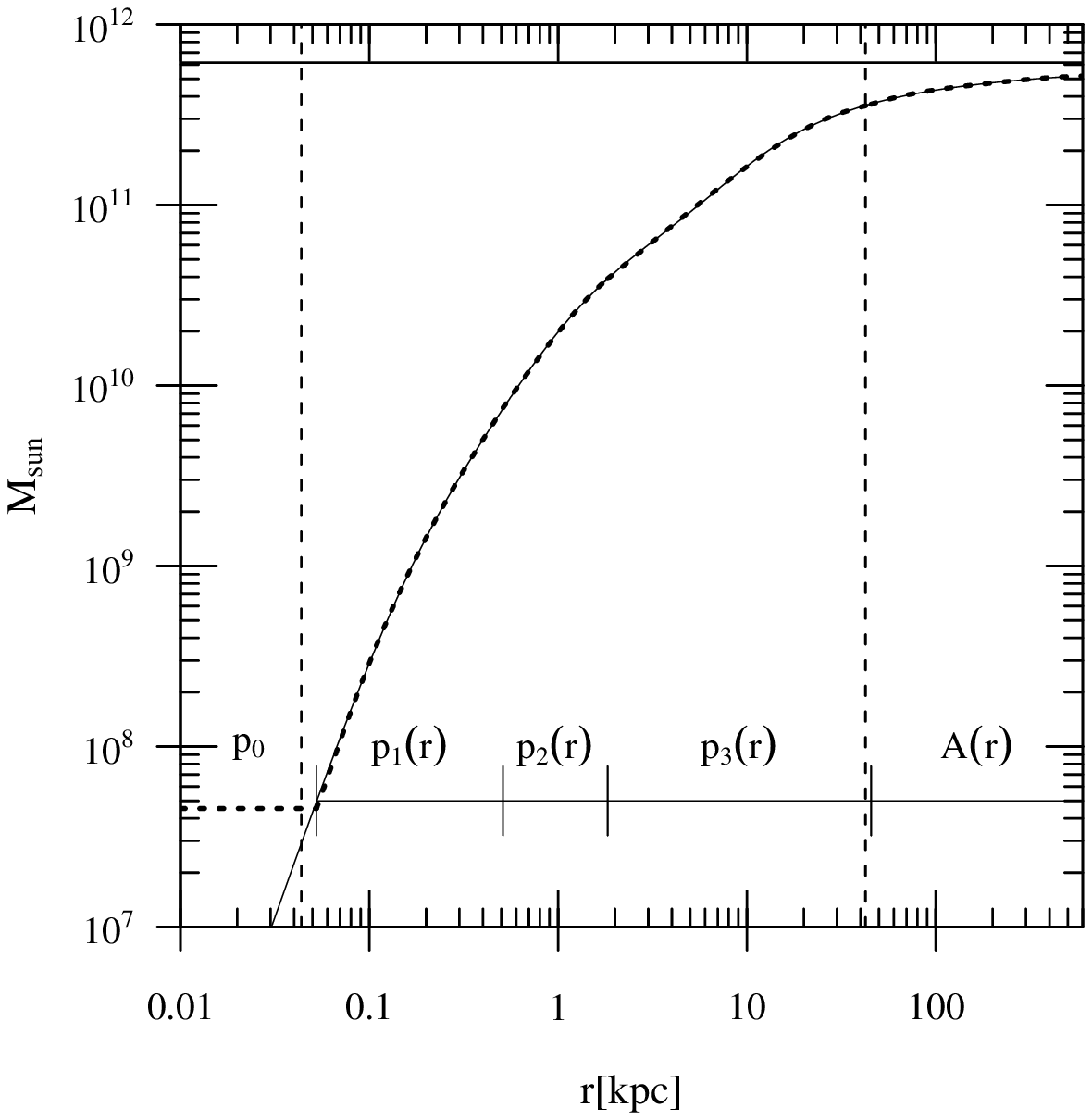}
\caption[]{NGC\,4636 stellar mass profile. The thin solid curve is
  Eq.~\ref{eq:massInt} for $\Upsilon_{\star,R}=5.8$ (obtained through
  numerical integration). The thick dotted curve is the piecewise
  approximation used in our modelling. The radial range of the
  respective pieces is indicated by the bars, and the functions and
  coefficients are given in Appendix\,\ref{sect:masspoly}. The vertical
  dashed lines indicate the radial range of the photometric data by
  D+05. The horizontal line at $6.15\times10^{11}M_\odot$ shows the
  asymptotic value, i.e.~$M(r=\infty)$.}
\label{fig:masspoly}
\end{figure}

\subsection{The stellar mass--to--light ratio}
\label{sect:stellarMtoL}

In Paper\,I, we used an $R$--band $\Upsilon_\star=6.8$. This was
derived from the dynamical estimate for the $B$--band given by
\cite{kronawitter00}. Having adopted a distance of 17.5\,Mpc for our
current analysis, the value from Paper\,I which was based on a
distance of 15\,Mpc is reduced to $\Upsilon_{\star,R}=5.8$. We will
adopt this value for our dynamical modelling of the NGC\,4636 GCs. 

%===== Figure (16) ==========
\begin{figure}
\includegraphics[width=0.49\textwidth]{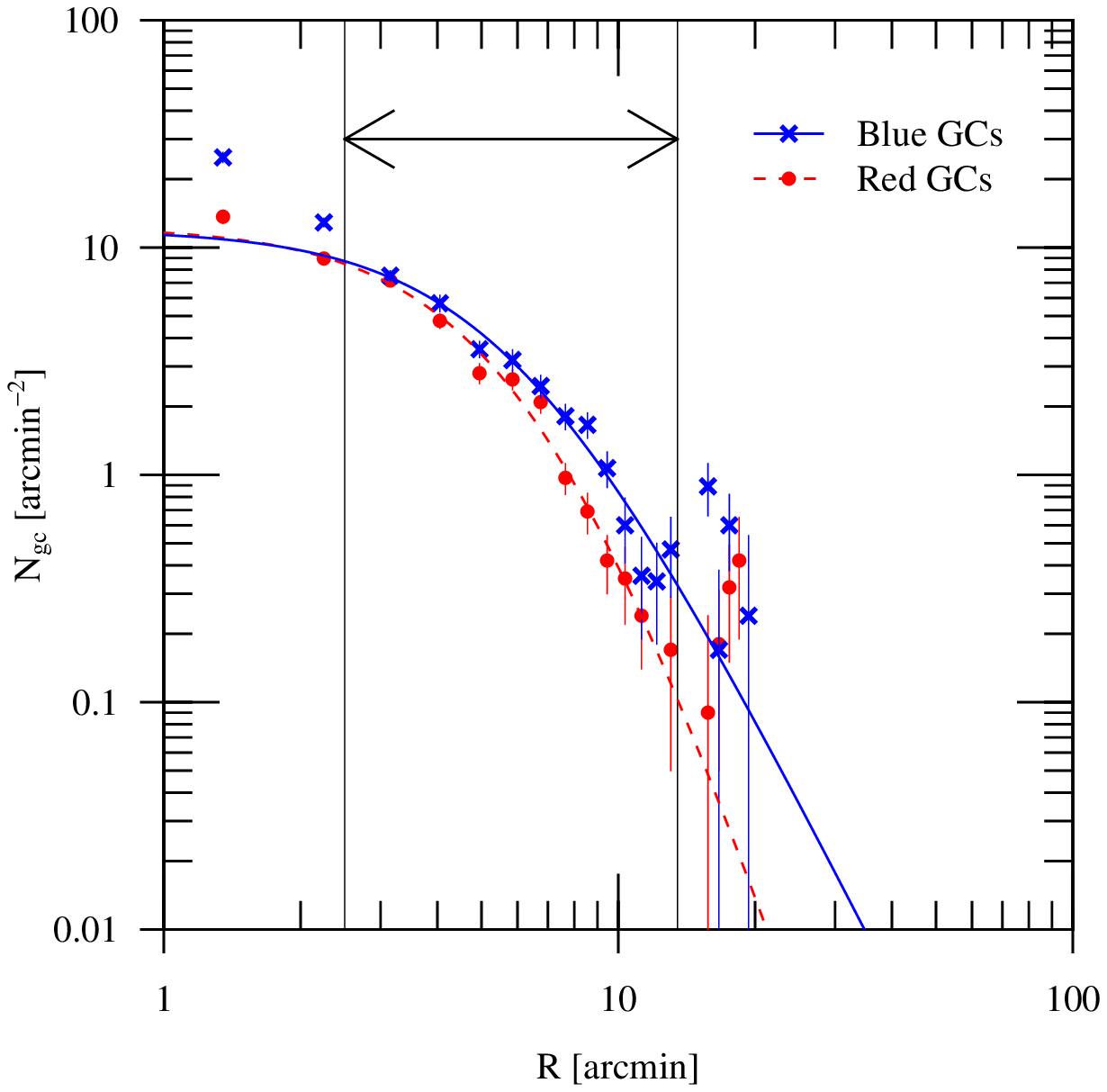}
  \caption{Number density profile of the blue and red GCs. The data
from D+05 (their Table\,A3) are shown as crosses and dots for the blue
and red GCs, respectively. The curves are the fits
(cf.~Eq.~\ref{eq:numdens}, Table\,\ref{tab:numdens}) to the data, and
the radial range $2\farcmin5\leq R \leq 13\farcmin5$ is indicated by
the arrow and the vertical lines.
}
\label{fig:4636Numdens}
\end{figure}

\subsection{Globular cluster number density profiles}
\label{sect:numdens}
Below we present the fits to the number density profiles of the GC
subpopulations as listed in Table\,A3 of D+05 and the analytical
expressions for the deprojections.
As in our study of the NGC\,1399 GCS \citep{schuberth10} we
parametrise the two--dimensional number density profiles in terms of a
Reynolds--Hubble law:
\begin {equation}
N(R) = N_{0} 
\left( 1 + \left(\frac{R}{R_{\mathrm{0}}}\right)^2 \right)^{-\alpha}\,,
\label{eq:numdens}
\end{equation}
where $R_0$ is the core radius, and $2\cdot\alpha$ is the slope of the
power--law in the outer region.  For the above expression, the Abel
inversion has an analytical solution and the
three--dimensional number density profile reads:
\begin{equation}
\ell(r) = \frac{N_{0}}{R_0} 
\frac{1}{\mathcal{B} \left(\frac{1}{2},\alpha \right)} 
\cdot
\left( 1 + 
\left(\frac{r}{R_{\mathrm{0}}}\right)^2 \right)^{-\left(\alpha+\frac{1}{2}\right)}\,,
\label{eq:depropl}
\end{equation}
where $\mathcal{B}$ is the Beta function.
For both subpopulations, the fits are performed for the radial range
$2\farcmin5 \leq R \leq 13\farcmin5$ where the lower boundary is the
minimum radius where blue and red GCs can be separated. The upper
boundary corresponds to the radius where the GC counts reach the
background level (D+05).  The parameters obtained for the blue and red
GCs are given in Table\,\ref{tab:numdens}.
Figure\,\ref{fig:4636Numdens} shows the data and the fitted
profiles. Note that the profile of the red GCs is significantly
steeper than that of the blue GCs.

\begin{table}
\caption[]{Fit parameters for the luminosity density profile of NGC\,4636 and the GC number density profiles
} \centering
\begin{tabular}{lr@{$\pm$}lr@{$\pm$}lr@{$\pm$}l c } 
\hline  \hline
& \multicolumn{2}{c}{$N_0$} & \multicolumn{2}{c}{$R_{0}$} 
& \multicolumn{2}{c}{$\alpha$} 
& \multicolumn{1}{c}{$\mathcal{B}\left(\frac{1}{2},\alpha \right)$} 
\\ 
& \multicolumn{2}{c}{} 
& \multicolumn{1}{c}{}
\\ 
\hline
Lumprof\,1 & \multicolumn{2}{l}{$2.07\times10^{-7}$} & \multicolumn{2}{l}{$3.47\times10^{-2}$} & \multicolumn{2}{c}{1.1}& 1.887 \\
Lumprof\,2 & \multicolumn{2}{l}{$9.73\times10^{-8}$} &  \multicolumn{2}{l}{$1.47\times10^{-1}$} & \multicolumn{2}{c}{1.1} & 1.887 \\
Lumprof\,3 & \multicolumn{2}{l}{$5.45\times10^{-9}$} &  \multicolumn{2}{l}{$1.41$} & \multicolumn{2}{c}{1.3} & 1.708\\
\hline
Blue \rule{0ex}{2.5ex} &12.0 &1.1 &6.0 & 0.3 & 2.0&0.1  & $\frac{4}{3}$ \\
Red \rule[-1.2ex]{0ex}{3.8ex} & 12.4&1.1 &6.8 &0.3  & 3.0&0.1  & $\frac{16}{15}$ \\ 
\hline \hline
\end{tabular}
\label{tab:numdens}
\note{The first three rows give the parameters for the three--component Hubble--Reynolds profile fit to the luminosity density profile (Eq.~\ref{eq:fitylva4636}). The last two rows are the parameters found for the GCs (Eq.~\ref{eq:numdens})}
\end{table}

\subsection{The dark matter halo}

All three dark matter halos considered in this study have two free
parameters, allowing a direct comparison of the results.

\subsubsection{The NFW profile}
\label{sect:nfw}

The mass profile of the NFW halo reads:
\begin{equation}
M_{\rm{NFW}}(r)= 4 \pi \varrho_s r_s^3 \cdot \left( 
\ln \left( 1 + \frac{r}{r_s} \right) -
\frac{\frac{r}{r_s}}{1+ \frac{r}{r_s}}
\right) \ ,
\label{eq:massnfw}
\end{equation}
where $\varrho_s$ and $r_s$ are the characteristic density and scale
radius, respectively. \par To express the halo parameters in terms of
concentration and virial mass, we use the definitions from
\cite{bullock01} and define the virial radius $R_{\rm{vir}}$ such that
the mean density within this radius is $\Delta_{\rm{vir}}=337$ times
the mean (matter) density of the universe (i.e.~$0.3\,\rho_c$), and
the concentration parameter is defined as
$c_{\rm{vir}}=R_{\rm{vir}}/r_s$.

\subsubsection{The Burkert halo}
\label{sect:burkert}

The density profile for the cored halo which \cite{burkert95}
introduced (to represent the dark matter halo of dwarf galaxies) reads:
\begin{equation}
\varrho(r) = \frac{\varrho_0}{\left( 
  1+ \frac{r}{r_{0}}\right) \left(1+ \frac{r^2}{{r_0^2}} \right)} \; ,
\label{eq:rhoburkert}
\end{equation}
and the cumulative mass is given by the following expression:
\begin{eqnarray}
\lefteqn{M(r) =} \nonumber\\
  & &  4 \pi \varrho_{\rm{0}} {r_{0}^3} \left( 
\frac{1}{2} \ln \left(  1 + \frac{r}{r_0}\right) +
\frac{1}{4} \ln \left( 1 + \frac{r^2}{r_0^2}\right) -
\frac{1}{2} \arctan \left( \frac{r}{r_0}\right) 
\right) \nonumber \; .
\\
\label{eq:MassBurkert}
  \end{eqnarray}

\subsubsection{The logarithmic potential}
\label{sect:logpot}

The logarithmic potential (see \citealt{binney81,binneytremaine}), by
construction, yields (asymptotically) flat rotation curves. In
contrast to the NFW profile, it has a finite central density. The
spherical logarithmic halo has two free parameters, the asymptotic
circular velocity $v_0$, and a core radius $r_0$. The mass profile
reads:
\begin{equation}
M_{\textrm{Log}}(r)= \frac{1}{G} \cdot 
\frac{r \cdot v_0^2}{1+\left(\frac{r_0}{r}\right)^2} \; ,
\label{eq:massLogPot}
\end{equation}
where $G$ is the constant of gravitation. 

\subsection{Modelling the velocity dispersion profiles}
\label{sect:modelling}

To find the parameters that best describe the observed GC velocity
dispersion data, we proceed as described in \cite{schuberth09}: For a
given tracer population and anisotropy $\beta \in \{-0.5, 0, +0.5\}$,
we create a grid of models where the density (or $v_0$ in case of
the logarithmic halo) acts as free parameter while the radii have
discrete values, i.e. $r_{\mathrm{dark}} \in \{1, 2, 3, . . . ,
100\}\,\rm{kpc}$.  For each point of this grid, the line--of--sight
velocity dispersion (Eq.~\ref{eq:siglos}) is computed using the
expressions given in \cite{mamonlokas05}, where the upper limit of the
integral in Eq.~\ref{eq:siglos} is set to 600\,kpc.  \par To find the
joint solution for the different tracer populations (labelled $a$ and
$b$), we determine the combined parameters by minimising the sum
$\chi^2 = \chi^2_a+\chi^2_b$. 
The confidence level (CL) contours
are calculated using the definition by \cite{avni76}, i.e. using
the difference $\delta \chi^2$ above the minimum $\chi^2$ value. With two
free parameters, e.g. ($\rm r_{dark}, \rho_{dark}$) the 68, 90, and 99 per
cent contours correspond to $\chi^2$ = 2.30, 4.61, and 9.21, respectively.
The results for the three dark matter halos (NFW, Burkert and
logarithmic potential) are presented in the following section.

\section{The mass profile of NGC\,4636}

We model the observed line--of--sight
dispersions for the final samples of the red and blue GCs shown in
Fig.~\ref{fig:4636modelcompare} and listed in Table\,\ref{tab:4636DispMod}
for three different parametrisations of the dark halo.

%===== Figure (17) ==========
\begin{figure*}
\centering
\includegraphics[width=0.99\textwidth]{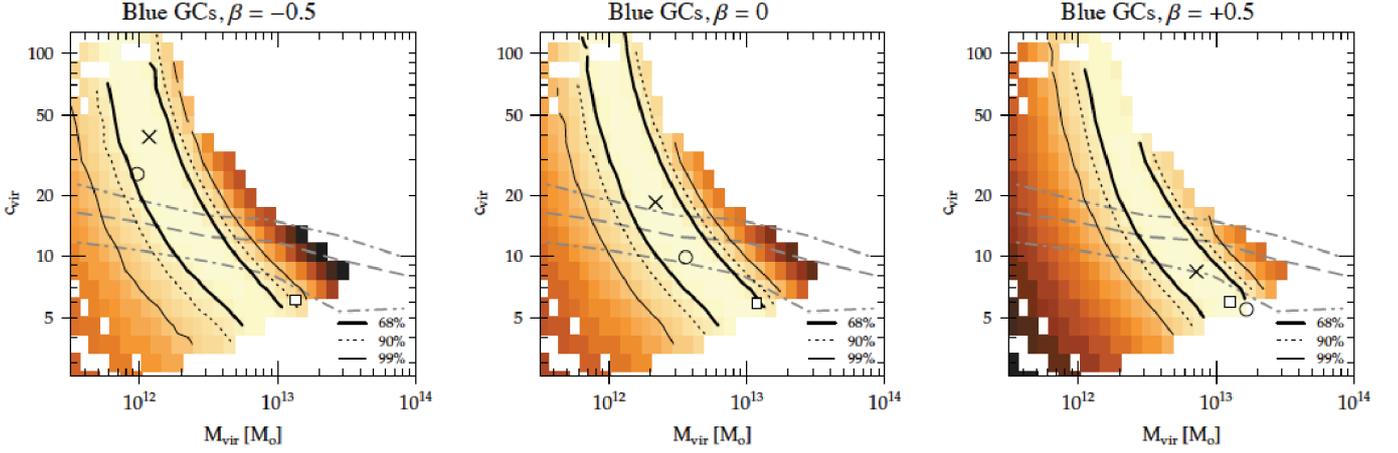}
\caption[] {Jeans models for the blue GCs ({\it{BlueFinal}}) for an NFW--type
dark halo. From left to right, the panels show the results for $\beta=-0.5,
0$ and $+0.5$. The parameters are shown in the
$(M_{\textrm{vir}},c_{\textrm{vir}})$--plane. The thick solid, dashed and
thin solid lines indicate the $68, 90$, and $99$ per cent confidence limits.
The colour map is the same for all panels, and the cross indicates the
location of the minimum $\chi^2$ value. The square shows the minimum $\chi^2$
for the joint models obtained from the blue GCs and an isotropic model for the
NGC\,4636 stellar dispersion profile. The circle indicates the respective
best--fit value from Paper\,I. All model parameters are listed in
Table\,\ref{tab:4636nfwpars}. In all panels, the long--dashed (dash--dotted)
lines show the median ($68$ per cent values) for simulated NFW halos as found
by \cite{bullock01}.}
\label{fig:4636isopars}
\end{figure*}

\subsection{Jeans models for an NFW halo}
\subsubsection{NFW Halo: Results for the  blue GCs} 

In Fig.~\ref{fig:4636isopars}, we show the NFW models for the blue GCs
after transforming the parameters to the
$(M_{\mathrm{vir}},c_{\mathrm{vir}})$ plane using the definitions in
\cite{bullock01}. In all panels, the respective best fit value as
given in Table\,\ref{tab:4636nfwpars} is shown as a cross. Circles are
the corresponding values from Paper\,I. In all cases, these values lie
within the 68\% CL contour of the present study. \par From the very
elongated shape of the confidence level contours it is apparent that
while the GCs can be used to estimate the total mass of the halo, 
the concentration is only poorly constrained. As will
be shown in Sect.~\ref{sect:jointNFW}, this degeneracy can be 
partially overcome
by considering models for the stellar velocity dispersion profile of
NGC\,4636.

The best-fit dispersion profiles for the three values of the
anisotropy parameters $(\beta \in \{-0.5, 0, +0.5 \})$ are shown as
thin black lines in the left panel of
Fig.~\ref{fig:4636modelcompare} and the corresponding parameters of the
NFW halos are listed in Table\,\ref{tab:4636nfwpars} (Cols.~4--9).
For all three values of $\beta$, a very good agreement between data and
models can be achieved, and the differences between the $\chi^2$ values
is marginal. The models diverge at small radii ($R\la 2\farcmin5
\approx 13\,\rm{kpc}$) where the velocity dispersion and the shape of
the number density profile of the blue GCs cannot be well constrained.
For comparison, we plot (as dot--dashed line) the velocity dispersion
curve expected if there were no dark matter and the only mass were
that of the stars. 

As expected, the best-fit halo derived assuming a tangential
orbital anisotropy ($\beta = -0.5$, {\it B.tan}, long-dashed line) is less
massive than the one obtained in the isotropic case ({\it B.iso}, solid
line) and the model for a radial bias ($\beta= +0.5$, {\it B.rad}, 
short-dashed line) returns the most massive dark halo. This is also
illustrated by the bottom left panel of Fig~\ref{fig:4636modelcompare} where 
the corresponding mass profiles (thin black lines) are shown in terms of
the circular velocity.

\subsubsection{NFW halo: Results for the red GCs:} 

The resulting halo parameters for the red clusters are listed in 
Table\,\ref{tab:4636nfwpars} (Cols.~4--9) and illustrated in the middle
panel of  Fig.~\ref{fig:4636modelcompare}.
Here, the agreement between data and models is worse than in the case of the 
blue GCs. A considerable part of the uncertainty is caused by the  
curiously low value at $\sim$16 kpc. In spite of this, the resulting circular
velocities (shown as thick grey lines in the bottom left of 
Fig.~\ref{fig:4636modelcompare}) of the 
different halo models are not dramatically different from those of the blue 
GCs. For all three values of $\beta$, the circular velocity stays 
approximately constant within 40 kpc.   

\subsubsection{Model for the stars}
\label{sect:4636stellarmod}

We use the stellar velocity dispersion measurements presented by
\cite{BSG94} to constrain the concentration parameter of the NFW halo.
The right panel of Fig.~\ref{fig:modstellar} shows the
($M_{vir}$-$c_{vir}$) plane for the isotropic case. High
concentrations are excluded, since adding large amounts of dark matter
in the central parts of NGC\,4636 would severely overestimate  the
velocity dispersion profile of the stellar component.

%===== Figure (18) ==========
\begin{figure}
\includegraphics[width=0.49\textwidth]{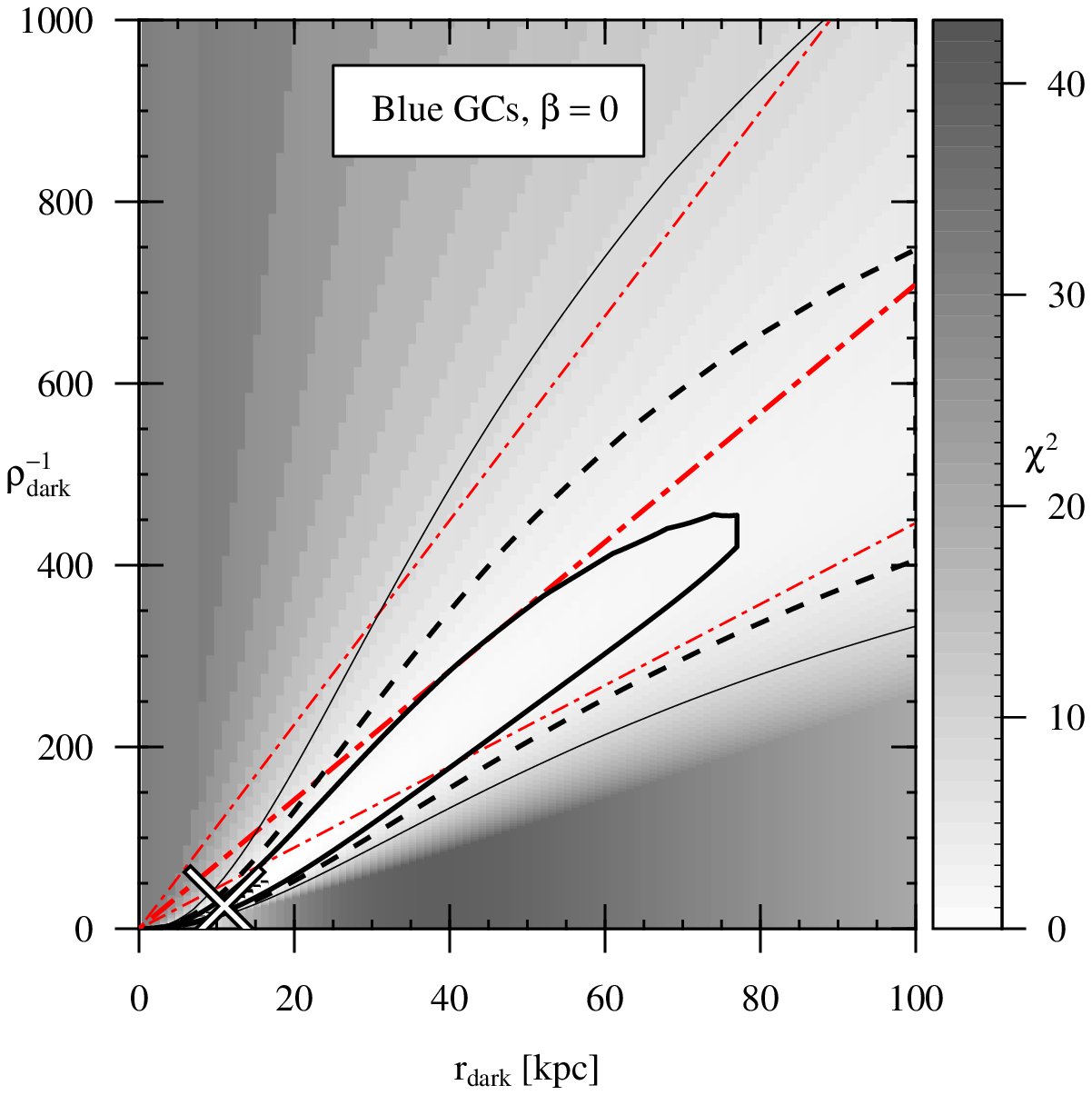}
\caption[]{Jeans models for the blue GCs (sample {\it{BlueFinal}},
  $\beta=0$) for a \citeauthor{burkert95} dark matter halo.  The cross
  marks the best--fit parameters
  (cf.~Table\,\ref{tab:4636nfwpars}). The thick solid, dashed and thin
  solid lines indicate the 68, 90 and 99 per cent confidence limits.
  The thick dot--dashed line indicates the \cite{donato09} central
  dark matter surface density relation $\log\mu_{0D}=2.15\pm0.2
  \,[\log(M_\odot \rm{pc}^{-2} )]$  (the thin dot--dashed lines
  show the uncertainties). }
\label{fig:burkert}
\end{figure}

\subsection{Jeans models for a Burkert halo}
%X\label{sect:burkert}

Figure\,\ref{fig:burkert} shows the parameter space explored to find
the best-fit isotropic Jeans model for the blue GCs for a Burkert-type dark
halo. The best-fit Burkert models for the GCs are shown in the middle panels
of Fig. \ref{fig:4636modelcompare}, and the parameters are given in Cols.
11-12 of Table \ref{tab:loghalopars}. The circular velocities corresponding
to the different mass distributions are compared in the bottom
middle panel of Fig. \ref{fig:4636modelcompare}.
The discrepancies between the best-fit models for the blue
GCs (shown as thin black lines) and the models for the red GCs
(thick grey curves) do not permit to prefer any specific halo model.

\subsection{Jeans models for a logarithmic potential}
%\label{sect:logpot}

The results are summarised in Table\,\ref{tab:loghalopars}, and the model 
grids solutions in the $(r_0,v_0)$--plane for the blue GCs (for $\beta = -0.5$ 
and 0) are shown in Fig.~\ref{fig:4636LogHaloBlue}.
Again, one notes a strong degeneracy: The asymptotic velocity $v_0$ (and hence
the total mass) is well constrained, while the scale radius $r_0$ is not.

%===== Figure (19) ==========
\begin{figure*}
\centering
\includegraphics[width=0.99\textwidth]{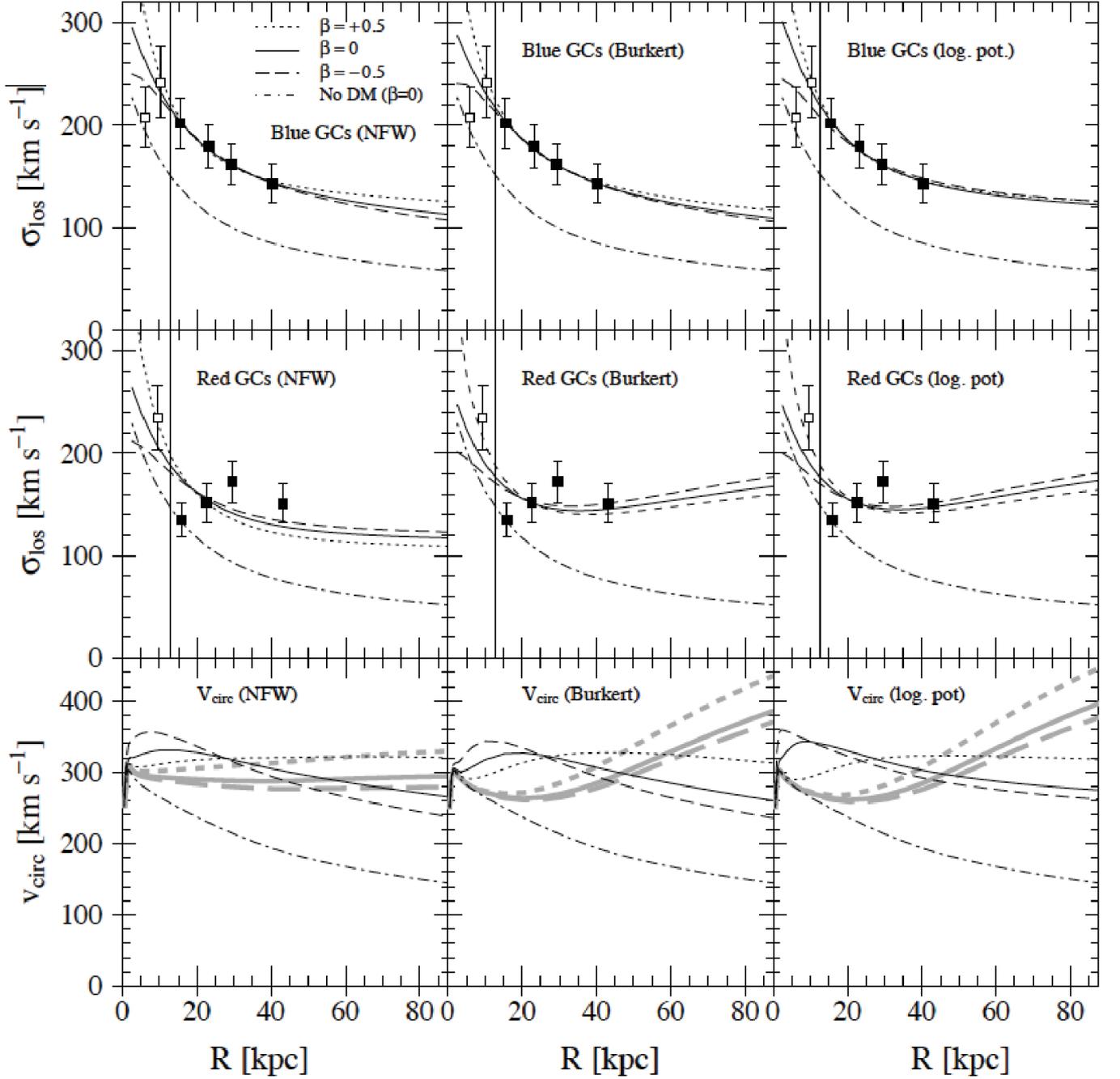}%&
\caption[NGC\,4636: Observed and modelled velocity dispersion
profiles]{Observed and modelled GC velocity dispersion profiles. \textbf{Top
row:} Models for the blue GCs (sample {\it BlueFinal}). From left to right,
the panels show the best-fit models for an NFW halo, Burkert halo and the
logarithmic potential. The solid lines are the isotropic models, dashed and
short-dashed lines are the tangential ($\beta = -0.5$) and radial ($\beta =
+0.5$) models, respectively. The dash-dotted line is the (isotropic) model
without dark matter. The thin vertical line at $\simeq$13 kpc indicates the
radial range inside which blue and red GCs cannot be distinguished. The data
points used in the modelling are shown as filled squares (see also Table 7).
The model parameters are listed in Table \ref{tab:4636nfwpars}.
\textbf{Middle row:} The same for the red GCs ({\it RedFinal}).
\textbf{Bottom row:} Circular velocity curves for the best-fit models. Again,
from left to right, the results for the NFW halo, Burkert halo and the
logarithmic potential are shown. The line styles are the same as in the upper
graphs, with thin black lines for the blue GCs while the respective models
for the red GCs are shown as thick grey lines.}
\label{fig:4636modelcompare}
\end{figure*}

\subsection{Joint solutions}

To find a joint solution describing the velocity dispersion profiles
of the three tracer populations, we combine the $\chi^2$ values of the
corresponding  models and obtain the solution by finding the minimum in the 
co-added $\chi^2$  maps (cf. Sect\ref{sect:modelling}).

\subsubsection{Models for the blue GCs and the stellar velocity
dispersion profile}
\label{sect:jointNFW}

Since the best agreement between models and data can be achieved for the 
blue GCs and the stellar velocity dispersion profile (see 
Fig.\,\ref{fig:4636modelcompare} and the $\chi^2$-values given in Table 
\ref{tab:loghalopars}), we will first combine these two tracer populations to
obtain a joint model. For the blue GCs, the anisotropy parameters $\beta$ 
takes the values $-$0.5, 0, $+$0.5, while the stellar models are isotropic. 
The parameters for corresponding joint models (labelled {\it S.B.tan, S.B.iso}
and {\it S.B.rad}) are given in Table \ref{tab:loghalopars}. The best-fit 
joint (isotropic) models are shown in the right panel of 
Fig.\,\ref{fig:modstellar} (lower sub-panel).
The agreement between data and model is best for the two cored
halo parametrisations: The velocity dispersions of both the stars
and the blue GCs are very well reproduced by a Burkert halo with 
$\rm \rho_0 = 4.89 \times 10^{-2} M_\odot pc^{-3}, r_0 =$ 10 kpc or a 
spherical logarithmic potential with $r_0$ = 8 kpc and $v_0$ = 237 
$\rm km s^{-1}$. The best-fit joint
NFW halo, on the other hand, has a very large scale radius and
over-estimates the velocity dispersion of the blue GCs in the last
bin (although model and data still agree within the uncertainties).

%===== TABLE (9) ==========
\begin{table*}
\caption[NGC\,4636: Jeans modelling best--fit NFW halos]{NGC\,4636 Jeans modelling best--fit NFW profiles,  Burkert halos and logarithmic potentials}
\centering
%\small
\resizebox{0.99\textwidth}{!}{
\begin{tabular}{llrrlrlrrrrrrll}\hline \hline
\multicolumn{3}{l}{}&
\multicolumn{6}{|c|}{NFW dark halo}& 
\multicolumn{3}{c}{Burkert halo\rule[-1ex]{0ex}{3.5ex}} &
\multicolumn{3}{|c}{Log.~potential}
\\ \hline
ID &Sample & $\beta$ &
 $r_{\textrm{s}}$ & 
 $\varrho_{\textrm{s}}$ & 
$M_{\textrm{vir}}$ & 
$R_{\textrm{vir}}$ & 
$c_{\textrm{vir}}$ & 
 $\chi^2$& 
% $\chi^2/\nu$\\
 $r_{\textrm{0}}$ & 
 $\varrho_{\textrm{0}}$ & 
 $\chi^2$&
 $r_{\textrm{0}}$ & 
 $v_{\textrm{0}}$ & 
 $\chi^2$\\
 &&  &
 $[\textrm{kpc}]$ & 
 $[M_{\odot}\,\textrm{pc}^{-3}]$ & 
$[10^{12} M_{\odot}]$ & 
$[\textrm{kpc}]$ & 
$$ & &
 $[\textrm{kpc}]$& 
  $[M_{\odot}\,\textrm{pc}^{-3}]$ & 
&
 $[\textrm{kpc}]$& 
$[\textrm{km\,s}^{-1}]$
\\
(1) & (2)  & (3) & (4) & (5) & (6) & (7) & (8)& 
 (9)& (10) &(11) &(12) &(13) & (14) & (15) \\
\hline
%#    RDARK     RHO    MVIR    CVIR    RVIR       CHI
%#302     8 0.07348 1.22696 34.4107 275.286 0.0796409%
{\it B.tan} &Blue final & 
$-0.5$ &
$7$ &  
$1.01\times10^{-1}$ & 
$1.2$ & 
$271$ & 
$ 38.7$ & $<0.1$ &
$6$ & $1.33\times10^{-1}$&$<0.1$
%# log halo:
& $1^\star$ & $219$ & \multicolumn{1}{r}{$0.15$}
 \\
%-------------- BLUE ISO
%#  RDARK    RHO    MVIR    CVIR    RVIR      CHI
%#  18 0.0147 2.17911 18.5209 333.375 0.122667
{\it B.iso} &Blue final & 
$0$ &
$18$ &  
$1.47\times10^{-2}$ & 
$2.2$ & 
$333$ & 
$18.5$ & 
$<0.1$ & 
$11$ & $4.04\times10^{-2}$ & $<0.1$ &
% log halo
$4$ & $234$ & \multicolumn{1}{r}{$<0.10$}
\\
%-------------- BLUE RAD
% RDARK       RHO    MVIR    CVIR    RVIR      CHI
%2341    59 0.0020685 7.19943 8.41566 496.524 0.248363
{\it B.rad} &Blue final & 
$+0.5$ &
$59$ &  
$2.07\times10^{-3}$ & 
$7.2$ & 
$497$ & 
$8.4$ & 
$0.1$ & 
$22$ & $1.39\times10^{-2}$ &$<0.1$ &
% log halo
$21$ & $292$ & \multicolumn{1}{r}{$<0.10$}
\\
\hline
%-------------- RED TAN
%    RDARK       RHO    MVIR   CVIR   RVIR     CHI
%   100 0.0005707 6.72672 4.8541 485.41 8.89991
{\it R.tan} &Red final & 
$-0.5$ &
$100^{\star}$ &  
$5.71\times10^{-4}$ & 
$6.7$ & 
$485$ & 
$4.9$ & 
$8.9$ & 
$100^\star$  & $1.94\times10^{-3}$&$5.4$ &
% log halo
$100^\star$ & 528 & 5.2 
\\
%    RDARK       RHO    MVIR   CVIR   RVIR     CHI
%   100 0.0006585 8.12312 5.1691 516.91 10.9168
%-------------- RED ISO
{\it R.iso} &Red final & 
$0$ &
$100^{\star}$ &  
$6.59\times10^{-4}$ & 
$8.1$ & 
$516$ & 
$5.2$ & 
$10.9$ & $100^\star$ & $2.13\times10^{-3}$ & $6.5$ &  
% log halo
$100^\star$ & 560 & 6.2 
\\
%-------------- RED RAD
%  RDARK      RHO    MVIR   CVIR   RVIR     CHI
%   100 0.000878 11.8181 5.8572 585.72 14.3694
{\it R.rad} &Red final & 
$+0.5$ &
$100^{\star}$ &  
$8.78\times10^{-4}$ & 
$11.8$ & 
$586$ & 
$5.9$ & 
$14.4$ & 
$100^\star$ & $2.81\times10^{-3}$ & $8.4$  & 
% log halo 
$100^\star$ & 640 & 8.0 
\\ 
%#     RDARK       RHO    MVIR    CVIR    RVIR     CHI
%#3981   100 0.0009219 12.5872 5.98159 598.159 2.21768
\hline
{\it S.iso} & Stars (BSG94)& 0& $100^\star$ & $9.219\times10^{-4}$ & $12.6$ & $598$ &$6.0$ & $2.2$ & 
% BUR
%STARS   1       100000  0.019   1.90281
% $17?$ & $3.82\times10^{-2}$ & $1.97$ &
$100^\star$ & $1.90\times10^{-2}$ & 1.9 & 
% log halo
$100^\star$ & 1725 & 1.9  
\\
%STARS   1       100000  1725    1.89755
\hline
\multicolumn{8}{l}{Joint solutions for blue GCs and Stars:\rule{0ex}{2.2ex}}\\ 
%#     RDARK       RHO    MVIR    CVIR    RVIR      CHI
%#4262   100 0.0009658 13.3651 6.10237 610.237 2.232177
\hline
{\it S.B.tan} & \multicolumn{2}{l}{Blue GCs $(\beta=-0.5)$,  Stars $(\beta=0)$}
&$100^\star$&$9.66\times10^{-4}$&$13.4$&$610$ &$6.1$ &$2.2$&
$8$ & $6.95\times10^{-2}$&$2.1$ &
% log halo
7 & 222 & 2.3 \\
%     RDARK      RHO    MVIR   CVIR   RVIR      CHI
%4700   100 0.000878 11.8181 5.8572 585.72 2.247505
{\it S.B.iso} & \multicolumn{2}{l}{Blue GCs $(\beta=0)$,  Stars $(\beta=0)$} 
&$100^\star$&$8.78\times10^{-4}$&$11.8$&$586$ &$5.9$ &$2.2$&
$10$ & $4.89\times10^{-2}$&$2.1$  
% log halo
& 8 & 237 & 2.2
 \\
%    RDARK       RHO    MVIR    CVIR    RVIR      CHI
%6341   100 0.0009219 12.5872 5.98159 598.159 2.322554
{\it S.B.rad} & \multicolumn{2}{l}{Blue GCs $(\beta=+0.5)$,  Stars $(\beta=0)$}
&$100^\star$&$9.22\times10^{-4}$&$12.6$&$599$ &$6.0$ &$2.3$&
$13$ & $3.71\times10^{-2}$&$2.4 $ & 
% log halo
10 & 270 & 2.4
\\
\hline \hline
\end{tabular}
}
\normalsize
\label{tab:4636nfwpars}
\label{tab:loghalopars}
\note{In all models, the $R$--band stellar mass--to light ratio is
$\Upsilon_{\star,R}=5.8$. Col.~(1) labels the models. The second column 
specifies the dispersion profile to which the Jeans models are fit. Col.~(3) 
gives the anisotropy parameter $\beta$. Cols.~(4)--(8) list the NFW 
(Eq.~\ref{eq:massnfw}) and virial parameters (see text for details).  
Column~(9) gives the $\chi^2$ value. The parameters for the \cite{burkert95}
halos (\ref{eq:MassBurkert}), i.e.~$r_0$, $\varrho_0$ and the $\chi^2$ value 
of the best fit model are given in Cols~(10--12). Cols.~(13--15) are the 
values for the logarithmic potential (Eq.~\ref{eq:massLogPot}). Asterisks  
indicate that the corresponding value is located  at the edge of the model 
grid.}
\end{table*}

%===== Figure (20) ==========
\begin{figure*}
\centering
\includegraphics[width=0.99\textwidth]{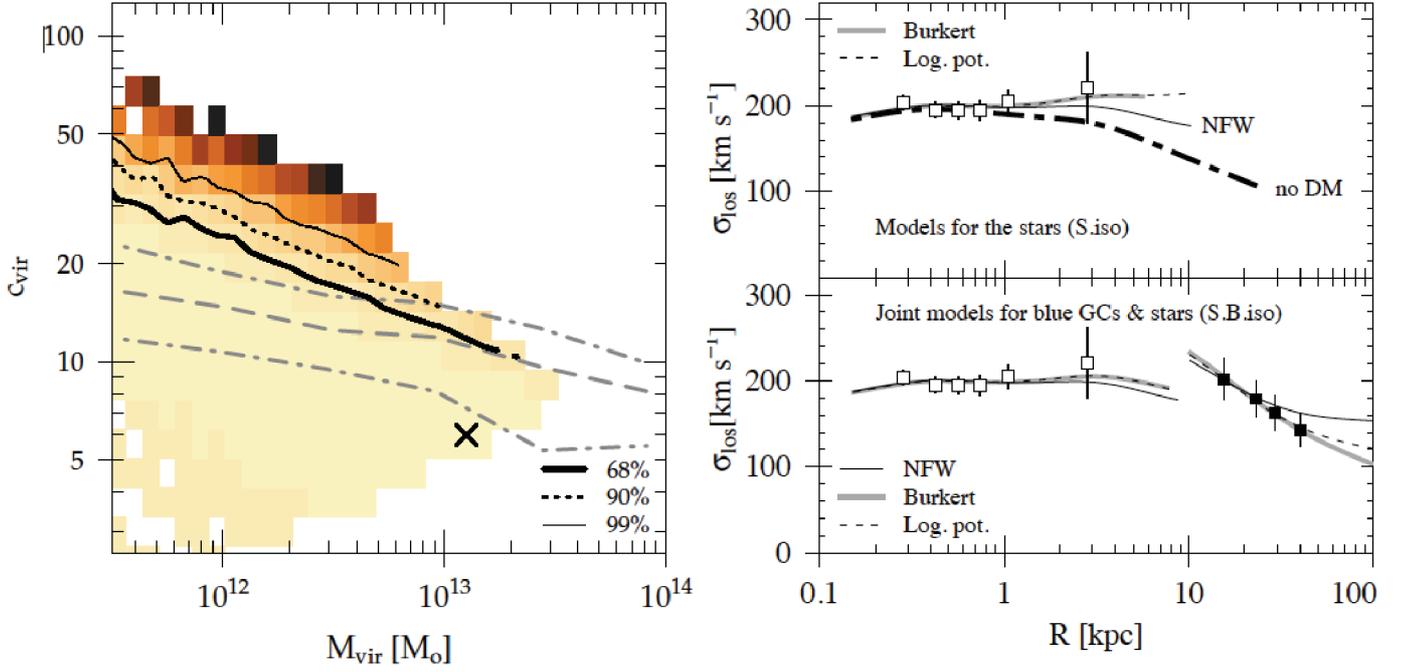}
\caption[]{Modelling the stellar velocity dispersion profile. \textbf{Left:}
Jeans models for the stars (Bender et al. 1994 data) for an NFW-type dark halo
and $\beta = 0$. The parameters are shown in the 
$(M_{\textrm{vir}},c_{\textrm{vir}})$--plane. The thick solid, dashed and 
thin solid lines indicate the 68, 90, and 99 per cent confidence limits. The 
cross indicates the location of the minimum $\chi^2$ value. The model 
parameters are listed in Table\,\ref{tab:4636nfwpars}. The long-dashed 
(dash-dotted) lines show the median (68 per cent values) for simulated NFW
halos as found by \cite{bullock01}. \textbf{Right:} Modelled velocity 
dispersion profiles for the stars (Bender et al. 1994 data, shown as unfilled 
squares). The upper sub-panel shows the best-fit isotropic models for the 
stars. The thin solid line shows the model for an NFW-type dark halo, and the 
thin dashed line is the model for the logarithmic potential. The Burkert halo
is shown as thick grey line. The thick dot-dashed line is the model without 
dark matter. The lower sub-panel shows the joint models for the blue GCs and
the stellar velocity dispersion profile (models {\it S.B.iso}). The 
line-styles are the same as in the upper panel. The black squares show the 
velocity dispersion profile for the blue GCs (sample {\it BlueFinal}). The 
halo parameters are given in Table\,\ref{tab:4636nfwpars}.}
\label{fig:modstellar}
\end{figure*}

%===== Figure (21) ==========
\begin{figure*}
\centering
\includegraphics[width=0.98\textwidth]{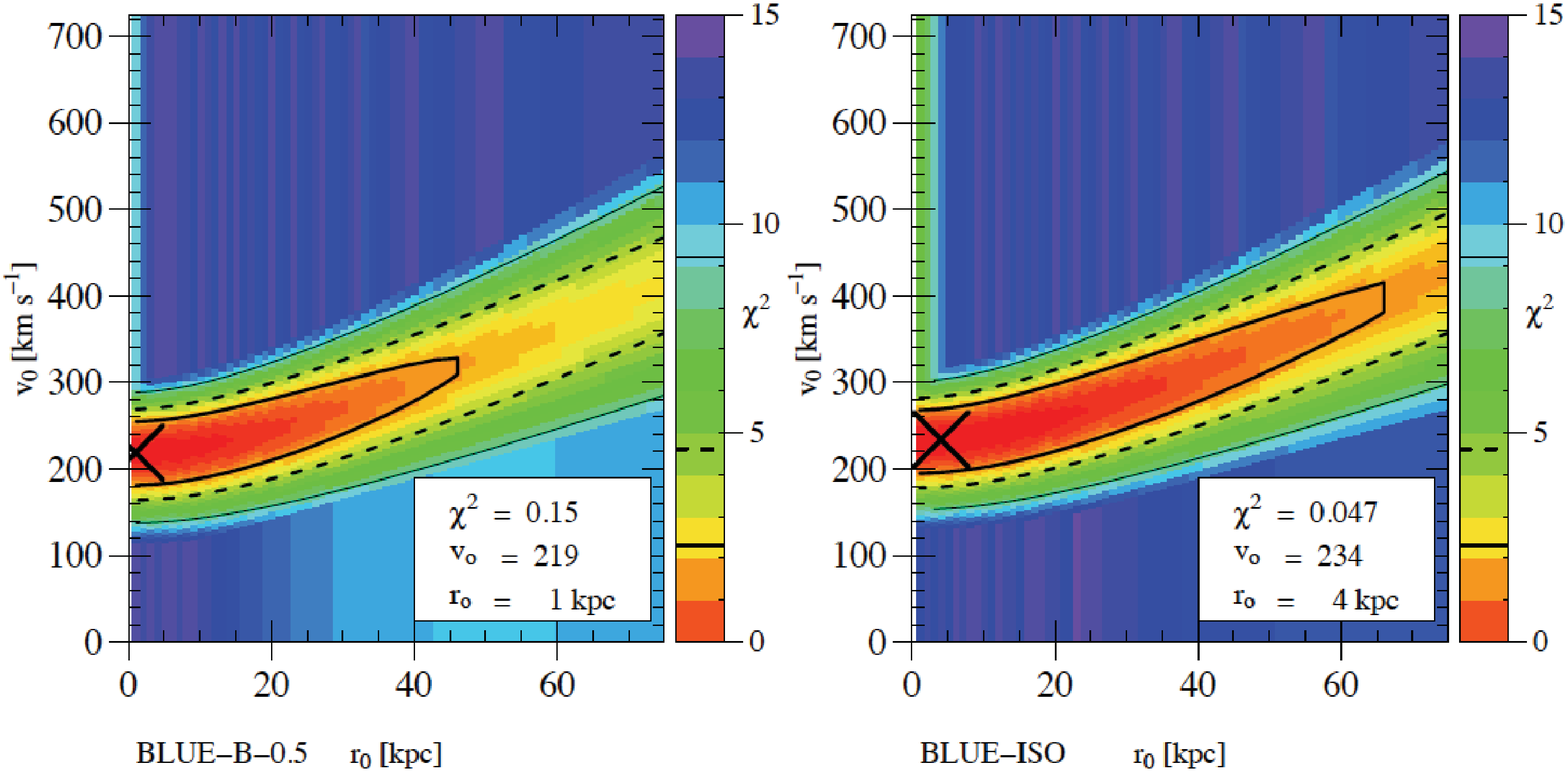}
\caption[]{NGC\,4636 Jeans models the blue GCs (final sample) where
the dark matter component is represented by a logarithmic potential
Eq.~\ref{eq:massLogPot}). 
{\bf Left}: Models for a mild tangential bias with $\beta = -0.5$. 
{\bf Right}: Isotropic models ($\beta = 0$).
In both panels, the best--fit solution is marked by a cross, and the thick
solid, dashed and thin solid lines show the $68$, $90$, and $99$ per
cent contour levels, respectively.  The parameters of the halos shown
here are also listed in Table\,\ref {tab:loghalopars}.}
\label{fig:4636LogHaloBlue}
\end{figure*}

\section{Discussion}
\label{sect:4636discussion}

\subsection{Comparison to the analysis by \citeauthor{chakra08}}
\label{sect:chakra08}

\cite{chakra08} used the GC kinematic database presented in Paper\,I
to study the dark matter content of NGC\,4636 using the
non--parametric inverse algorithm CHASSIS \citep{chassis}. Their main
finding was that the dark halo required to explain the GC kinematics
is {very concentrated}.  While a high concentration parameter
$c_{\rm{vir}}>9$ as derived by \citeauthor{chakra08} is consistent with
our isotropic Jeans models for the blue GCs (which allow for a wide
range of concentration parameters), their estimate for the total mass
exceeds ours: The circular velocity curve shown in their Fig. ~6 (left
panel) rises to about $450\,\rm{km\,s}^{-1}$ at $\sim10\rm{kpc}$ and
then declines, reaching a value of $\sim370\,\rm{km\,s}^{-1}$ at
40\,kpc. Our mass models (shown in Fig.~\ref{fig:4636modelcompare}) ,
however, translate to significantly lower values of $v_c$ with
maximal values around $360\,\rm{km\,s}^{-1}$ (at
$R\simeq10\,\rm{kpc}$) and $300 \la v_c \la
340\,\rm{km\,s}^{-1}$ at 40 kpc. Recently, their work has been
complemented by an X--ray study which we discuss in the following
section.

\subsection{Comparison to the analysis by \citeauthor{johnson09}}

In their recent work on the X--ray halo of NGC\,4636 \cite{johnson09}
use a very detailed analysis of deep (80\,ks) archival Chandra data to
derive a mass profile which they compare to the dynamical modelling by
\cite{chakra08}. Again, the concentration parameters derived for the
NFW dark halo models are high, with values between 18 and 20.  A key
finding of their analysis is that the derived mass profile depends
strongly on whether the metal abundance gradient of the X--ray halo is
taken into account. While the overall shape of the mass profile
remains the same, the inclusion of the abundance gradient reduces the
mass at all radii by a factor of about 1.6 (see their Fig.~4).
Moreover, both models show the same behaviour for large radii where
the enclosed mass rises as $r^{1.2}$, a feature that was also found by
\cite{loewenstein03}.
To compare our dispersion measurements to the NFW profiles derived by
\cite{johnson09}, we proceed as follows: We calculate the velocity
dispersion profiles expected for the blue GCs for the isotropic case
($\beta=0$), adopting the NFW parameters given in their Sect.~4.2.\par
\citeauthor{johnson09} parametrise their NFW halos in terms of  concentration
 $c$ and the scale radius $r_s$.  Table\,\ref{tab:johnson}
lists their values together with the corresponding density $\rho_S$
and the virial parameters\footnote{Note that \cite{johnson09} use a
different definition of the virial parameters, i.e~$R_{200}=c\cdot
r_s$, where $R_{200}$ is the radius within which the mean density
equals 200 times the critical density of the Universe.}.

\begin{table}
\caption[]{Parameters of the NFW halos derived by \cite{johnson09}}
\centering
\resizebox{0.49\textwidth}{!}{
\begin{tabular}{llllll} \hline \hline
Model & $r_s$ & $c$ &$R_{200}$ & $\varrho_s$ & $M_{200}$\\
 & [kpc] &  &[kpc] & $[M_\odot\,\rm{pc}^{-3}]$ & $[10^{12}\,M_\odot]$\\
(1) & (2) & (3) &(4) &(5) & (6)\\\hline
%J0 & $33.7\pm3.7$& $9.0$ & $303\pm33$ & $0.00481$ & $3.2^{+1.2}_{-0.9}$\\
J1 & $21.8\pm0.9$ & $20.1\pm 0.8$& $438\pm25$ & $0.0359\pm 0.0036$& $9.8^{+2.6}_{-2.2}$ \rule{0ex}{2.5ex}\\
J2 &$24.6\pm0.9$ & $18.0\pm0.6$ & $443\pm22$ & $0.0270\pm0.0023$ &$10.0^{+2.4}_{-1.9}$ \rule{0ex}{2.5ex} \\
J3 & $26.1\pm1.0$ & $14.4\pm0.4$ & $376\pm18$ & $0.0154\pm0.0011$ & $6.2^{+1.3}_{-1.1}$\rule{0ex}{2.5ex} \rule[-1ex]{0ex}{0.5ex}\\ \hline \hline
\end{tabular}
} \note{The first Col. labels the models in order of their appearance
in Sect.~4.2 of \cite{johnson09}, where
%J0 is the model from GC
%analysis (total mass) by \cite{chakra08} (no uncertainty quoted for
%$c$);
J1: X--ray data, total mass; J2: X--ray data, stars subtracted;
J3: X--ray profile including metal abundance gradient, stars
subtracted.  Columns\,2 and 3 are the parameters quoted by these
authors, Col.~4 gives $R_{200}$ in units of kpc. The corresponding
values for the density $\varrho_s$ and $M_{200}$, i.e.~the enclosed
mass at $R_{200}$ are given in Cols.~5 and 6.  }
\label{tab:johnson}
\end{table}

\par
For consistency, we adopt for these calculations a distance of
$16\,\rm{Mpc}$, i.e.~the value used by \citeauthor{chakra08} and
\citeauthor{johnson09}.  Using the NFW parameters given in
Table\,\ref{tab:johnson}, we compute the expected velocity dispersion
profiles. \par These models for the blue GCs are compared to the
observations in the upper panel of Fig.~\ref{fig:johnson}. Since
\citeauthor{johnson09} assumed a very low stellar mass--to light
ratio, the difference between models J1 and J2, (i.e.~the X--ray mass
estimate without abundance gradient) before and after the subtraction of
the stellar component is small. The corresponding velocity
dispersions lie well above the data points.  
\par A much better
agreement between X--ray and GC based mass estimates is achieved when
the metal abundance gradient of the X--ray gas is taken into
account. Model J3 agrees, within the uncertainties, with the GC data
out to about 30\,kpc. For the abundance gradient corrected mass
profile shown in Fig.~4 of \citeauthor{johnson09}, one obtains a very
similar velocity dispersion profile\footnote{For this calculation we used 
a piecewise fit to the data in Fig.~4 of \citet{johnson09}}. The $r^{1.2}$
rise for large radii, however, leads to an almost constant velocity
dispersion profile for $R\ga40\,\rm{kpc}$. For reference, we also
plot, in Fig.~\ref{fig:johnson}, the best--fit isotropic NFW model derived
for the blue GCs assuming a distance of 16\,Mpc and the adjusted
$\Upsilon_{\star,R}=6.4$.

%===== Figure (22) ==========
\begin{figure}
\includegraphics[width= 0.49\textwidth]{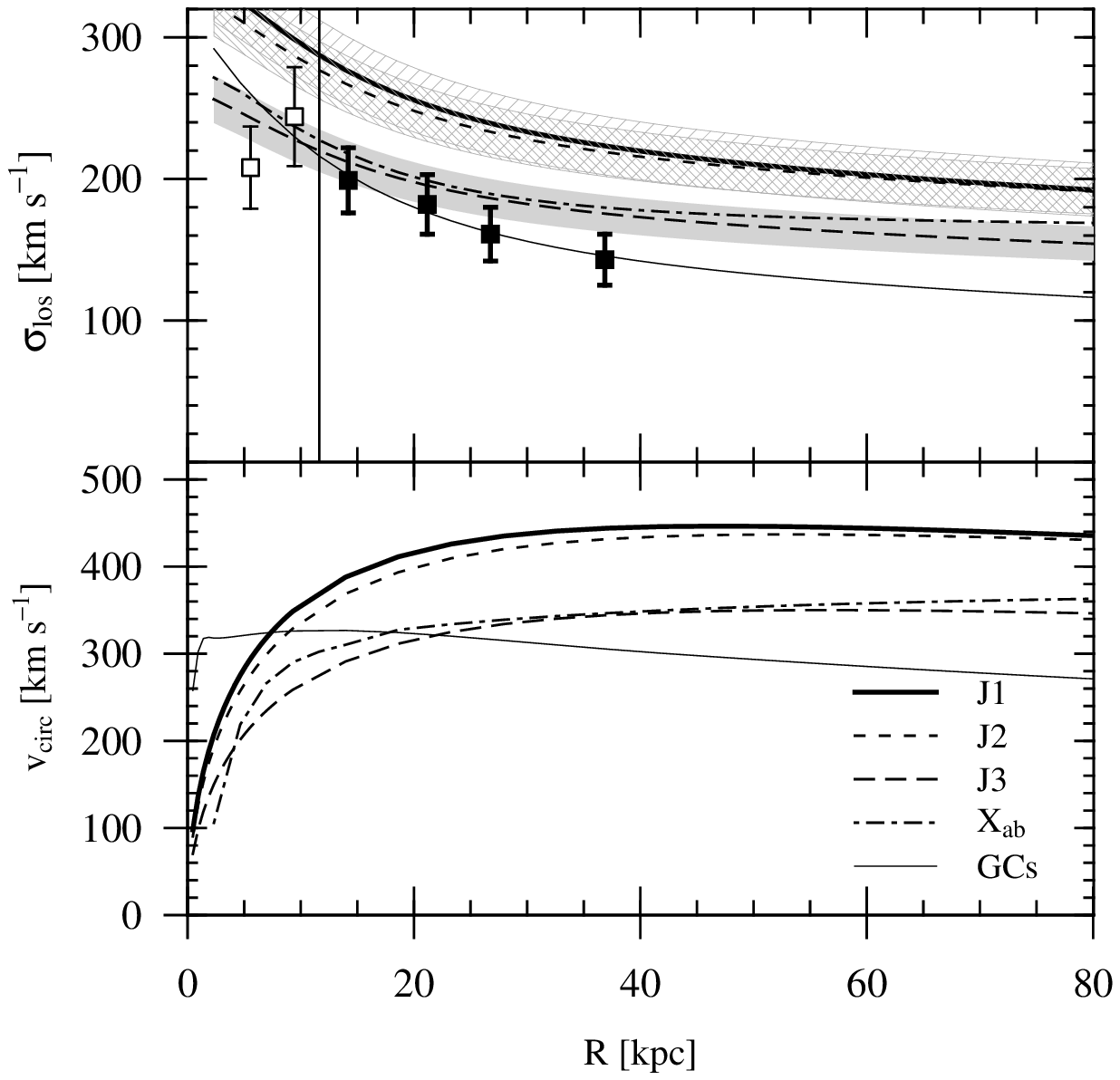}
\caption[]{Comparison to the NFW halos derived by \cite{johnson09} with the 
parameters listed in Table\,\ref{tab:johnson}. {\textbf{Upper panel:}} 
Velocity dispersion profiles. The data points show our final blue GC sample 
(same as in Fig.~\ref{fig:4636modelcompare}, left panel but for a distance of
16\,Mpc). The thick solid line shows model J1; model J2 is shown as 
short--dashed line, the dashed regions show the corresponding uncertainties. 
The long--dashed line is model J3, and the grey area shows the uncertainties. 
The dash--dotted line (labelled $X_{\rm ab}$) corresponds to the mass profile 
(incorporating the abundance gradient) shown in Fig.~4 of 
\citeauthor{johnson09} The thin solid line is the best--fit model for the GCs 
($r_s=20\,\rm{kpc}, \varrho_s=0.012$, $\Upsilon_{\star,R}=6.4$). 
{\textbf{Lower panel:}} Circular velocity curves. The line styles are the 
same as in the upper panel.}
\label{fig:johnson}
\end{figure}

\subsection{Are all GCs bound to NGC\,{4636}?}
\label{sect:bound}

Objects with velocities in excess of the escape velocity are probable
interlopers. Due to the logarithmic divergence of the NFW potential,
the escape velocity is not defined. But, in any spherical potential
bound particles travel on planar orbits, and energy and angular
momentum conservation are used to derive the following expression:
\begin{equation}
{v_p}^2 =  \frac{2\,{r_a}^2 (\Phi(r_p) -\Phi(r_a))}{{r_a}^2-{r_p}^2}\,,
\label{eq:rperi}
\end{equation}
where $v_p$ is the pericentric velocity, $r_p$ and $r_a$ are the
pericentre and apocentre distances, respectively.\\
The gravitational potential $\Phi(r)$ given by
\begin{equation}
\Phi(r) = -4 \pi \left[\frac{1}{\,r} \int^{r}_{0} \varrho(s) s^2\, {\mathrm{d}}s 
+ \int^{\,\infty}_{r} \varrho(s) s \,{\mathrm{d}}s 
\right]\,,
\label{eq:potential}
\end{equation}
where  $\varrho = \varrho_{\rm{stars}} + \varrho_{\rm{DM}}$ is
numerically integrated using the NGC 4636 stellar mass profile and the 
NFW halo dark
matter density profile obtained from the blue GCs (model {\it B.iso}).
\par  Objects
outside a given curve have apocentric distances larger than the
corresponding value of $r_a$.
The set of curves shown
in Fig.~\ref{fig:4636caust} 
is obtained from Eq.~\ref{eq:rperi} by
fixing 
\mbox{$r_{a}\in \{ 40,60,100,150,200,300 \}\rm{kpc})$}
\par
For the two blue GCs (objects 3.1:69 and 3.2:65) 
with good velocity measurements ($\mathrm v_{helio} = 1428\pm37 $ and
1441 $\pm$28$\,\rm{km\,s}^{-1}$, respectively)  at a galactocentric distance of $\approx$ 34 kpc, we find apocentric distances of
more than 150 kpc. Given that these conditions are extreme, an unknown
population of GCs with large apogalactic distances may be present in the
bulk of velocities.  However, the question whether these GCs are bound or 
unbound, cannot be answered. Recall that, even in the Milky Way system, some 
GCs have  Galactocentric distances of more than 100 kpc. 
How do these objects compare to the GCs with surprisingly
high relative velocities in the NGC 1399 GCS identified by
\citet{richtler04}? These authors show, in their Fig. 20, that
the objects in their sample of (about a dozen) GCs with heliocentric
velocities below 800 $\rm{km\,s}^{-1}$  (which corresponds to velocities
of at least 640 $ \rm{km\,s}^{-1}$ with respect to NGC 1399) have apogalactic
distances between 100 to 200 kpc (with one GC even featuring
ra Å 400 kpc). However, as shown in \citet{schuberth08},
blue, velocity-confirmed NGC 1399 GCs are still found at distances
of about 200 kpc. Thus, while the large number of GCs
with apogalactic distances of $r_a  \approx$  200 kpc might be surprising,
it is well possible that these objects belong to the very extended
NGC 1399 GCS. Another scenario (cf. \citealt{schuberth10}) is
that these metal-poor GCs were stripped from infalling galaxies
Ñwhich is not unlikely for a galaxy such as NGC 1399 which is
the central galaxy in a relatively dense cluster. In the NGC 1399
GCS, the most extreme combination of radial distance and velocity
is that of gc381.7 (from the catalogue of \citealt{bergond07}) which would 
have an apogalactic distance of the order 0.5
to 1\,Mpc (Schuberth et al. 2008), i.e. of at least twice the Fornax
cluster core radius.

%===== Figure (23) ==========
\begin{figure}
\centering
\includegraphics[width=0.49\textwidth]{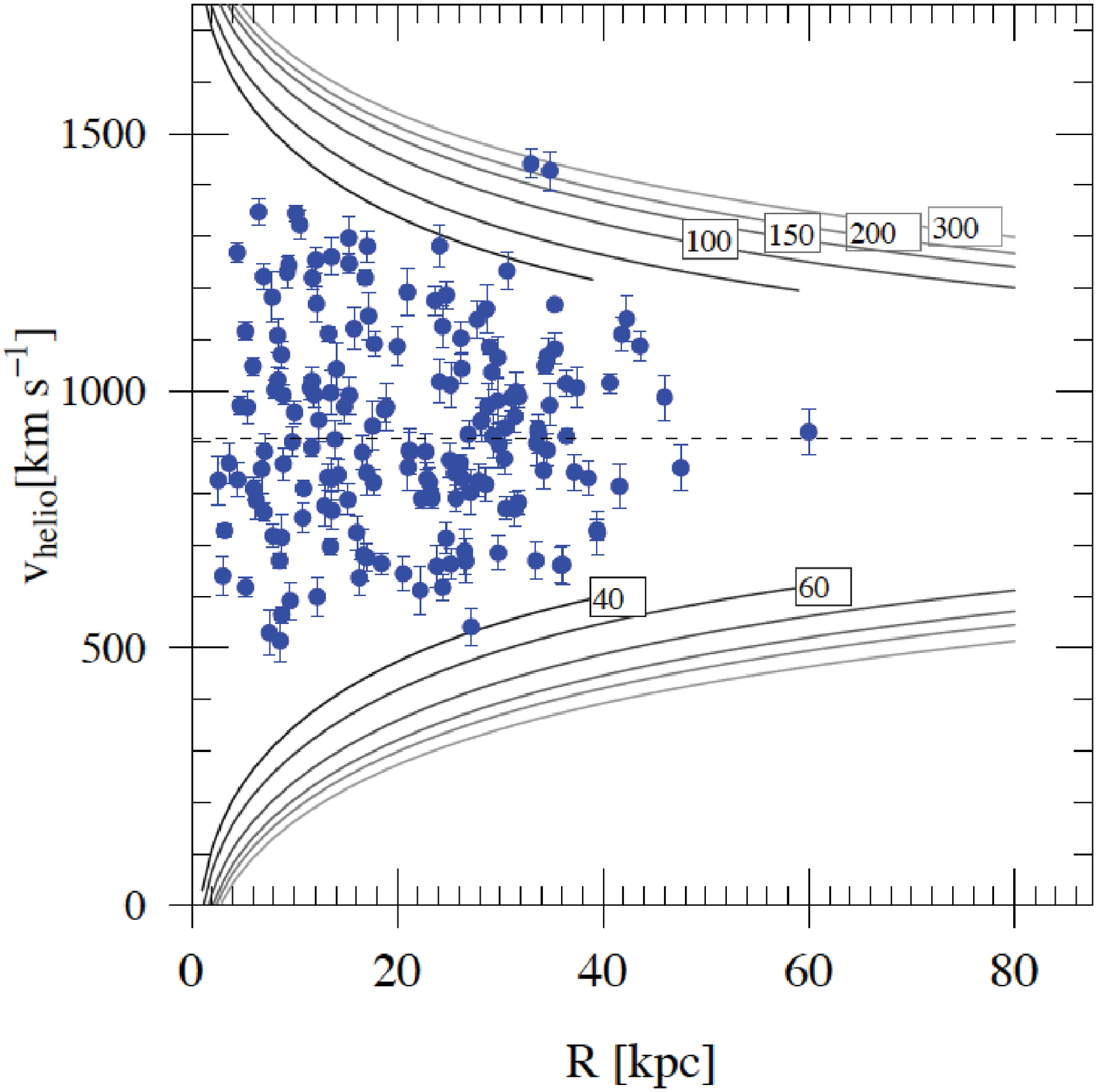}%&
\caption[]{Radial velocity versus galactocentric distance for blue GCs with
velocity uncertainties $\Delta v < 50\,\textrm{km\,s}^{-1}$. The curves,
calculated from Eq.~\ref{eq:rperi} for the total mass as given in model
{\it B.iso} (Table\,\ref{tab:4636nfwpars} NFW halo), indicate pericentric
velocities for fixed apocentric distances. Objects outside a given line have
apocentric distances of at least the value for which the curve was calculated.
}
\label{fig:4636caust}
\end{figure}

In the case of NGC 4636, the two blue GCs with extreme
velocities are remarkable. Although the estimated apogalactic
distances are smaller than those of the extreme GCs near
NGC 1399, one has to take into account that the NGC 4636 GCS
appears to be truncated, and that almost no GCs are found beyond
$\approx$ 60 kpc. Moreover, NGC 4636 is relatively isolated and
does not show any signs of recent major mergers \citep{tal09}.
This would make these objects candidates for a population
of `vagrant' GCs belonging to the Virgo cluster rather than
being genuine members of the NGC 4636 GCS.

\subsection{Stellar mass-to-light ratio} 

Our value of the stellar
$M/L_R$-value of $5.8$, which we need under isotropy to model the
central velocity dispersion data, is not directly supported by
theoretical single stellar population synthesis models.
%population synthesis of single stellar populations.
Depending on the adopted stellar initial mass function (IMF), models
predict values around 3.5--4 for metal-rich old populations 
\citep{bc03,thomas03,maraston03,percival09}
and can reach up to 4.7 for super--solar abundance even for IMFs with
a gentle slope in the mass--poor domain like the \cite{kroupa01} IMF.

\par Of course, one could surely find an appropriate radial bias which
at a given mass enhances the projected velocity dispersion in the
central regions and thus would permit a lower $M/L$--value. This sort
of fine--tuning is somewhat artificial and moreover is not supported
by observational evidence.  NGC\,4636 exhibits in the analysis of
\cite{kronawitter00} a tangential bias, though at larger radii, and is
isotropic in its inner region.

A good reference to stellar $M/L$ -values of the inner regions of
elliptical galaxies is the study of \cite{cappellari06}.  These
authors used the SAURON integral--field spectrograph to compare the
$I$--band dynamical $M/L$ with the ($M/L_{\textrm{pop}}$) obtained
from stellar population models for a sample of 25 early--type (E/S0)
galaxies. \newline For the 24 sample galaxies which lie in the
$I$--band magnitude range $-20 \la M_I \la -24$ they find a
correlation between the dynamical $M/L$ and the galaxy luminosity, in
the sense that $M/L$ weakly increases with luminosity as
$L_I^{\,0.32}$ (see.~their Fig.~9, and Eq.~9 for the fit). \par To
address the question whether these observed $M/L_{\textrm{dyn}}$
variations are due to a change in the stellar populations or to
differences in the dark matter fraction, \citeauthor{cappellari06} plot in
their Fig.~17 the dynamical $M/L_{\textrm{dyn}}$ as a function of
$M/L_{\textrm{pop}}$. Both quantities are correlated but while
$M/L_{\textrm{dyn}}\geq M/L_{\textrm{pop}}$ for all sample
galaxies\footnote{$M/L_{\textrm{pop}}$ was estimated using the
\cite{vazdekis96,vazdekis99} stellar population models assuming a
\cite{kroupa01} IMF, and the dynamical $M/L$ was
obtained from Schwarzschild modelling of the SAURON data.}, the data
points clearly lie off the one--to--one relation. The authors consider
the lower luminosity fast rotators and the high luminosity slowly
rotating galaxies separately (NGC\,4636 would belong to the latter
group). For old (age $> 7\,\textrm{Gyr}$) galaxies they find
that the dark matter fraction within one effective radius increases
from zero to about 30 per cent as the dynamical mass to light ratio
increases from 3 to 6. At a given ($M/L_{\textrm{pop}}$), the massive
slow rotators have higher dynamical $M/L$ values than the less massive
fast rotators.  How does NGC\,4636 fit into this picture?  \par For
NGC\,4636 the \cite{maraston03} SSP model predicts an
$M/L_{\textrm{pop},I}=3.27$ (for solar metallicity and an age of
$13\,\textrm{Gyr}$). Converting this to the $B$ and $R$ bands, one
obtains $M/L_{\textrm{pop},B}= 6.6$ and $M/L_{\textrm{pop},R}= 3.9$,
respectively. Gerhard et al. quote an even lower value
$M/L_{\textrm{pop},B}= 5.9$.  
\par What would we expect from the
relations given by \citeauthor{cappellari06}?

From Table 1, one obtains $M_{I,4636}=-23.3$, and (for $M_{I,\odot}=4.08$ 
\cite{lang99}) from Eq.~9 in \citeauthor{cappellari06}, we thus would expect 
a dynamical $M/L_I=4.7$, corresponding to $M/L_{\textrm{dyn},R}=5.7$.  
 
However, the models of \cite{cappellari06} have, by definition, a
radially constant M/L, while our M/L depends on radius and reaches
$M/L_{\textrm{dyn},R} \approx 8$ at the effective radius.

\citeauthor{cappellari06} speculate on the possibility that the
difference between dynamical and population $M/L$ is due to a higher
dark matter content of more luminous galaxies, but the general
question is whether it is appropriate to apply SSP models to composite
stellar systems. Let us consider $\omega$ Centauri, probably the
dynamically best investigated stellar system, which is unrelaxed and
composed of different populations. \cite{vandeven06} quote a $V$--band
M/L of $2.5 \pm 0.1$. The metallicity distribution of stars in
$\omega$ Cen has a maximum at [Fe/H]$\approx$-1.7 with a broad tail
towards higher metallicities (e.g. \citealt{hilker04,calamida09}).
The more metal-rich populations are probably
also younger by a few Gyr. From the population synthesis market, we
cite \cite{percival09} who quote $2.3$ as the value for a population
with [Fe/H]$=-1.7$ dex and an age of 13.5 Gyr, and $2.0$ for a population
with [Fe/H]$=-1.3$ dex and an age of 10 Gyr, adopting a Kroupa
IMF. Without aiming at precision, the composite `population' $M/L$
will probably not reach the dynamical value of $2.5$, unless there are
old metal--rich populations, for which there is no evidence, so $\omega$
Centauri is at least a mild example without dark matter, where the
dynamical mass is larger than the population mass.

However, an elliptical galaxy is a composite system with a long and
complicated star formation history.  If the IMF in a local star
formation event is universal, there is no guarantee that the final
mass function in a galaxy bears the same universality. Star formation
occurs in star clusters and a galaxy's field population is composed of
dissolved star clusters.  If the mass spectrum of star clusters is a
power-law like $m^{-2}$ then the dissolved population is the result of
adding up many low-mass clusters, but fewer high-mass clusters, where
the full stellar mass spectrum can be expected. \cite{weidner06}
showed that, if the maximum stellar mass within a cluster depends on
the clusters' mass, the resulting stellar mass function can be even
steeper than a Salpeter mass function. There are no simulations of the
final $M/L$ of an elliptical galaxy available, but since a
Salpeter-like mass function increases the $M/L$ by factor of roughly
1.4 (e.g. \citealt{cappellari06}), it is plausible that there is not a
strict universality of stellar mass functions among galaxies, but that
the stellar mass function of an old elliptical galaxy may depend on the
history of its assembly.
In conclusion, a stellar $M/L_R$-ratio of $5.8$ might well represent the
stellar population.

Another consideration may be worthwhile:

If we require the M/L-values to agree with the SSP predictions, we need a
$M/L_R$ = 4 or smaller, lets say, 3.7. The dark halo, represented by a 
logarithmic halo, would assume parameters like $r_0$ = 1 kpc and $v_0$ = 250 
km/s. The central density of dark matter then is 3.5 $M_\odot/pc^3$ under 
isotropy, and equality of stellar mass and dark mass is reached already at a 
radius of about 3.5 kpc. The central projected velocity dispersion is
170 km/s for the stellar mass alone and 192 km/s for the total mass.
If that would be typical for elliptical galaxies (of which there is no 
evidence), scaling relations like the fundamental plane would dynamically be 
dominated by dark matter and the `conspiracy' between dark and luminous 
matter would reach a level even more difficult to understand than it is now.

Finding such a high central dark matter density prompts us to consider an 
older argument brought forward by \cite{gerhard01}: the dark halos of 
elliptical galaxies in their sample turned out to exhibit a central density
which is higher by a factor of at least 25 than those of spiral galaxies of 
similar luminosity, and also that the phase space densities are higher. 
Since in collisionless merging ev ents phase space densities cannot grow, 
\citeauthor{gerhard01} argued that it is unlikely that dark halos of 
ellipticals formed by the merging of dark halos of present-day spirals.

Gerhard et al.'s expression for the phase space density reads $f_h =
2^{3/2} \rho_h/ v_h^3$, being $\rho_h$ the central density and $v_h^3$
the characteristic halo velocity of a logarithmic potential.

The above hypothetical dark matter density is about a factor 1000 higher than 
that of spirals (see Fig.\,18 of \citealt{gerhard01}) and the phase space 
density of the corresponding halo is $6.7 \times 10^{-7}$ in units solar 
masses, pc, km/s, much higher than those of spirals. We conclude that with 
our example low M/L, it might not be possible to reach these densities and 
phase space densities by collionless accretion of spiral-like halos, and one 
has to resort to dark halos resembling those of dwarf spheroidals.

\subsection{MOND related issues}

The question whether elliptical galaxies fulfill the predictions of
Modified Newtonian Dynamics (MOND, see e.g. Milgrom 2009, Sanders \& McGaugh 
2002) obviously is a fundamental one. Ellipticals have so far been less in 
the focus of MOND than disk galaxies. The most compelling case of an 
apparently  MONDian elliptical galaxy is the E4 galaxy NGC\,2974
\citep{weijmans08}
where the extended H\textsc{i} disk permits a secure determination of
the circular velocity which is constant out to 20\,kpc.
\par In Paper\,I, we noted already that NGC\,4636 
seems to be consistent with being MONDian.
Here, we plot, in the lower subpanel of Fig.\ref{fig:donato} the MOND circular
velocity curve (shown as thick solid line) for the stellar mass profile of 
NGC\,4636 (for $\rm M/L_{\star,R}$=5.8). The MOND circular velocity curve is
obtained from the Newtonian one via the following equation:
\begin{equation}
V^2_{circ,M} = \frac{V^2_{circ,N}}{2} + \sqrt{ \frac{V^4_{circ,N}}{4} +V^2_{circ,N} \cdot a_0
\cdot r }
\label{eq:mond}
\end{equation}
where $\rm V_{circ,N}$ is the Newtonian velocity. For $\rm a_0$ we adopt the 
value recommended by \cite{famaey07}: $\rm 1.35 \times 10^{-8} cm s^{-2}$.
Within the central $\approx$ 40 kpc, i.e. the radial range, for which we have 
data, the MOND circular velocity curve agrees fairly well with the best-fit 
joint model for stars and the blue GCs (models S.B.iso, Burkert halo, shown 
as thin dashed line).
 
Below we put NGC 4636 into the context of the more recent literature.

\subsubsection{Comparison to the $\mu_{0D}$--relation by \citeauthor{donato09}}

Recently \cite{donato09}, in their extension of the work by
\cite{kormendy04}, confirmed that the central surface density of
galaxy dark matter halos is nearly constant and almost independent of
galaxy luminosity. MOND-related aspects of this finding have been discussed by
\cite{gentile09} and \cite{milgrom09}.
\citeauthor{donato09}  assume that the dark matter
halos of the galaxies are described by \cite{burkert95} halos
(cf.~Eq.~\ref{eq:rhoburkert} and \ref{eq:MassBurkert}).

For the DM surface density $\mu_{0D}\equiv r_0 \varrho_0$,
\citeauthor{donato09} find the following relation to hold for galaxies
in the magnitude range $-8 \geq M_B \geq -22$:
\begin{equation}
\log\left(\frac{\mu_{0D}}{M_\odot\rm{pc}^{-2}}\right)=2.15\pm0.2 \;.
\label{eq:donato}
\end{equation}
How does NGC\,4636 fit into this picture? 
 From the values given
in Table\,\ref{tab:4636nfwpars}, it appears that the dark matter
density of our Burkert halo models (for the blue GCs and the stars) 
($2.50\la\log(\mu_{0D}/M_\odot\rm{pc}^{-2})\la 2.90$) is too
high with respect to the above value.

\par Going back to
Fig.~\ref{fig:burkert}, however, one sees that for $r_0\ga
20\,\rm{kpc}$ the 68 per cent CL contour (thick solid line) of our
isotropic models for the blue GCs lies within the range of values from
Eq.~\ref{eq:donato} (shown as dot--dashed lines). In the upper panel
of Fig.~\ref{fig:donato} we show, as an example, isotropic Jeans
models for the blue GCs and the stellar component for
$r_0=20\,\rm{kpc}$, $\varrho_0=1.115\times 10^{-2}$,
$\log\mu_{0.D}=2.35$ (solid lines) , i.e.~a \citeauthor{burkert95}
halo which is consistent (within the uncertainties) with the relation
from \citealt{donato09}. For reference, the best--fit joint model is
shown with dashed lines. The lower panel of Fig.~\ref{fig:donato}
shows the corresponding circular velocity curves. Regarding the stars,
we find $\chi^2$ values of $\chi^2_{\rm{joint}}=2.0$ and
$\chi^2_{\rm{donato}}=2.9$ for the joint model and the model which is
consistent with Eq.~\ref{eq:donato}, respectively. For the blue GCs,
the respective values are $\chi^2_{\rm{joint}}< 0.1$ and
$\chi^2_{\rm{donato}}=0.7$. \par We conclude that the GC and stellar
dynamics of NGC\,4636 do not contradict the constant dark matter
density relation, but we caution that the halos of 
ellipticals in the \citeauthor{donato09} sample are constrained only by weak
lensing shears, which do not probe the inner regions.

\citeauthor{gerhard01}, for their sample of ellipticals, implicitly found
(multiplying their equations (6) and (8)) $\rho_0 \cdot r_0 = 635
L_{11}^{-0.1} h_{0.65}$, which for all practical purposes is constant
(they used logarithmic halos instead of Burkert halos). This fits
better to our value and the question, whether disk galaxies and
ellipticals really show the same surface density of dark matter, 
remains open.

%===== Figure (24) ==========
\begin{figure}
\includegraphics[width=0.49\textwidth]{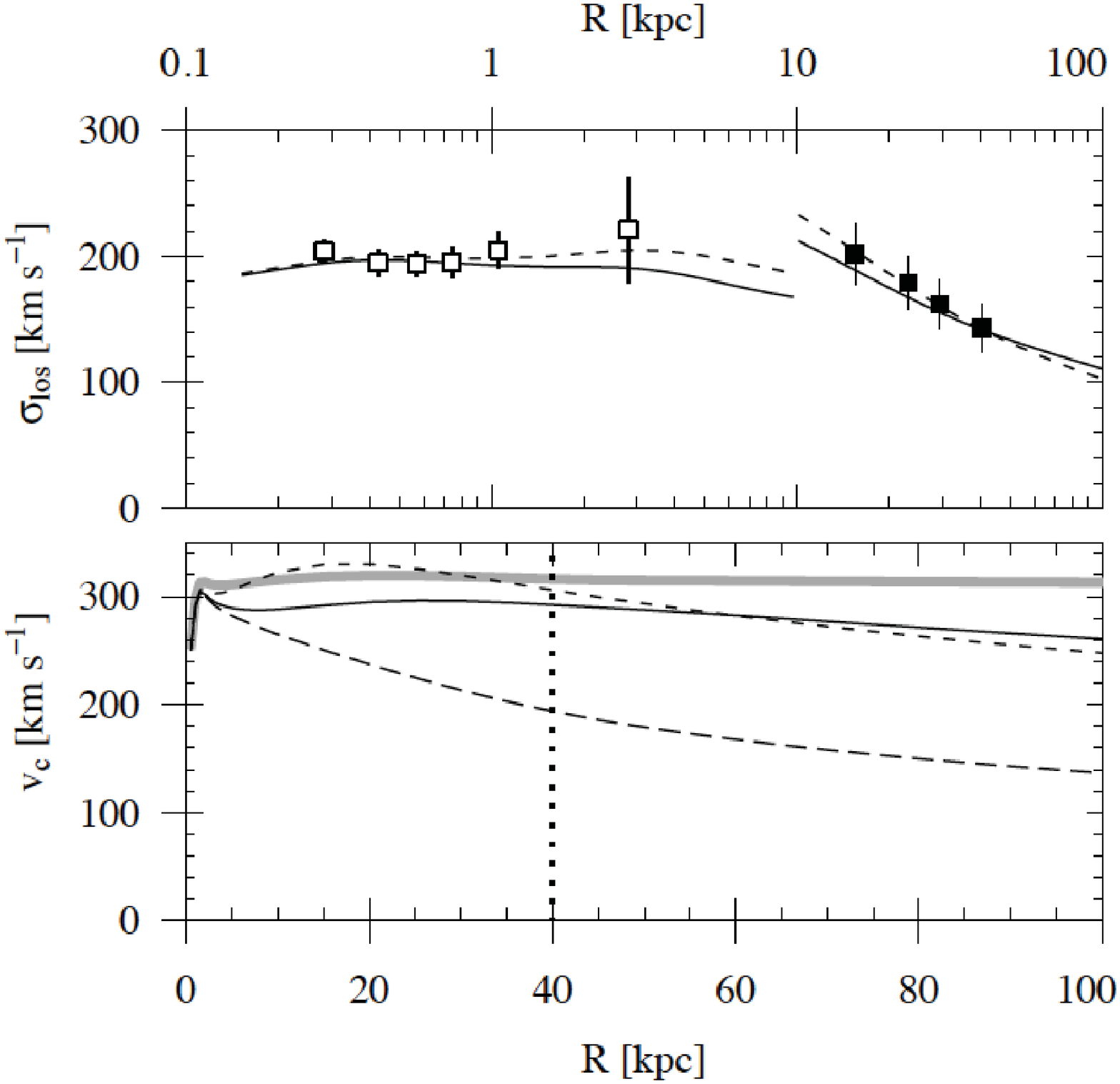}
\caption[]{Comparison to the results from \cite{donato09}. {\textbf{Upper 
panel:}} Open squares show the stellar velocity dispersion profile (BSG94), 
filles squares are the blue GCs ({\it{BlueFinal}}). The velocity dispersion 
profile for the joint isotropic model (blue GCs and stars, {\it S.B.iso}, 
Burkert halo) is shown as dashed line. The solid line shows the dispersion 
for a Burkert dark matter halo $r_0=20\,\rm{kpc}$, $\varrho_0=1.115\times 
10^{-2}$ (i.e.~$\log\mu_{0.D}=2.35$) which is consistent with the relation
by \cite{donato09} (see text for details). {\textbf{Lower panel:}} Circular 
velocity curves. Again, the thin solid line shows the curve for the halo with
$\log_{0D}=2.35$, and the dashed line is the joint model. For reference, the 
circular velocity curve for the stars alone ($\Upsilon_{\star,R}=5.8$) is 
shown as long--dashed line. The MOND circular velocity curve 
(Eq.\,\ref{eq:mond}) is shown as a thick solid line. The dotted vertical line 
at 40\,kpc indicates the location of the outermost velocity dispersion bin for
the blue GCs.}
\label{fig:donato}
\end{figure}

\subsubsection{The baryonic Tully--Fisher relation and NGC 4636}

A severe irritation to the $\Lambda$CDM paradigm on galactic scales is
the existence of a baryonic Tully--Fisher relation (BTFR) among
spiral galaxies with an astonishingly small scatter, covering five
orders of magnitude in mass, which reads \citep{mcgaugh05}:
\begin{equation}
 M_{\textrm{bar}} =
 50 \cdot v_{\textrm{flat}}^4 \; ,
\label{eq:BTFR}
\end{equation}
 where  $v_{\textrm{flat}}$ is the
 circular velocity (in units of $\textrm{km\,s}^{-1}$) at a radius
 where the rotation curve becomes flat, and 
 $M_{\textrm{bar}}$ is the total mass in
 baryons (in units of $M_\odot$).  Such a relation would
 naturally result from Modified Newtonian Dynamics (MOND,  
 e.g.~\citealt{milgrom83}, \citealt{sanders02}).
In the deep
 MOND regime: 
\begin{equation}
M = \frac{v_c^4 \cdot a_0}{G}\;,
\end{equation}
 $M$ being the total mass, $G$ the gravitational constant, and $a_0$
 the MOND constant, which has the value
 $1.35^{+0.28}_{-0.42}\times10^{-8} \textrm{cm\,s}^{-2}$
 \citep{famaey07}. The factor of 50 in the above relation
 (Eq.~\ref{eq:BTFR}) corresponds to a somewhat higher value of
 $a_0=1.5\times10^{-8}\,\textrm{cm\,s}^{-2}$ which still lies within
 the uncertainties of the value quoted above. 
The recent work of \cite{stark09} and \cite{trachternach09} confirmed
this relation, which in  \citeauthor{stark09} formally reads 
$M_{\textrm{bar}} =
 61 \cdot v_{\textrm{flat}}^{3.94}$ \, agreeing even better with the canonical
value for $a_0$.   

There also exists a BTFR for elliptical galaxies
\citep{gerhard01,magorrian01,thomas07}, but most elliptical galaxies
lie off the spiral relation \citep{gerhard01}.  Since the outermost
radii in these dynamical studies may be still on the declining part of
the circular velocity curves, it is interesting to put NGC 4636 into
this picture.  Although we cannot strictly distinguish between a
constant circular velocity curve and a slightly declining one, the
model with the flattest rotation curve has a circular velocity of
$300\,\textrm{km\,s}^{-1}$ and thus we expect a total baryonic mass of
$4 \times 10^{11} M_\odot$. With the data from Table 1, we have 
$M_R = -22.6$, thus $5.5 \times 10^{10} L_\odot$ and a total baryonic mass 
of $3.2 \times 10^{11} M_\odot$,
which would place NGC\,4636 a bit below the relation for spirals.
However, given the uncertainties in adopting distances, M/L-ratio and even
the absolute solar R-magnitude, we are reluctant to assess these value as
a clear displacement and repeat our conclusion from Paper I that NGC 4636
is consistent with the MONDian prediction.

In any way it is of fundamental interest to investigate  more elliptical 
galaxies at large radii. If ellipticals and spirals would follow the same 
BTF-relation in spite of very different formation histories, the challenge 
for the Cold Dark Matter paradigm of galaxy formation would be considerable.

\section{Conclusions}

We revisit the dynamics of the globular clusters system of NGC 4636 on the
basis of 289 new globular cluster (GC) velocities.
Including the data from \citeauthor{schuberth06} (\citeyear{schuberth06}, Paper\,I), our total sample now consists of
460 GC velocities, one of the largest sample
obtained until now for a non-central elliptical galaxy. In addition we
present  new kinematical stellar data extending in radius the analysis by
\cite{kronawitter00}.

We model the total mass profile by the sum of the stellar mass plus a dark halo,
for which we adopt different analytical forms. With our distance of 17.5 Mpc, we
need a stellar $M/L_R$-value of 5.8 to satisfactorily reproduce the projected
stellar velocity dispersion near the center under isotropy. This value is higher than
values predicted from canonical population synthesis of an old, metal-rich population,
resembling the results from the SAURON collaboration \citep{cappellari06}.
We argue, however that the actual stellar mass function of an elliptical galaxy might
be somewhat steeper than IMFs of local young star clusters.
%which sugges universality of
%the IMF. 

We perform  spherical Jeans-analyses independently for the red and the blue cluster
population and fit dark matter profile parameters for NFW-profiles, for Burkert profiles
and for logarithmic halos for different anisotropies. The fits using the red cluster
populations are consistently worse than using the blue populations, a finding, which differs
from our previous study of the central cluster galaxy NGC 1399 and for which we
do not have a good explanation. Our recommended 
joint logarithmic halo which uses the stellar light and the blue GCs, has the parameters $r_0$ = 8 kpc and $v_0$ = 237 km/s.

The higher moments of the velocity distributions of the blue and red GC subpopulations are not
really stable against the sample selections and thus do not permit to seriously constrain
possible GC orbit anisotropies. However, they are consistent with the GC orbits being isotropic to
a good approximation.

We compare our results with the mass profile, derived from X-ray
analysis, of \cite{johnson09}. When the element abundance gradient
of the X-ray gas is not taken into account, the X-ray mass profile
exceeds the GC mass profile by a significant factor. If the abundance
gradient is accounted for, the agreement is good out to 30 kpc, but
the X-ray mass profile still exceeds our mass profile beyond this
radius.  The might be a general problem of X-ray analyses, if strong
abundance gradients are present, for example in galaxy clusters.

NGC 4636 almost falls onto the baryonic Tully-Fisher relation for spirals and behaves
more or less MONDian, as already noted in PaperI. However, when modelled with a logarithmic
halo, its halo surface  density
resembles that of other elliptical galaxies.
 
\acknowledgements{
We thank an anonymous referee for constructive remarks and the editor for
support. This research has made use of the NASA/IPAC
Extragalactic Database (NED) which is operated by the Jet Propulsion
Laboratory, California Institute of Technology, under contract with the National
Aeronautics and Space Administration. T.R. acknowledges support from the
Chilean Center of Astrophysics FONDAP No. 15010003 and from FONDECYT 
project No. 1100620}
%-----------------------------------------------------------------------

\bibliographystyle{aa} % style aa.bst
\bibliography{schuberth12.bib} % your references Yourfile.bib

\appendix

\section{The velocity dispersion profile of NGC\,4636}
\label{sect:4636table}

\begin{table}
\caption{NGC\,4636 velocity dispersion profile from FORS\,2 data}
\label{tab:4636table}
\centering
\begin{tabular}{lrr@{$\pm$}l}
Slit ID         &\multicolumn{1}{c}{$R$}        &\multicolumn{2}{c}{$\sigma$}   \\ \hline \hline
        &$[\arcsec]$    &\multicolumn{2}{c}{$[\rm{km\,s}^{-1}]$}\\  
104     &67.9   &185    &24 \\
103     &64.7   &142    &25 \\
100     &54.8   &147    &22  \\
99      &51.3   &186    &18 \\
98      &47.9   &197    &20 \\
97      &44.7   &213    &17 \\
96      &41.2   &178    &13   \\
95      &37.6   &199    &11 \\
94      &34.1   &179    &14 \\
89      &13.7   &197    &10  \\
86      &4.8    &201    &7    \\
85      &0.7    &233    &7  \\
84      &-2.5   &199    &7   \\
83      &-5.5   &192    &8 \\
82      &-8.7   &186    &8 \\
81      &-12.2  &199    &9 \\
80      &-15.3  &196    &9 \\
79      &-18.2  &201    &9 \\
78      &-21.1  &202    &10 \\
75      &-30.8  &212    &10 \\
74      &-33.7  &192    &16 \\
73      &-36.5  &198    &16 \\
70      &-47.5  &182    &13 \\
69      &-51.0  &207    &23 \\
68      &-54.5  &202    &20 \\
67      &-57.9  &171    &20 \\
66      &-61.1  &207    &21 \\
65      &-64.6  &178    &18 \\
64      &-67.4  &145    &23 \\ \hline \hline
\end{tabular}
\note{Column\,(1) gives the slit number on
  the FORS\,2 Mask\,1\_1 from Paper\,I. The galactocentric distance
  $R$ (in units of arcseconds) is given in the second column. Positive
  and negative values of $R$ refer to positions to the north and south
  of the centre of NGC\,4636 respectively. Column\,(3) lists the
  velocity dispersion (in $\rm{km\,s}^{-1}$)}.
\end{table}

\section{The stellar mass profile of NGC\,4636}
\label{sect:masspoly}
Here we give the piecewise approximation of the stellar mass profile
of NGC\,4636 (see Sect.~\ref{sect:dep4636}) used in our modelling.  In
the following expressions, $x$ is in units of parsec, $D$ is the
distance of NGC\,4636 in Mpc (assumed to be D=17.5 in our modelling),
and $\Upsilon_{\star,R}$ ($=5.8$) is the $R$--band stellar
mass--to--light ratio.
%uber0 <- 3.0
%uber1 <- 29.18354816
%uber2 <- 104.4893857
%uber3 <- 2597.930969
%] "uber1"
%[1] "29.184"
%[1] "uber2"
%[1] "104.49"
%[1] "uber3"
%[1] "2597.9"

for $0<x\leq 3.0\cdot D$: 
\begin{equation}
p_o= \Upsilon_{\star,R}  \left(\frac{D}{15}\right)^2 \cdot 5.74\times10^6\;, %\sum_{i=0}^4 a_i \left(\frac{x}{D}\right)^i\
\end{equation}
%where $const=$

for $3.0\cdot D<x\leq 29.184  \cdot D$: 
\begin{equation}
p_1(x)= \Upsilon_{\star,R}  \left(\frac{D}{15}\right)^2 \cdot \sum_{i=0}^4 a_i \left(\frac{x}{D}\right)^i\;,
\end{equation}
where $a_0= 1.322\times10^7$,
$a_1=-1.118\times10^7$,
$a_2=3.158\times10^6$,
$a_3=-9.127\times10^4$, and
$a_4= 1.164\times10^3$.
%[1] "1.322e+07"
%[1] "-1.118e+07"
%[1] "3.158e+06"
%[1] "-9.127e+04"
%[1] " 1164"

for $29.184\cdot D<x\leq  104.49 \cdot D$: 
\begin{equation}
p_2(x)= \Upsilon_{\star,R}  \left(\frac{D}{15}\right)^2 \cdot \sum_{i=1}^4 b_i \left(\frac{x}{D}\right)^i \;,
\end{equation}
where
$b_1=9.616\times10^6$,
$b_2=1.034\times10^6$,
$b_3=-9.133\times10^3$, and
$b_4=25.78$.

%[1] "b1"
%[1] "9.616e+06"
%[1] "b2"
%[1] "1.034e+06"
%[1] "b3"
%[1] "-9133"
%[1] "b4"
%[1] "25.78"
for $ 104.49 \cdot D<x\leq 2597.9 \cdot D$: 
\begin{equation}
p_3(x)= \Upsilon_{\star,R}  \left(\frac{D}{15}\right)^2 \cdot \sum_{i=0}^3 c_i \left(\frac{x}{D}\right)^i \;, 
\end{equation}
where
$c_0=6.101\times10^8$, 
$c_1=4.31\times10^7$, 
$c_2-1.507\times10^4$, and
$c_3=1.996$

%1] "c0"
%[1] "6.101e+08"
%[1] "c1"
%[1] "4.31e+07"
%[1] "c2"
%[1] "-1.507e+04"
%[1] "c3"

%[1] "1.996"
for $x>  2597.9 \cdot D$
\begin{equation}
% d1*atan(%
%                                          (%
%                                           (x/DIST)/abs(d0))+ 
%                                          d2*pow(((x/DIST)/d0),2.0) +
%                                          d3*pow(((x/DIST)/abs(d0)),3.0)%
%                                          )    
A(x)=d_1\cdot\arctan\left(  \left( \frac{x/D}{d_0} \right) +d_2\cdot \left(\frac{x/D}{d_0} \right)^2 
+d_3\cdot \left(\frac{x/D}{d_0} \right)^3 
\right)
\end{equation}
where 

$d_0=1.194\times10^3$, 
$d_1=4.221\times10^{10}$,
$d2=-0.0668$,
and $d3=0.003774$.

%[1] "++++++++++++"
%[1] "d0"
%[1] " 1194"
%[1] "d1"
%[1] "4.221e+10"
%[1] "d2"
%[1] "-0.0668"
%[1] "d3"
%[1] "0.003774"

\section{Globular cluster velocities}

%===== TABLE (ANHANG) ==========

\onltab{2}{
\begin{landscape}
\begin{longtable}{lllllllr@{$\pm$}lr@{$\pm$}llll}
%\begin{table*}
%\centering
\caption{\label{tab:veltab}
Catalogue of globular cluster velocities. Column (1) gives the spectrum
identifier ({\it mask:slit}). Note that the new data have the prefix ``f''. For 
the purpose of cross-referencing, the data from Paper\,I are also included in 
this table.  Columns (2) and (3) are the right ascension and declination, 
respectively (for objects with Washington photometry, the coordinates are those 
from Dirsch et al. 2005; for objects not in the photometric catalogue, the 
values were estimated from the pre--images). The projected distance from the 
centre of NGC\,4636 (in units of arcminutes) is given in Col. (4). Column (5) 
is the position angle, (measured North over East). Columns (6) and (7) are the 
magnitude and colour as given by Dirsch et al. (2005). For the new objects, 
Column (8) gives the heliocentric velocity measured for a given spectrum, for 
the data from Paper\,I, the final velocity given there is quoted. Column (9) 
gives the final velocity used in this study, for objects for which more than 
one spectrum was obtained, the error--weighted mean is given. Column (10) 
indicates the subsample (``R.25'' denotes GCs with projected distances 
$R\leq2\farcmin5$). Column (11) gives the cross--identifications within this 
list. The entry in Col. (12) is `1' for the first appearance of an object in 
this table. All subsequent entries for the same object are labelled `0'.}\\
%\label{tab:veltab}
%\begin{tabular}{lllllllr@{$\pm$}lr@{$\pm$}llll} 
\hline\hline
ID & R.A. & Dec. & $R_{\rm{4636}}$ & PA & $m_R$ & $C-R$ & \multicolumn{2}{c}{$v_{\rm{spec}}$} & \multicolumn{2}{c}{$v_{\rm{final}}$} & Sample & Cross-Identification(s) & 1$^{\rm{st}}$ \\
 & (J2000) & (J2000) & [arcmin] & [deg] & [mag] & [mag] & \multicolumn{2}{c}{$[\rm{km\,s}^{-1}]$} & \multicolumn{2}{c}{$[\rm{km\,s}^{-1}]$} & & & \\
(1) & (2) & (3) & (4) & (5) & (6) & (7) & \multicolumn{2}{c}{(8)} & \multicolumn{2}{c}{(9)} & (10) & (11) & (12) \\
\hline
\endfirsthead
\caption{continued.}\\
\hline\hline
ID & R.A. & Dec. & $R_{\rm{4636}}$ & PA & $m_R$ & $C-R$ & \multicolumn{2}{c}{$v_{\rm{spec}}$} & \multicolumn{2}{c}{$v_{\rm{final}}$} & Sample & Cross-Identification(s) & 1$^{\rm{st}}$ \\
 & (J2000) & (J2000) & [arcmin] & [deg] & [mag] & [mag] & \multicolumn{2}{c}{$[\rm{km\,s}^{-1}]$} & \multicolumn{2}{c}{$[\rm{km\,s}^{-1}]$} & & & \\
\hline
\endhead
\hline
\endfoot
f01-03 & 12:43:04.83 & +02:33:06.8 & 8.97 & 155.4 & 22.01 & 1.22 & 584&66  &  584&66  & Blue & & 1 \\
f01-07 & 12:42:58.79 & +02:33:30.4 & 8.07 & 164.0 & 22.20 & 1.77 & 871&34  &  899&53  & RedFinal & f06-15 & 1 \\
f01-09 & 12:43:02.89 & +02:33:46.0 & 8.17 & 156.6 & 21.14 & 1.08 & 814&43  &  814&43  & BlueFinal & & 1 \\
f01-11 & 12:43:07.49 & +02:34:04.5 & 8.43 & 148.5 & 21.55 & 1.75 & 573&18  &  573&18  & RedFinal & & 1 \\
f01-14 & 12:43:02.07 & +02:34:33.2 & 7.37 & 155.6 & 21.88 & 1.17 & 847&40  &  859&74  & Blue & f06-25 & 1 \\
f01-15 & 12:43:03.79 & +02:34:43.3 & 7.41 & 152.0 & 21.26 & 1.79 & 840&21  &  840&21  & RedFinal & & 1 \\
f01-21 & 12:43:06.53 & +02:35:26.3 & 7.16 & 144.5 & 20.63 & 1.37 & 912&15  &  912&15  & BlueFinal & & 1 \\
f01-24 & 12:43:03.56 & +02:36:02.6 & 6.24 & 146.8 & 22.14 & 1.89 & 995&36  &  995&36  & RedFinal & & 1 \\
\end{longtable}
\end{landscape}
}

\section{Foreground star velocities}

\onltab{3}{
\begin{landscape}
\begin{longtable}{lllllllr@{$\pm$}lr@{$\pm$}lll}
%\begin{table*}
%\centering
\caption{\label{tab:startab}
Catalogue of foreground star velocities. Column (1) gives the spectrum
identifier ({\it mask:slit}). Note that the new data have the prefix ``f''. 
For the purpose of cross-referencing, the data from Paper\,I are also 
included in this table. Columns (2) and (3) are the right ascension and 
declination, respectively (for objects with Washington photometry, the 
coordinates are those from Dirsch et al. 2005; for objects not in the 
photometric catalogue, the values were estimated from the pre-images). 
The projected distance from the centre of NGC\,4636 (in units of 
arcminutes) is given in Col. (4). Column (5) is the position angle, 
(measured North over East). Columns (6) and (7) are the magnitude and 
colour as given by Dirsch et al. (2005). For the new objects, Column (8) 
gives the heliocentric velocity measured for a given spectrum, for the data 
from Paper\,I, the final velocity given there is quoted. Column (9) gives 
the final velocity used in this study, for objects for which more than one 
spectrum was obtained, the error-weighted mean is given. Column (10) gives 
the cross-identifications within this list The entry in Col. (11) is `1' 
for the first appearance of an object in this table. All subsequent entries 
for the same object are labelled `0'.}\\
%\label{tab:startab}
%\begin{tabular}{lllllllr@{$\pm$}lr@{$\pm$}lll} 
\hline\hline
ID & R.A. & Dec. & $R_{\rm{4636}}$ & PA & $m_R$ & $C-R$ & \multicolumn{2}{c}{$v_{\rm{spec}}$} & \multicolumn{2}{c}{$v_{\rm{final}}$} & Cross-Identif
ication(s) & 1$^{\rm{st}}$ \\
 & (J2000) & (J2000) & [arcmin] & [deg] & [mag] & [mag] & \multicolumn{2}{c}{$[\rm{km\,s}^{-1}]$} & \multicolumn{2}{c}{$[\rm{km\,s}^{-1}]$} & & \\
(1) & (2) & (3) & (4) & (5) & (6) & (7) & \multicolumn{2}{c}{(8)} & \multicolumn{2}{c}{(9)} & (10) & (11) \\
\hline
\endfirsthead
\caption{continued.}\\
\hline\hline
ID & R.A. & Dec. & $R_{\rm{4636}}$ & PA & $m_R$ & $C-R$ & \multicolumn{2}{c}{$v_{\rm{spec}}$} & \multicolumn{2}{c}{$v_{\rm{final}}$} & Cross-Identif
ication(s) & 1$^{\rm{st}}$ \\
 & (J2000) & (J2000) & [arcmin] & [deg] & [mag] & [mag] & \multicolumn{2}{c}{$[\rm{km\,s}^{-1}]$} & \multicolumn{2}{c}{$[\rm{km\,s}^{-1}]$} & & \\
\hline
\endhead
\hline
\endfoot
f01-01 & 12:43:07.15 & +02:32:54.3 &   9.41 &  152.7 & 19.91 & 0.79 & 249&35  & 249&35 & & 1 \\
f01-13 & 12:43:09.22 & +02:34:27.0 &   8.36 &  144.7 & 22.04 & 3.46 & -68&53  & -68&53 & & 1 \\
f01-18 & 12:43:07.75 & +02:35:02.3 &   7.66 &  144.4 & 21.91 & 2.46 & -167&58 & -167&58 & & 1 \\
f01-19 & 12:43:06.68 & +02:35:08.9 &   7.42 &  145.5 & 21.86 & 2.70 & 86&41   & 86&41 & & 1 \\
f01-23 & 12:43:07.00 & +02:35:39.2 &   7.06 &  142.7 & 21.86 & 1.72 & -23&29  & -23&29 & & 1 \\
f01-27 & 12:43:10.05 & +02:36:25.4 &   6.99 &  133.9 & 18.55 & 1.24 & -44&13  & -51&21 & 6.2:9 & 1 \\
f01-29 & 12:42:59.66 & +02:36:40.3 &   5.21 &  152.0 & . . . & . . .&  -83&46 & -46&55 & f06-44 & 1 \\
f01-39 & 12:43:10.33 & +02:37:39.9 &   6.25 &  125.2 & 22.05 & 0.67 & 68&58   & 22&98 & f12-06 & 1 \\
\end{longtable}
\end{landscape}
}

\end{document}